\documentclass[journal]{IEEEtran}

\usepackage{amsmath,amsfonts}
\usepackage{algorithmic}
\usepackage{algorithm}
\usepackage{array}
\usepackage[caption=false,font=normalsize,labelfont=sf,textfont=sf]{subfig}
\usepackage{textcomp}
\usepackage{stfloats}
\usepackage{url}
\usepackage{verbatim}
\usepackage{graphicx}
\usepackage{cite}

\usepackage[graphicx]{realboxes}

\usepackage[acronym]{glossaries}
\newacronym{1L}{1L}{One-layer}
\newacronym{2D}{2D}{Two Dimentional}
\newacronym{2L}{2L}{Two-layer}
\newacronym{5G}{5G}{Fifth Generation}
\newacronym{6G}{6G}{Sixth Generation}
\newacronym{ADC}{ADC}{Analog-to-Digital Converter}
\newacronym{AF}{AF}{Amplify-and-Forward}
\newacronym{AIRS}{AIRS}{Aerial Inelligent Reflecting Surfaces}
\newacronym{AO}{AO}{Alternating Optimization}
\newacronym{AOP}{AOP}{Average Outage Probability}
\newacronym{ASTARS}{ASTARS}{Active Simultaneously Transmitting/Reflecting Surface}
\newacronym{AWGN}{AWGN}{Additive White Gaussian Noise}
\newacronym{B5G}{B5G}{Beyond 5G}
\newacronym{BC}{BC}{broadcast channel}
\newacronym{BCD}{BCD}{Block Coordinate Descent}
\newacronym{BPCU}{BPCU}{bits per channel use }
\newacronym{BS}{BS}{Base Station}
\newacronym{C-RAN}{C-RAN}{Cloud Radio Access Network}
\newacronym{C-RSMA}{C-RSMA}{Cognitive Rate Splitting Multiple Access}
\newacronym{CSMA}{CSMA}{Carrier Sense Multiple Access}
\newacronym{CBS}{CBS}{Cognitive Base Station}
\newacronym{CCU}{CCU}{Central Control Unit}
\newacronym{CDF}{CDF}{Cumulative Distribution Function}
\newacronym{CDMA}{CDMA}{Code Division Multiple Access}
\newacronym{SCMA}{SCMA}{Sparse Code Multiple Access}
\newacronym{PDMA}{PDMA}{Pattern Division Multiple Access}
\newacronym{MUSA}{MUSA}{Multi User Shared Access}
\newacronym{LDS}{LDS}{Low Density Spreading}
\newacronym{PGAM}{PGAM}{Projected Gradient Ascent Method}
\newacronym{CD}{CD}{Code Domain}
\newacronym{CEE}{CEE}{Channel estimation error}
\newacronym{CE}{CE}{Cross Entropy}
\newacronym{CEG}{CEG}{Cell-edge Group}
\newacronym{CEU}{CEU}{cell-edge user}
\newacronym{CMD}{CMD}{Common Message Decoding}
\newacronym{CPS}{CPS}{Coherent Phase Shift}
\newacronym{CRS}{CRS}{Cooperative Rate Splitting}
\newacronym{CS}{CS}{Compressed Sensing}
\newacronym{CS1}{CS1}{Constraints Set 1}
\newacronym{CSCG}{CSCG}{Circularly Symmetric Complex Gaussian}
\newacronym{CSI}{CSI}{Channel State Information}
\newacronym{CSIT}{CSIT}{Channel State Information at the Transmitter}
\newacronym{CSIR}{CSIR}{Channel State Information at the Receiver}
\newacronym{CSP}{CSP}{Constraint-Satisfaction-Processing}
\newacronym{CSR}{CSR}{Constraint Satisfaction Ratio}
\newacronym{CU}{CU}{Cognitive User}
\newacronym{DAC}{DAC}{Digital-to-Analog Converter}
\newacronym{DC}{DC}{Difference-of-Convex}
\newacronym{DDPG}{DDPG}{Deep Deterministic Policy Gradient}
\newacronym{DEP}{DEP}{Detection Error Probability}
\newacronym{DF}{DF}{Decode-and-Forward}
\newacronym{DFAPN}{DFAPN}{Deep unFolding Active Precoding Network}
\newacronym{DFT}{DFT}{Discrete Fourier Transform}
\newacronym{DL}{DL}{Downlink}
\newacronym{DNN}{DNN}{Deep Neural Network}
\newacronym{DRL}{DRL}{Deep Reinforcement Learning}
\newacronym{DPC}{DPC}{Dirty Paper Coding}
\newacronym{DoF}{DoF}{Degrees of Freedom}
\newacronym{EE}{EE}{Energy Efficiency}
\newacronym{EH}{EH}{Energy Harvesting}
\newacronym{EM}{EM}{Electromagnetic}
\newacronym{ES}{ES}{Energy Splitting}
\newacronym{FBL}{FBL}{Finite Block-Length}
\newacronym{FC}{FC}{Fully-Connected}
\newacronym{FD}{FD}{Full-Duplex}
\newacronym{FDMA}{FDMA}{Frequency Division Multiple Access}
\newacronym{FIPSA}{FIPSA}{Feasible Initial Point Search Algorithm}
\newacronym{FTS}{FTS}{First-Order Taylor Series}
\newacronym{FU}{FU}{Far-User}
\newacronym{FA}{FA}{False Alarm}
\newacronym{GA}{GA}{Genetic Algorithm}
\newacronym{GBD}{GBD}{Generalized Bender’s Decomposition}
\newacronym{GMMV}{GMMV}{Generalized MMV}
\newacronym{HD}{HD}{Half-Duplex}
\newacronym{HMD}{HMD}{Head Mounted Device}
\newacronym{HRS}{HRS}{Hierarchal RS}
\newacronym{IBL}{IBL}{Infinite Block-Length}
\newacronym{ICSI}{ICSI}{Imperfect CSI}
\newacronym{IFBL}{IFBL}{Infinite Block Length}
\newacronym{IR}{IR}{Information Receiver}
\newacronym{IRS}{IRS}{Intelligent Reflecting Surface}
\newacronym{ISAC}{ISAC}{Integrated Sensing And Communication}
\newacronym{IoT}{IoT}{Internet of Things}
\newacronym{JFI}{JFI}{Jain's Fairness Index}
\newacronym{LEO}{LEO}{Low Earth Orbit}
\newacronym{LP}{LP}{Linear Precoding}
\newacronym{LPC}{LPC}{Long Packet Communication}
\newacronym{LoS}{LoS}{Line-of-Sight}
\newacronym{LoSC}{LoSC}{Level of Supportive Connectivity}
\newacronym{MA}{MA}{Multiple Access}
\newacronym{MAC}{MAC}{Multiple Access Channel}
\newacronym{MAR}{MAR}{Minimum Achievable Rate}
\newacronym{MD}{MD}{Missed Detection}
\newacronym{MDP}{MDP}{Markov Decision Process}
\newacronym{MEMS}{MEMS}{Micro electro-mechanical systems}
\newacronym{MIMO}{MIMO}{multiple-input multiple-output}
\newacronym{MINLP}{MINLP}{Mixed-Integer Non-Linear Programming}
\newacronym{MISO}{MISO}{multiple-input single-output }
\newacronym{MLP}{MLP}{MultiLayer Perceptron}
\newacronym{MM}{MM}{Majorization-Minimization}
\newacronym{MMF}{MMF}{Max-Min Fairness}
\newacronym{MMSE}{MMSE}{Minimum Mean Square Error}
\newacronym{MRC}{MRC}{Maximum Ratio Combining}
\newacronym{MRT}{MRT}{Maximum Ratio Transmission}
\newacronym{MR}{MR}{Minimum Rate}
\newacronym{MS}{MS}{Mode Switching}
\newacronym{MSD}{MSD}{Mean Square Distance}
\newacronym{MSE}{MSE}{Mean Square Error}
\newacronym{MU}{MU}{Multi-User}
\newacronym{NCCS}{NCCS}{Non-Central Chi-Square}
\newacronym{NDF}{NDF}{Non-regenerative Decode-and-Forward}
\newacronym{NGMA}{NGMA}{Next-Generation Multiple Access}
\newacronym{NIRS}{NIRS}{No IRS}
\newacronym{NLoS}{NLoS}{Non Line-of-Sight}
\newacronym{NMS}{NMS}{Nelder-Mead Simplex}
\newacronym{NMSE}{NMSE}{Normalized Mean Square Error}
\newacronym{NOMA}{NOMA}{Non-Orthogonal Multiple Access}
\newacronym{NOUM}{NOUM}{Non-Orthogonal Unicast and Multicast}
\newacronym{NU}{NU}{Near-User}
\newacronym{NoRS}{NoRS}{No RS}
\newacronym{OFDM}{OFDM}{Orthogonal Frequency Division Multiplexing}
\newacronym{OFDMA}{OFDMA}{Orthogonal Frequency-Division Multiple Access}
\newacronym{OMA}{OMA}{Orthogonal Multiple Access}
\newacronym{OP}{OP}{Outage Probability}
\newacronym{OPSD}{OPSD}{optimal Phase-Shift Design}
\newacronym{ORS}{ORS}{Opportunistic Rate Splitting}
\newacronym{PARAFAC}{PARAFAC}{PARAllel FACtor}
\newacronym{PBS}{PBS}{Primary Base Station}
\newacronym{PD}{PD}{Power Domain}
\newacronym{PDF}{PDF}{Probability Density Function }
\newacronym{PPO}{PPO}{Proximal Policy Optimization}
\newacronym{PRIS}{PRIS}{Passive Reconfigurable Intelligent Surface}
\newacronym{PS}{PS}{Power Splitting}
\newacronym{PSD}{PSD}{Positive Semidefinite}
\newacronym{PSO}{PSO}{Particle Swarm Optimization}
\newacronym{PSTARS}{PSTARS}{Passive Simultaneously Transmitting/Reflecting Surface}
\newacronym{PU}{PU}{Primary User}
\newacronym{QoS}{QoS}{Quality-of-Service}
\newacronym{RIS}{RIS}{Reconfigurable Intelligent Surface}
\newacronym{RE}{RE}{Resource Efficiency}
\newacronym{RO}{RO}{Reflecting-only}
\newacronym{RPSD}{RPSD}{Random Phase-Shift Design}
\newacronym{RRN}{RRN}{RIS Reflecting Network}
\newacronym{RRU}{RRU}{Remote Radio Unit}
\newacronym{RS}{RS}{Rate-Splitting}
\newacronym{RSI}{RSI}{Residual-Self Interference}
\newacronym{RSMA}{RSMA}{Rate Splitting Multiple Access}
\newacronym{RZF}{RZF}{Reqularized Zero Forcing}
\newacronym{SAA}{SAA}{Sample Average Approximation}
\newacronym{SAC}{SAC}{Soft Actor-Critic}
\newacronym{SAS}{SAS}{Safe Action Shaping}
\newacronym{SC-SIC}{SC-SIC}{superposition coding with successive interference cancellation}
\newacronym{SCA}{SCA}{Successive Convex Approximation}
\newacronym{SDMA}{SDMA}{Space-Division Multiple Access}
\newacronym{SDP}{SDP}{Semidefinite Programming}
\newacronym{SDR}{SDR}{Semidefinite Relaxation}
\newacronym{SE}{SE}{Spectral Efficiency}
\newacronym{SIC}{SIC}{Successive Interference Cancellation}
\newacronym{SISO}{SISO}{Single-input single-output}
\newacronym{SNR}{SNR}{Signal-to-Noise Ratio}
\newacronym{SOP}{SOP}{Secrecy Outage Probability}
\newacronym{SPC}{SPC}{Short Packet Communication}
\newacronym{SPCA}{SPCA}{Sequentially Parametric Convex Approximation}
\newacronym{SR}{SR}{Secrecy Rate}
\newacronym{SRE}{SRE}{Smart Radio Environment}
\newacronym{SROCR}{SROCR}{Sequential Rank-One Constraint Relaxation}
\newacronym{SSA}{SSA}{Salp Swarm Algorithm}
\newacronym{SSR}{SSR}{Sum Secrecy Rate}
\newacronym{STAR}{STAR}{Simultaneously Transmitting and Reflecting}
\newacronym{STARS}{STARS}{Simultaneously Transmitting/Reflecting Surface}
\newacronym{SU}{SU}{Secondary User}
\newacronym{SWIPT}{SWIPT}{Simultaneous Wireless Information and Power Transfer}
\newacronym{TARC}{TARC}{Transmission and Reflection Coefficients}
\newacronym{TD}{TD}{Temporal Difference}
\newacronym{TDMA}{TDMA}{Time Division Multiple Access}
\newacronym{THz}{THz}{Terahertz}
\newacronym{TIN}{TIN}{Treating Interference as Noise}
\newacronym{TISR}{TISR}{Terrestrial IRS}
\newacronym{TMA}{TMA}{Time Modulation Array}
\newacronym{TO}{TO}{Transmitting-only}
\newacronym{TRIS}{TRIS}{Transmissive Reconfigurable Intelligent Surface}
\newacronym{TS}{TS}{Time Switching}
\newacronym{UAV}{UAV}{Unmanned Aerial Vehicle}
\newacronym{UER}{UER}{Untrusted Energy Receiver}
\newacronym{UL}{UL}{Uplink}
\newacronym{UM}{UM}{Ultra Massive}
\newacronym{ULA}{ULA}{Uniform Linear Array}
\newacronym{UPA}{UPA}{Uniform Planar Array}
\newacronym{URA}{URA}{Unsourced Random Access}
\newacronym{WCSSR}{WCSSR}{Worst-case Sum Secrecy Rate}
\newacronym{WESR}{WESR}{Weighted Ergodic Sum Rate}
\newacronym{WMMSE}{WMMSE}{Weighted Minimum Mean Square Error}
\newacronym{WMSE}{WMSE}{Weighted Mean Square Error}
\newacronym{WSR}{WSR}{Weighted Sum-Rate}
\newacronym{ZF}{ZF}{Zero Forcing}
\newacronym{ZFBF}{ZFBF}{Zero-Forcing Beamforming}
\newacronym{mMIMO}{mMIMO}{Massive Multiple-input Multiple-output}
\newacronym{mmWave}{mmWave}{Millimeter Wave}
\newacronym{cmWave}{cmWave}{Centimeter Wave}
\newacronym{TP}{TP}{Time Partitioning}
\newacronym{PP}{PP}{Power Partitioning}
\newacronym{IGS}{IGS}{Improper Gaussian Signaling}
\newacronym{IQI}{IQI}{I/Q Imbalance}
\newacronym{HMIMO}{HMIMO}{Holographic MIMO}
\newacronym{LWA}{LWA}{Leaky-Wave Antennas}
\newacronym{TCA}{TCA}{Tightly Coupled antenna Arrays}
\newacronym{MOOP}{MOOP}{Multi-Objective Optimization Problem}
\newacronym{FWb}{FWb}{Frank-Wolfe-based}
\newacronym{PGS}{PGS}{Proper Gaussian Signaling}
\newacronym{eMBB}{eMBB}{Enhanced Mobile Broadband}
\newacronym{URLLC}{URLLC}{Ultra-Reliable Low-Latency Communication}
\newacronym{mMTC}{mMTC}{Massive Machine-Type Communication}
\newacronym{RF}{RF}{Radio Frequency}
\newacronym{RCG}{RCG}{Riemannian Conjugate Gradient}
\newacronym{BLER}{BLER}{BLock Error Rate}
\newacronym{SINR}{SINR}{Signal-to-Interference-Plus-Noise Ratio}
\newacronym{FP}{FP}{Fractional Programming}
\newacronym{QCQP}{QCQP}{Quadratically Constrained Quadratic Programming}
\newacronym{ER}{ER}{Energy Receiver}
\newacronym{VR}{VR}{Virtual Reality}
\newacronym{PEP}{PEP}{Packet Error Probability}
\newacronym{CoMP}{CoMP}{Coordinated Multiple Points}
\newacronym{PLS}{PLS}{Physical Layer Security}
\newacronym{LU}{LU}{Legitimate User}
\newacronym{DCO}{DCO}{Differentiable Convex Optimzation}
\newacronym{SL}{SL}{Supervised Learning}
\newacronym{UE}{UE}{User Equipment}
\newacronym{AWMMSE}{AWMMSE}{Approximate Weighted Minimum Mean Square Error}

\newcommand{\allarticles}[0]{\cite{01Yang2020,02Bansal2021miso,03Fu2021,04Weinberger2021,05Bansal2021,06Jolly2021,07Fang2022,08Shambharkar2022,09Weinberger2022csi,10Sena2022,11LiuP2022,12Dhok2022,13Li2022,14Weinberger2022,15Zhao2022,16Chen2022,17Hashempour2022,18Katwe2022clustring,19Lima2022,20Shambharkar2022edge,21Gao2022,22Camana2022,23Katwe2022shortPacket,24Pang2022,25Singh2022,26Bansal2023,27Li2023maxmin,28Gao2023,29Soleymani2023workshop,30You2023,31Aswini2023,32Darabi2023,33Tian2023,34TianY2023,35Katwe2023uplink,36Kim2023,37Soleymani2023signal,38Sena2023,39Elganimi2023,40Huang2023,41Huang2023j,42Li2023,43Niu2023,44Wu2023,45Zhang2023,46Karim2023,47Liu2023cognitive,48Khisa2023,49Liu2023,50Sun2023,51Katwe2023,52Yang2023,53LiuP2023,54Singh2023,55Xie2023,56Mohamed2023,57Wang2023multiF,58Ge2023,59Lotfi2023,60Hua2023,61LiB2023,62Zhao2023,63Huroon2023,64Xiao2023,65Chen2024,66Tang2024,67Meng2024,68Pala2024}}
\newcommand{\downlinkarticles}[0]{\cite{01Yang2020,02Bansal2021miso,03Fu2021,04Weinberger2021,05Bansal2021,06Jolly2021,07Fang2022,08Shambharkar2022,09Weinberger2022csi,10Sena2022,12Dhok2022,13Li2022,14Weinberger2022,15Zhao2022,16Chen2022,17Hashempour2022,19Lima2022,20Shambharkar2022edge,21Gao2022,22Camana2022,23Katwe2022shortPacket,24Pang2022,25Singh2022,26Bansal2023,27Li2023maxmin,28Gao2023,29Soleymani2023workshop,31Aswini2023,32Darabi2023,33Tian2023,34TianY2023,36Kim2023,37Soleymani2023signal,38Sena2023,39Elganimi2023,40Huang2023,41Huang2023j,42Li2023,43Niu2023,44Wu2023,45Zhang2023,46Karim2023,48Khisa2023,49Liu2023,52Yang2023,53LiuP2023,54Singh2023,55Xie2023,56Mohamed2023,57Wang2023multiF,58Ge2023,59Lotfi2023,61LiB2023,62Zhao2023,63Huroon2023,64Xiao2023,65Chen2024,66Tang2024,67Meng2024,68Pala2024}}
\newcommand{\uplinkarticles}[0]{\cite{11LiuP2022,18Katwe2022clustring,30You2023,35Katwe2023uplink,47Liu2023cognitive,50Sun2023,51Katwe2023,60Hua2023}}

\begin{document}
	
	\title{Resource Management in RIS-Assisted Rate Splitting Multiple Access for Next Generation (xG) Wireless Communications: Models, State-of-the-Art, and Future Directions}
	
	\author{Ibrahim Aboumahmoud,~\IEEEmembership{Graduate Student Member, IEEE},  Ekram Hossain,~\IEEEmembership{Fellow, IEEE}, and \\Amine Mezghani,~\IEEEmembership{Member, IEEE}
		\thanks{The authors are with the Department of Electrical and Computer Engineering at the University of Manitoba, Winnipeg, Canada (emails: aboumahi@myumanitoba.ca, ekram.hossain@umanitoba.ca, and amine.mezghani@umanitoba.ca).}
	}

	\maketitle
	
	\begin{abstract}
		Next generation wireless networks require more stringent performance levels.
		New technologies such as Reconfigurable intelligent surfaces (RISs) and rate-splitting multiple access (RSMA) are candidates for meeting some of the performance requirements, including higher user rates at reduced costs.
		RSMA provides a new way of mixing the messages of multiple users, and the RIS provides a controllable wireless environment.
		This paper provides a comprehensive survey on the various aspects of the synergy between reconfigurable intelligent surfaces (RISs) and rate splitting multiple access (RSMA) for next-generation (xG) wireless communications systems. 
		In particular, the paper studies more than 60 articles considering over 20 different system models where the RIS-aided RSMA system shows performance advantage (in terms of sum-rate or outage probability) over traditional RSMA models. 
		These models include reflective RIS, simultaneously transmitting and reflecting surfaces (STAR-RIS), as well as transmissive surfaces. 
		The state-of-the-art resource management methods for RIS-assisted RSMA communications employ traditional optimization techniques and/or machine learning techniques. 
		We outline major research challenges and multiple future research directions.
	\end{abstract}
	
	\begin{IEEEkeywords}
		Reconfigurable intelligent surfaces (RIS), Rate splitting multiple access (RSMA), rate splitting, simultaneously transmitting and reflecting surfaces, transmissive RIS
	\end{IEEEkeywords}
	
\section{Introduction}

\IEEEPARstart{T}hree Pillars formed the target of development in \gls{5G}: \gls{eMBB}, \gls{URLLC}, and \gls{mMTC}.
The first pillar targets faster data rates to enable high resolution video streaming and augmented reality, with an average of 50 Mbps and a peak of 10 Gbps.
The second pillar is for applications like autonomous driving and it targets latency on the order of a millisecond.
The third pillar specifies a support for a million low-power devices per square kilometre. 
Newer generations, such as \gls{6G}, aspire to reach higher requirements that necessitate the development of newer technologies that not only satisfy the connectivity and rate requirements, but also achieve better spectral and energy efficiencies.
These include new multiple access techniques such as \gls{RSMA}, in addition to \gls{RIS} which enables unprecedented level of control on the channel.
These technologies are at the intersection of three foundations described in \cite{Saad2020}, namely: Communications foundations, Large intelligent surfaces foundations, and optimizations foundations.

Although orthogonal transmission techniques like \gls{OFDMA} are still prevalent, \gls{RSMA} is a promising non-orthogonal transmission technique, which can lead the proliferation of \gls{NGMA}. In addition to the use of \glspl{RIS}, newer multiple access techniques are currently being considered for the \gls{B5G} and the upcoming \gls{6G}. Exploring the effect of the coexistence of these two technologies is an important discovery endeavour. In the following, we review the basics of \gls{RSMA} and \gls{RIS}.

\IEEEpubidadjcol

\subsection{Rate Splitting Multiple Access}
It was suggested that the mere classification of multiple access systems into orthogonal and non-orthogonal schemes is incomplete \cite{Clerckx2023rsmaTut}.
Instead, one should think about how the interference among multiple users is managed, and \gls{RSMA} can provide a general framework to answer that question. 
\gls{RSMA} has many variations \cite{Mao2018}, all are based on the idea of \gls{RS} which 
splits the message at the transmitter into one or more pieces, and the splitting ratio must be known at the receiver \cite{Mao2022}.
More details on \gls{RS} schemes can be found in \cite{Mao2018,Clerckx2023rsmaTut}.

\begin{figure}
	\centering
	\includegraphics{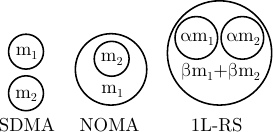}
	\caption{A conceptual view of \gls{SDMA}, \gls{NOMA}, and \gls{1L}-\gls{RS} for a 2-user \gls{DL} system. For \gls{NOMA} and \gls{RS}, an ellipse denotes a \gls{SIC} step.}
	\label{fig:rs1l_2users}
\end{figure}

\begin{figure}
	\centering
	\includegraphics{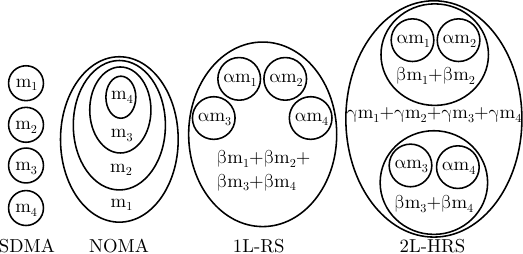}
	\caption{A conceptual view of \gls{SDMA}, \gls{NOMA}, \gls{1L}-\gls{RS}, and \gls{2L}-\gls{HRS} for a 4-user \gls{DL} system. For \gls{NOMA} and \gls{RS}, an ellipse denotes a \gls{SIC} step.}
	\label{fig:rs1l_4users}
\end{figure}

In the \gls{DL} \gls{SDMA}, or \gls{MU}-\gls{LP}, the \gls{BS} will send a data vector $\mathbf{s}=[s_1, \dots, s_K]^\top$ where $s_k$ is intended to the $k$th user, and is only decoded by the $k$th user while treating other interfering messages as noise.
If \gls{NOMA} is used, each user decodes $K-1$ messages then subtracts them using \gls{SIC} to access its private message. 
In other words, the first user treats all interfering messages as noise, and decodes its message, then the second user decodes the message of the first user first, then applies \gls{SIC} to cancel the interference of the that message, then decodes its message while treating the remaining messages as noise.
Fig.~\ref{fig:rs1l_2users} shows a conceptual view of the two systems just described (\gls{SDMA} and \gls{NOMA}) in addition to \gls{1L}-\gls{RS} described in the next paragraph. 
Fig.~\ref{fig:rs1l_4users} shows the same for a four-user system.

\begin{figure*}
	\centering
	\includegraphics[width=0.8\textwidth]{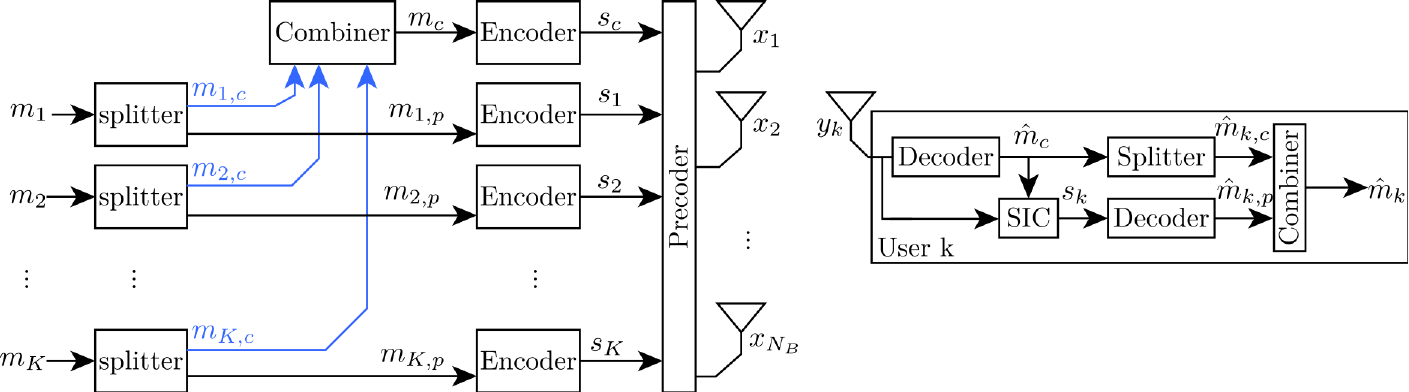}
	\caption{System diagram of a K-user \gls{MISO} system implementing \gls{1L}-\gls{RS}.}
	\label{fig:rs1l_draw}
\end{figure*}

\gls{1L}-\gls{RS} is a simple \gls{RS} scheme where a single \gls{SIC} layer is required at each receiver.
In the \gls{DL}, the message of each user is split into two (usually not equal) parts, the first part of each user is joined to form the common part $s_c$, and the other part remains private to each user after decoding the common part $s_c$. 
Hence, the \gls{BS} sends the data vector $\mathbf{s}=[s_c,s_1, \dots, s_K]^\top$, and the $k$th user decodes $s_c$ first then decodes its private message $s_k$. 
It is assumed that the codebook of the common stream
is shared by all users, 
and that $\text{tr}(\mathbf{s}\mathbf{s}^H)=1$. 
Note that in the example of Fig.~\ref{fig:rs1l_2users}, each user has splits its message into two parts, one of them is within the common message that is decoded first by both receivers, that is $\beta m_1 + \beta m_2$ in Fig.~\ref{fig:rs1l_2users}. In other words, the full message intended to user 1 will be composed of the portion in the common part in addition to the private part.
This implies that if a user is unable to decode either the common or the private part, then that user will be in outage.
Outage performance is discussed in Section~\ref{section-outage-analysis}.
\gls{NOMA} is a special case of \gls{1L}-\gls{RS} where there is no private message to the first user, and the message of the first user is the common message.
That is, by setting $\alpha m_1 =0$, and $\beta m_2 = 0$.
In case of no \gls{QoS} constraints per users, e.g. no minimum rate requirements, then only one user has to split its rate \cite{Joudeh2016tcomm}.
Splitting the messages of multiple users was found useful in the cases of \gls{QoS} constraints, such as rate requirements, max-min fairness and \gls{WSR} \cite{Joudeh2016tcomm, Joudeh2016tsigp, Joudeh2017}.
Note that the common rate is limited by the rate supported by the user with the least-favourable condition. 
One possible way to mitigate this limitation is the cooperative-\gls{RS} which was proposed in \cite{Mao2020coopRS} where the user with the better channel volunteers to deliver the common message to the other user having a weaker channel. 
The \gls{RIS}-assisted cooperative-\gls{RSMA} system has been shown to lower the energy consumption of the system \cite{48Khisa2023}.
A different view of the \gls{1L}-\gls{RS} system is presented in Fig.~\ref{fig:rs1l_draw} where a \gls{BS} serves $K$-users through $N_B$ antennas.
Only a single receiver is shown since all receivers follow the same logic.
In the notation of Fig.~\ref{fig:rs1l_2users}, $\beta m_k = m_{k,c}$ and $\alpha m_k = m_{k,p}$.

Another way of \gls{RS} is called \emph{multicommon \gls{RS}}, where each user splits its message into two parts, just like \gls{1L}-\gls{RS}, but the common messages of users are not combined into a single common message. 
Instead, each user gets a common and a private message that are encoded separately from the messages of other users. 
In other words, there are $2K$ messages for $K$ users.
The messages can then be transmitted from the same or different \glspl{BS}. 
Different \glspl{BS} are used in \cite{04Weinberger2021,14Weinberger2022}. 
A third way is the \emph{dual-polarized \gls{RS}}, which is similar to \gls{1L}-\gls{RS} but uses different polarizations to send the common and the private parts. 
It is assumed that the receivers also use polarized antennas. This idea was proposed in \cite{Sena2022}, 
 and is used by \cite{38Sena2023} in the \gls{RIS}-assisted \gls{RSMA} system.

\begin{figure*}
	\centering
	\includegraphics[width=0.9\textwidth]{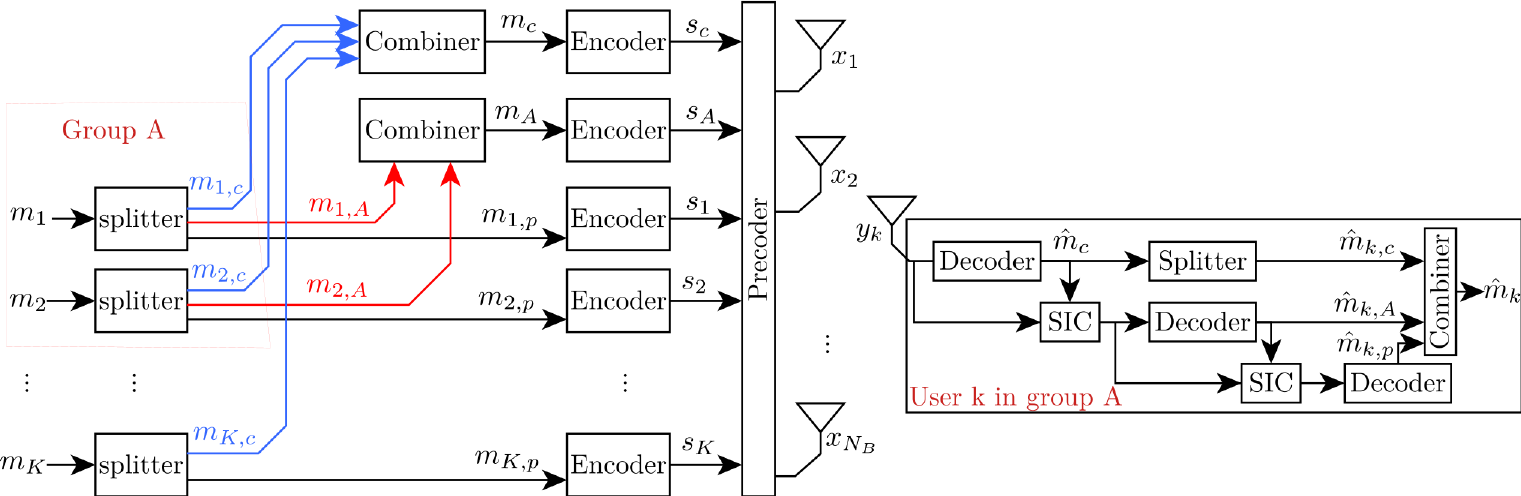}
	\caption{System diagram of a K-user \gls{MISO} system implementing \gls{2L}-\gls{HRS}. A single group is shown for simplicity.}
	\label{fig:rs2l_draw}
\end{figure*}

\gls{2L}-\gls{HRS} is a more general \gls{RS} scheme compared to \gls{1L}-\gls{RS}. \gls{2L}-\gls{HRS} boils down to \gls{1L}-\gls{RS} when users are not grouped. 
In \gls{2L}-\gls{HRS}, the message to each user is split into three parts: 1) a universal common part, 2) a group common part, and 3) a private part. 
These can be reworded as: 1) an inter-group part, 2) an intra-group part, and 3) a private part.
Consequently, each user is required to perform \gls{SIC} twice: Once for the universal common part, and once for the group common part. 
After that, the user gets to decode its private part.
Fig.~\ref{fig:rs1l_4users} shows an example with 4 users. Users 1 and 2 are in the same group while users 3 and 4 are in a different group.
Each user compiles its message once it successfully decodes all three parts of its message. 
For instance, the message of user 1 is $m_1 = \alpha m_1 + \beta m_2 + \gamma m_3$.
It is assumed that $\alpha + \beta + \gamma = 1$. 
Another view for \gls{2L}-\gls{HRS} is shown in Fig.~\ref{fig:rs2l_draw} for $K$-user \gls{MISO} system. 
Users can be grouped to perform an extra \gls{SIC} within the group.
The common message is still made up of portions of the messages of all users.
Similarly, the common message of the group is composed of portions of messages intended for users in the group.
In terms of the notation of Fig.~\ref{fig:rs1l_4users}, $\gamma m_k = m_{k,c}$, $\beta m_k = m_{k,A}$, $\alpha m_k = m_{k,p}$ for all $k$ in group A.

\subsection{Reconfigurable Intelligent Surface (RIS)}

The \gls{RIS} is an array of discrete elements that can be controlled over time.
A metamaterial is a candidate material for manufacturing \gls{RIS} elements.  
The term ``meta'' refers to the unusual properties of these materials, for instance, a negative or anisotropic refractive index \cite{Chen2016}.
Metamaterials are a periodic structure of metallic or dielectric units called meta-atoms, which resonate to the incident \gls{EM} wave \cite{Chen2016}.
If these metamaterials are layered in a planar structure, then they are referred to as a metasurface \cite{Chen2016}. 
Manufacturing metasurfaces can be achieved through lithography or nanoprinting \cite{Chen2016}.
Simpler methods can be used if the wavelength is not too small. 
For instance, metasurfaces for light waves are typically manufactured using lithography techniques. 
Another example is the possibility to use metal patches for unit cells in the microwave frequencies \cite{Yang2016}. 
The spacings between the meta-atoms are usually less than half wavelength \cite{Chen2016}.
Element dimensions are usually less than the wavelength of operation, e.g. $\lambda/5 \times \lambda /5$ \cite{Bjoernson2022}.
The \gls{RIS} can be passive or active. 
Passive surfaces do not amplify the incident signal, and the reflected signal can at best remain at the power of the incident signal.
On the other hand, active surfaces can introduce gain in the reflected signal without using an \gls{RF} chain or signal processing.
For example, the passive surface can be modelled as a passive filter, where each element can introduce a delay, attenuation, or a polarization change \cite{Bjoernson2022}. 
In addition, it can be classified in other ways like opaque vs. transparent (to the light wavelengths), or based on the connectivity modelling \cite{Shen2022}. 
The \gls{RIS} can be tuned to support communications, localization, power transfer, sensing, and physical layer security \cite{DiRenzo2020,Wymeersch2020}.
This article focuses on the papers \allarticles, and the some of the assumptions they consider for the \gls{RIS} are listed in Table~\ref{tab:model-number-of-elements-across-papers}.
The next paragraphs briefly introduce the reflective, transmissive, and \acrfull{STAR} surfaces.

\subsubsection{Reflective Surface}
\begin{figure}
	\centering
	\includegraphics{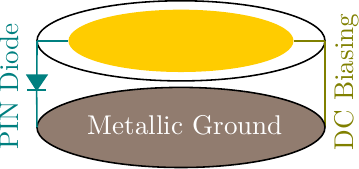}
	\caption{A unit-cell of a reflective surface with DC biasing, and binary control through a single diode. Adapted from \cite{Yang2016}.}
	\label{fig:unit-cell}
\end{figure}

The reflecting surface is sometimes referred to as the \gls{IRS}.
Fig.~\ref{fig:unit-cell} presents a sample unit cell of a metasurface.
The figure is adapted from \cite{Yang2016} but with a circular shape to illustrate that the unit cells need not be rectangular.
Consequently the detailed dimensions of the unit cells in \cite{Yang2016} are omitted. 
Details of the \gls{RIS} are beyond the scope of this paper, but three modelling techniques will be highlighted.
An \gls{RIS} with $N_R$ elements can be modelled by an $N_R$-port reconfigurable impedance network represented by a complex-valued $N_R\times N_R$ scattering matrix $\mathbf{\Phi}$ that relates the incident and reflected waves at the surface. 
When the impedance network is single-connected, $\mathbf{\Phi}$ is a diagonal matrix, and there is a single impedance to tune for each element, or a total of $N_R$ elements per surface. 
This model is quite common in the literature. 
Another model for the \gls{RIS} is the fully-connected model, which assumes that all the elements in the \gls{RIS} are interconnected by an impedance \cite{Shen2022}. 
For example, this model is adopted in \cite{07Fang2022}. 
In this model, each \gls{RIS} element is connected to all other elements, hence the matrix $\mathbf{\Phi}$ is no longer sparse, and a total of $\binom{N_R}{2} + N_R = N_R(N_R+1)/2$ scattering parameters need to be tuned (because each element still has one impedance connected to the ground).
A third model is the group connected model \cite{Shen2022}, used by \cite{36Kim2023}. 
Admittedly, the number of parameters to tune in the fully-connected model just-described can be very complicated, thus, a group-connected model limits connectivity between elements to a few neighbouring elements, and the resulting scattering matrix $\mathbf{\Phi}$ is block-diagonal.

\subsubsection{Transmissive Surface}
A quick analogy to describe the transmissive surface is to think of metalenses. 
A metalens is a thin metasurface structure that can be designed to manipulate light waves similar to traditional glass-based or plastic-based lenses.
In that sense, the light passes through the surface with its amplitude, phase, polarization, or all of them altered.
Similarly, a transmissive surface can modulate an \gls{EM} wave as it passes through it.
Compared to the reflective surface, however, the transmissive surface may be more difficult to manufacture if reconfigurability is desired.
For instance, the back plane of the reflective surface can carry internal components to support the reconfigurability of the surface, while the transmissive surface has to maintain the bulk of both surfaces clear for the passage of the \gls{EM} wave.
Design of transmissive surfaces can follow techniques such as those described in \cite{Brown2020, Niu2023}.

\subsubsection{Simultaneously Transmitting and Reflecting Surface}
\acrfull{STAR}-\gls{RIS} is a surface that can transmit and reflect (simultaneously or non-simultaneously) \lq portions\rq~of the incident signal. 
These \lq portions\rq~are determined according to three common protocols: 
\begin{enumerate}
	\item \gls{ES} where each element in the surface simultaneously transmits and reflects energy according to two parameters called \glspl{TARC}. This mode was used in \cite{17Hashempour2022, 55Xie2023, 67Meng2024}, and \cite{12Dhok2022} which also considers \gls{MS}.
	\item In \gls{MS}, each element of the surface is associated with either the transmit mode or the reflection mode, hence, the surface maintains a \lq simultaneous\rq~transmission and reflection but with lower gain compared to \gls{ES}. This mode was used in \cite{12Dhok2022} and \cite{64Xiao2023}.
	\item \gls{TS} flips all elements from transmit mode to reflect mode at orthogonal time intervals. This mode requires synchronization and is not common.
\end{enumerate}

Note that these signal splittings should not be confused with the signal splittings done by the \gls{RSMA} protocol at the transmitter. 
Note also that \lq transmission\rq~in the context of \gls{STAR}-\gls{RIS} is the mere passage of a portion of the incident signal to other half plane behind the surface, and does not imply signal generation at the surface. 
Thus, the \gls{STAR}-\gls{RIS} provides a full 360$^0$ coverage in the \gls{SRE}, 
i.e. it provides coverage to the two half planes before and behind the surface, and that is different from the \gls{RO}-\gls{RIS} which only covers the half plane before the surface, or 180$^0$.
\gls{STAR}-\gls{RIS} is sometimes called intelligent omni-surface \cite{Ahmed2023,Zhang2022} but may have a single tuning parameter for both transmission and reflection \cite{Liu2021}.
Controlling the transmission and reflection coefficients of the \gls{STAR}-\gls{RIS} independently require a semi-passive or active surface. 
Arbitrary independent assignment of the coefficients for the passive surface may be impractical \cite{Liu2022b}.

\subsection{Synergy Between RSMA and RIS}
Mutual gains have been reported for the \gls{RIS}-assisted \gls{RSMA} systems.
The gains can be in terms of reduced complexity (e.g. \gls{1L}-\gls{RS} is less complex compared to \gls{NOMA} for three or more users),
more energy/spectrum efficiency or more \gls{WSR}, etc. 
In addition, some weaknesses of one technology can be reduced by leveraging the other one. Many benefits of combining \gls{RIS} with \gls{RSMA} were explored in \cite{13Li2022} along with a brief on both technologies. For example, the role of the \gls{RIS} as a passive (low cost) assistant in communications poses difficulty in \gls{CSI} acquisition, which can be mitigated by the robustness of \gls{RSMA} to imperfect \gls{CSI}.

The idea of \gls{RS} originated decades ago \cite{Carleial1978,Han1981,Rimoldi1996}. 
Since \gls{RS} is good at handling interference in \gls{SISO} channels, further investigations into \gls{MISO} channels led to the development of \gls{RSMA}, which can be thought of as a generalization of both \gls{SDMA} and \gls{NOMA} \cite{Mao2018,Hao2015,Joudeh2016tcomm,Joudeh2016tsigp}. 
\gls{NOMA} achieves capacity for \gls{SISO}-\gls{BC}, but the complexity is large since the strong user will decode the messages of all other users. 
In contrast, \gls{RSMA} can match or surpass the performance of \gls{NOMA} with a single \gls{SIC} decoding step at each receiver, which is the case of \gls{1L}-\gls{RS}.
\gls{RSMA}, albeit simpler than \gls{DPC}, may outperform \gls{DPC} in the imperfect \gls{CSI} setting \cite{Mao2020}. 
With perfect \gls{CSIT} and \gls{CSIR} however, \gls{DPC} is the capacity-achieving strategy \cite{Costa1983,Weingarten2006,Caire2003}.

\gls{NOMA} is capacity-achieving for the degraded \gls{SISO}-\gls{BC} channel, and in that channel, \gls{RSMA} is effectively \gls{NOMA}. Nevertheless, \gls{RSMA} can achieve more than 90\% of the performance of \gls{NOMA} with a single \gls{SIC} layer \cite{Mao2018} \cite{Clerckx2023rsmaTut}. 
On the contrary, in the non-degraded \gls{SISO}-\gls{BC} channel, \gls{RSMA} always outperforms \gls{NOMA} in terms of rate region\cite{Clerckx2023rsmaTut}. 
\gls{BC} channels with \gls{RIS} are then expected to always benefit from \gls{RSMA} as \gls{RIS}-assisted channels are non-degraded.
In addition, since perfect \gls{CSIT} is almost never available, the channel will be non-degraded \cite{Clerckx2023rsmaTut}. 
Hence, based on the previous two reasons, in a \gls{SISO}-\gls{BC} setting, a system of \gls{RSMA} with \gls{RIS} is expected to always outperform that of \gls{NOMA} with \gls{RIS}. 
So far, the current literature mostly focused on \gls{MISO}-\gls{BC} systems (or its uplink dual), where each user is equipped with a single antenna. 
\gls{NOMA} incurs \gls{DoF} loss in \gls{MIMO} channels, and therefore it is not a capacity achieving scheme \cite{Mishra2022}. 
The capacity of \gls{RIS}-assisted \gls{RSMA} \gls{MU}-\gls{MIMO} is an open problem \cite{Mishra2022}.

The \gls{RIS} is opening up a new paradigm in the wireless networks design by providing the possibility of engineering the channel. This includes the assistance of an already-existing link, or providing the sole link in case of shadowing. Hence, an extra number of users can be served due to induced favourable propagation conditions to these extra users. 
In addition, in a wireless channel, some users may share similar propagation conditions, which diminish the gain of \gls{MA} schemes other than \gls{OMA}. The \gls{RIS} can help diversify the channels allowing improved gain of \gls{NOMA} over \gls{OMA} \cite{Ding2022}. 
In \gls{NOMA} (power-domain \gls{NOMA}), it is preferred to group users in small clusters to reduce the decoding complexity \cite{Ding2022}. This issue does not exist with \gls{1L}-\gls{RS} where only one \gls{SIC} layer is required at every user, or \gls{2L}-\gls{HRS} where only two \gls{SIC} layers are required at every user regardless of the number of users. 
This paper also considers papers with the \gls{STAR}-\gls{RIS} in the system model.
The \gls{STAR}-\gls{RIS} was shown to increase fairness among users and improve the system sum rate \cite{67Meng2024}. 
In addition, using \gls{RSMA} improved the system sum rate when compared to \gls{NOMA}, and both achieve higher sum rate than \gls{RO}-\gls{RIS} with \gls{RSMA} \cite{67Meng2024}.

The existence of the \gls{RIS} was also proven beneficial to the system. For instance, using multiple \glspl{RIS} with \gls{RSMA} was shown to outperform multiple \glspl{RIS} with \gls{OFDMA} in terms of \gls{EE} \cite{01Yang2020}, and also showed some improvement compared to \glspl{RIS} with \gls{NOMA} \cite{01Yang2020}.
Another direction was to examine the effect of fully-connected \gls{RIS} \cite{Shen2022}
which showed about 4.6\% better performance than the traditional single \gls{RIS} with \gls{RSMA} \cite{07Fang2022}, and 16.5\% increase compared to \gls{RSMA} without \gls{RIS} \cite{07Fang2022}.
Multiple optimization objectives have been considered for \gls{RIS}-assisted networks, where favourable performance improvement is achieved. This includes maximizing \gls{EE}, maximizing \gls{SE}, minimizing transmit power, etc.
In addition, multiple authors studied the outage performance of the system.
\gls{RIS}-assisted \gls{RSMA} networks is more energy efficient compared to \gls{RIS}-assisted \gls{NOMA} and \gls{RIS}-assisted \gls{SDMA} under \gls{SPC} constraints. In addition,  \gls{RIS}-assisted \gls{RSMA} \gls{SPC} can be more energy efficient than  \gls{RIS}-assisted  \gls{LPC} with \gls{NOMA} or \gls{SDMA} for some packet sizes \cite{23Katwe2022shortPacket}.
In a multi-user network, \gls{RSMA} alone can achieve nearly 95\% gain in terms of \gls{EE}, 
and with the \gls{RIS}, the combined improvement in \gls{EE} is higher than the sum of improvement by \gls{RSMA} or \gls{RIS} alone \cite{14Weinberger2022}.

\subsection{Scope and Related Surveys}
\begin{figure}
	\centering
	\includegraphics[width=0.6\columnwidth]{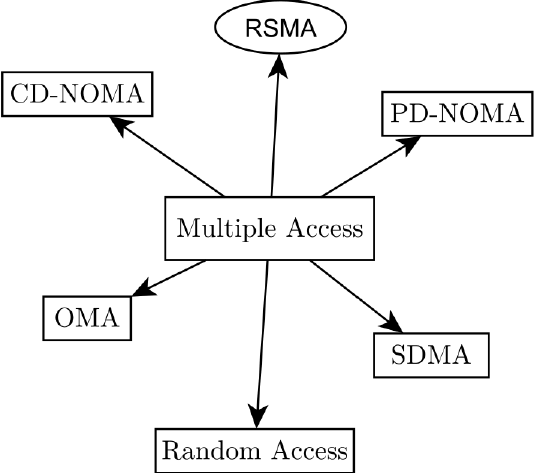}
	\caption{Various multiple access techniques.}
	\label{fig:ma}
\end{figure}

\acrfull{RSMA} is one of many possible techniques to manage the access of available resources for multiple users.
Fig.~\ref{fig:ma} shows many possible techniques for multiple access.
The focus of this article, \gls{RSMA}, is encircled.
Moreover, many articles in the literature compared \gls{RSMA} with \gls{SDMA} and \gls{PD}-\gls{NOMA}, mainly because \gls{RSMA} is a general framework that includes both techniques.
Throughout this article, \gls{PD}-\gls{NOMA} will simply be referred to as \gls{NOMA}.
The \gls{OMA} classification includes traditional multiple access techniques which are widely adopted like \gls{TDMA}, \gls{CDMA}, \gls{FDMA}, and \gls{OFDMA}, while \acrfull{SDMA} is a technique to separate users spatially, to allow reuse of available resources like time and frequency.
Furthermore, \gls{CD}-\gls{NOMA} \cite{Wang2015,Gamal2021} does not necessarily rely on \gls{SC-SIC} as in \gls{PD}-\gls{NOMA}, and it includes \gls{LDS}-\gls{CDMA}, \gls{LDS}-\gls{OFDM}, \gls{SCMA}, \gls{PDMA}, and \gls{MUSA}.
Lastly, the random access category includes schemes like ALOHA, and \gls{CSMA}.

This article examines papers that explore the use of the \gls{RIS} with \gls{RSMA} \allarticles. 
One paper was published in 2020 \cite{01Yang2020}, followed by five papers in 2021 \cite{02Bansal2021miso,03Fu2021,04Weinberger2021,05Bansal2021,06Jolly2021}, 19 papers in 2022 \cite{07Fang2022,08Shambharkar2022,09Weinberger2022csi,10Sena2022,11LiuP2022,12Dhok2022,13Li2022,14Weinberger2022,15Zhao2022,16Chen2022,17Hashempour2022,18Katwe2022clustring,19Lima2022,20Shambharkar2022edge,21Gao2022,22Camana2022,23Katwe2022shortPacket,24Pang2022,25Singh2022}, and more than 30 papers in 2023  \cite{26Bansal2023,27Li2023maxmin,28Gao2023,29Soleymani2023workshop,30You2023,31Aswini2023,32Darabi2023,33Tian2023,34TianY2023,35Katwe2023uplink,36Kim2023,37Soleymani2023signal,38Sena2023,39Elganimi2023,40Huang2023,41Huang2023j,42Li2023,43Niu2023,44Wu2023,45Zhang2023,46Karim2023,47Liu2023cognitive,48Khisa2023,49Liu2023,50Sun2023,51Katwe2023,52Yang2023,53LiuP2023,54Singh2023,55Xie2023,56Mohamed2023,57Wang2023multiF,58Ge2023,59Lotfi2023,60Hua2023,61LiB2023,62Zhao2023,63Huroon2023,64Xiao2023}. 
Table~\ref{intercitations} shows the inter-citations and the publication year for these articles in addition to the \gls{RS} scheme and the type of the \gls{RIS}. 
Selected keywords from each article are also listed.
The articles are ordered according to their publication date, yet the order here is not an exact order of the first appearance of the article.

\begin{figure*}
	\centering
	\includegraphics[width=\textwidth]{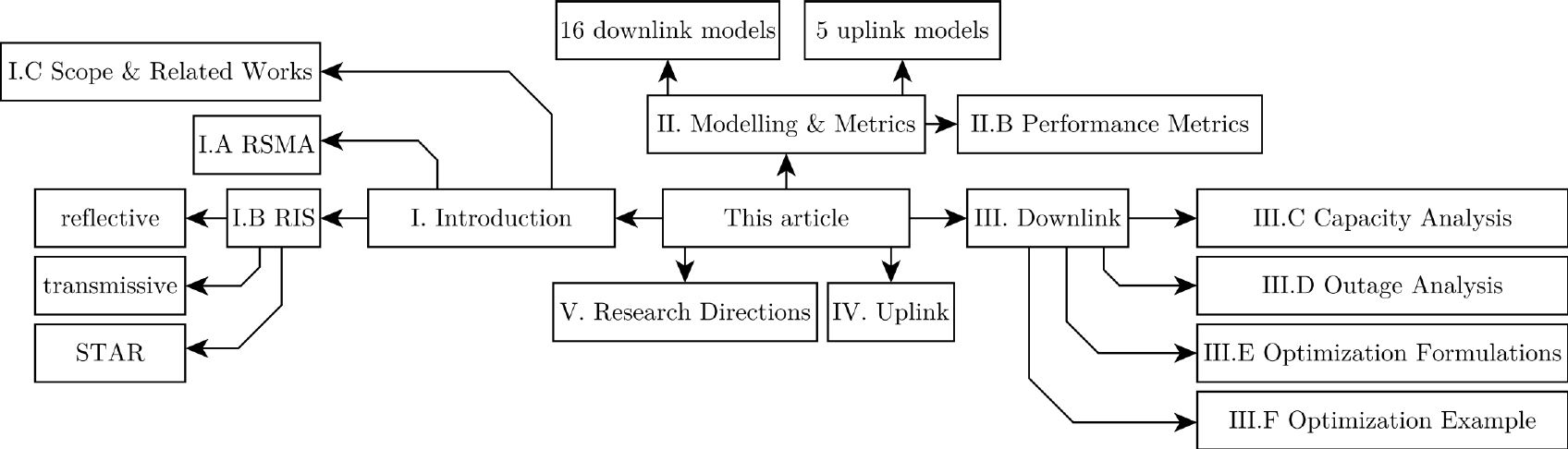}
	\caption{Brief outline of the sections in this article.}
	\label{fig:outline}
\end{figure*}

There are two publications close to the topic of this article \cite{10Sena2022,13Li2022}, and are described in Sections \ref{f} and \ref{j}, respectively.
About 200 abbreviations are used in this article, some of them are listed in Table~\ref{selectedAbbreviations}.
A survey on the \gls{RIS}-assisted \gls{NOMA} networks was published in 2022 \cite{Ding2022}, and it does not cite any of the articles \allarticles~in the focus of this paper. 
This article, however, focuses on \gls{RSMA}. 
Since some articles discussed herein discuss covert communications, a survey on covert communication might be instructive \cite{Chen2023covert}. 
Details on \gls{RIS} \cite{Liu2021Naofal} and \gls{STAR}-\gls{RIS} \cite{Liu2021,Ahmed2023} can be useful. 
An overview of the outline of this paper is depicted in Fig.~\ref{fig:outline}.

\begin{table*}
	\begin{center}
		\caption{Inter-citations of examined papers, along with selected keywords, \gls{RS} scheme, and \gls{RIS} type.}
		\label{intercitations}
\begin{tabular}{| c| c | rl|l|l|}
	\hline
	Cites                                                                                                                                                                              & Year & Ref                            & Keywords                              & \acrshort{RS} Scheme                                          & \gls{RIS} Type                         \\ \hline
	-                                                                                                                                                                                  & 2020 & \cite{01Yang2020}              & 3 Tx antennas, 3 \acrshortpl{RIS} & \acrshort{1L}-\acrshort{RS}                                   & passive reflective                     \\
	-                                                                                                                                                                                  & 2021 & \cite{02Bansal2021miso}        & cell-edge, on-off \acrshort{RIS}      & \acrshort{2L}-\acrshort{HRS}                                  & passive reflective                     \\
	\cite{01Yang2020}                                                                                                                                                                  & 2021 & \cite{03Fu2021}                & discrete \acrshort{RIS}                                      & \acrshort{1L}-\acrshort{RS}                                   & passive reflective                     \\
	-                                                                                                                                                                                  & 2021 & \cite{04Weinberger2021}        & \acrshort{C-RAN}, 4 \acrshortpl{BS}                                      & multicommon                                                   & passive reflective                     \\
\end{tabular}
	\end{center}
\end{table*}

\subsection{Notes on Terminology}

This subsection discusses general terms only. Some other acronyms that are specific to certain group of papers (like machine learning acronyms) are commented on later if necessary.
Similarly, optimization-related acronyms are discussed later in Section \ref{secOptTerminology}.
Starting with the two main acronyms, \acrfull{RSMA} is consistently used by all authors except the three papers by Weinberger et. al. \cite{04Weinberger2021,09Weinberger2022csi,14Weinberger2022} which never mention \gls{RSMA} and just use \gls{RS}.

The second main term, can be \gls{IRS} or \gls{RIS}. \gls{IRS} may suggest that only reflection capabilities are possible, while \gls{RIS} can be interpreted to support both reflecting and transmitting surfaces. 
In fact, many authors emphasize these special capabilities and other features of the surfaces through terms like ARIS where the A stands for ``Active'' \cite{55Xie2023} or for ``Aerial'' \cite{25Singh2022,54Singh2023}, \gls{PRIS} \cite{55Xie2023},  \gls{STAR} \cite{12Dhok2022,17Hashempour2022,18Katwe2022clustring,34TianY2023,51Katwe2023,67Meng2024,64Xiao2023}, \gls{STARS} \cite{55Xie2023}, \gls{ASTARS} \cite{55Xie2023}, \gls{PSTARS} \cite{55Xie2023}.
At any rate, in the focus of this paper, 36 papers use \gls{RIS} and 20 papers use \gls{IRS}, with a general trend of decreasing use of the latter over the duration of this study.
To add to the complication, \cite{67Meng2024} uses \acrshort{RO}-\acrshort{RIS} and \acrshort{TO}-\acrshort{RIS} to indicate reflecting-only \acrshort{RIS} and transmitting-only RIS. 
Moreover \gls{AIRS} \cite{19Lima2022} and \gls{NIRS} \cite{03Fu2021} have been used.

\acrshort{URA} is not used for uniform rectangular array except by \cite{18Katwe2022clustring}. 
Nonetheless, since \acrshort{URA} can indicate \acrlong{URA} as mentioned by \cite{66Tang2024}, it is probably better to keep \gls{UPA} for the planar \gls{RIS} structure (or for the planar array structure at the \gls{BS} as well).
\gls{NGMA} is a general term that can be used to include \gls{MA} terms like \gls{RSMA}, \gls{NOMA} \cite{Liu2022Ekram}, and \gls{URA} \cite{Che2023}.
A user is sometimes used to refer to a receiver terminal, or a pair of a transmitter and a receiver. 
This article uses the former definition.
Short block length codes can be called \acrfull{FBL} or \acrfull{SPC} \cite{12Dhok2022,23Katwe2022shortPacket}. 
Similarly, long block length codes can be called \acrfull{IBL} or \acrfull{LPC} \cite{23Katwe2022shortPacket}.
Throughout this article and all reviewed articles, \acrshort{RSMA} refers to \acrlong{RSMA}, and not Resource Spread Multiple Access.

\begin{table}
	\begin{center}
		\caption{Selected abbreviations used throughout the article}
		\label{selectedAbbreviations}
		\begin{tabular}{|>{\raggedright\arraybackslash}m{1cm}|l|}
			\hline
			CU    & Cognitive User                                       \\
			DDPG  & Deep Deterministic Policy Gradient                   \\
			DEP   & Detection Error Probability                          \\
			DFAPN & Deep unFolding Active Precoding Network              \\
			DF    & Decode-and-Forward                                   \\
			DFT   & Discrete Fourier Transform                           \\
			DNN   & Deep Neural Network                                  \\
			DoF   & Degrees of Freedom                                   \\
			DPC   & Dirty Paper Coding                                   \\
			DRL   & Deep Reinforcement Learning                          \\
			EE    & Energy Efficiency                                    \\
			EH    & Energy Harvesting                                    \\
			ES    & Energy Splitting                                     \\
			FBL   & Finite Block-Length                                  \\
			FC    & Fully-Connected                                      \\
			FD    & Full-Duplex                                          \\
			FDMA  & Frequency Division Multiple Access                   \\
			FU    & Far-User                                             \\
			HMIMO & Holographic MIMO                                     \\
			HRS   & Hierarchical RS                                        \\
			IFBL  & Infinite Block Length                                \\
			IQI   & I/Q Imbalance                                        \\
			IR    & Information Receiver                                 \\
			ISAC  & Integrated Sensing And Communication                 \\
			JFI   & Jain's Fairness Index                                \\
			LoSC  & Level of Supportive Connectivity                     \\
			LPC   & Long Packet Communication                            \\
			LP    & Linear Precoding                                     \\
			LWA   & Leaky-Wave Antennas                                  \\
			MAR   & Minimum Achievable Rate                              \\
			MLP   & MultiLayer Perceptron                                \\
			MMF   & Max-Min Fairness                                     \\
			MM    & Majorization-Minimization                            \\
			MMSE  & Minimum Mean Square Error                            \\
			MRC   & Maximum Ratio Combining                              \\
			MR    & Minimum Rate                                         \\
			MRT   & Maximum Ratio Transmission                           \\
			MSD   & Mean Square Distance                                 \\
			MSE   & Mean Square Error                                    \\
			MS    & Mode Switching                                       \\
			NOMA  & Non-Orthogonal Multiple Access                       \\
			OP    & Outage Probability                                   \\
			ORS   & Opportunistic Rate Splitting                         \\
			PGS   & Proper Gaussian Signalling                            \\
			PP    & Power Partitioning                                   \\
			PS    & Power Splitting                                      \\
			RE    & Resource Efficiency                                  \\
			RIS   & Reconfigurable Intelligent Surface                   \\
			RPSD  & Random Phase-Shift Design                            \\
			RRN   & RIS Reflecting Network                               \\
			RS    & Rate-Splitting                                       \\
			SE    & Spectral Efficiency                                  \\
			SR    & Secrecy Rate                                         \\
			SOP   & Secrecy Outage Probability                           \\
			SPC   & Short Packet Communication                           \\
			SRE   & Smart Radio Environment                              \\
			SSR   & Sum Secrecy Rate                                     \\
			STAR  & Simultaneously Transmitting and Reflecting           \\
			SWIPT & Simultaneous Wireless Information and Power Transfer \\
			TARC  & Transmission and Reflection Coefficients             \\
			TCA   & Tightly Coupled antenna Arrays                       \\
			TIN   & Treating Interference as Noise                       \\
			TP    & Time Partitioning                                    \\
			TRIS  & Transmissive RIS                                     \\
			TS    & Time Switching                                       \\
			UER   & Untrusted Energy Receiver                            \\
			UM    & Ultra Massive                                        \\
			UPA   & Uniform Planar Array                                 \\
			URA   & Unsourced Random Access                              \\
			WCSSR & Worst-case Sum Secrecy Rate                          \\
			WESR  & Weighted Ergodic Sum Rate                            \\
			WMMSE & Weighted Minimum Mean Square Error                   \\
			WMSE  & Weighted Mean Square Error                           \\
			WSR   & Weighted Sum-Rate                                    \\ \hline
		\end{tabular}
	\end{center}
\end{table}

\textit{Notations:} 
$\mathbf{H}_k$ ($\mathbf{h}_k$) is the matrix (vector) channel from the transmitter to the $k$th user, and 
$\mathbf{H}_k^\mathsf{H}$ ($\mathbf{h}_k^\mathsf{H}$) is its conjugate transpose (or hermitian).
$\mathbf{P}$ denotes the linear precoder at the transmitter, while $P$ denotes the power budget at the transmitter.
$\text{tr}(\mathbf{S})$ is the trace of the square matrix $\mathbf{S}$.
$\mathbb{E}\left[\cdot\right]$ denotes the mathematical expectation.
$\mathbb{C}^{N_B\times N_U}$ is the set of all complex-valued $N_B \times N_U$ dimensional matrices.
The six integers $M$, $L$, $K$, $N_B$, $N_R$, $N_U$ are explained before the beginning of Section~\ref{secModels}.
Special notation is adopted for Table~\ref{tab:optimizations-across-papers} and is explained before the table in Section~\ref{secSummaryTable}.

\section{System Modelling and Performance Evaluation Metrics}

\begin{figure*}
	\centering
	\includegraphics[width=\textwidth]{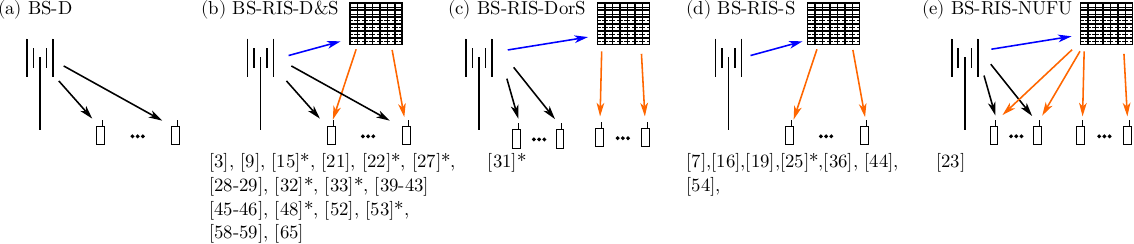}
	\caption[Models assuming a single RIS]{Models assuming a single \gls{RIS}. D\&S: Direct and Secondary. DorS: Direct or Secondary. NUFU: Near Users and Far Users.}
	\label{fig:models01-single-ris}
\end{figure*}

\begin{figure*}
	\centering
	\includegraphics{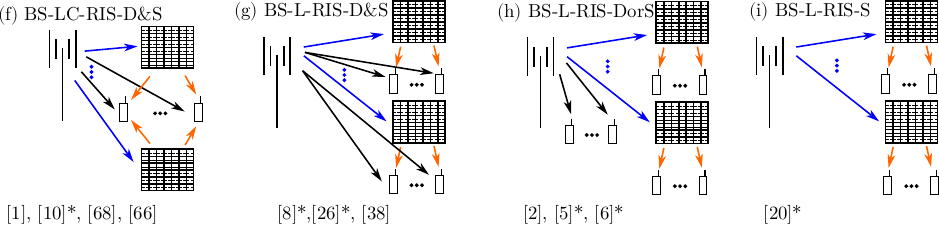}
	\caption[Models assuming multiple RISs]{Models assuming multiple \glspl{RIS}. L-RIS: Several RIS surfaces (L of them), LC-RIS: as before, and C stands for common (the same users are served by all RISs and BS).}
	\label{fig:models02-multi-ris}
\end{figure*}

\begin{figure*}
	\centering
	\includegraphics{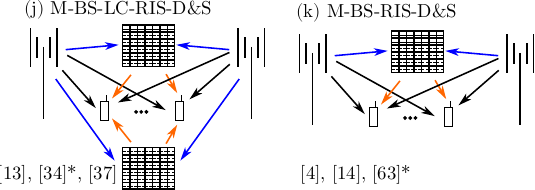}
	\caption[Models with multiple BS ]{Models with multiple \glspl{BS}. M-BS: Multiple BSs (M of them).}
	\label{fig:models03-multi-bs}
\end{figure*}

\begin{figure*}
	\centering
	\includegraphics{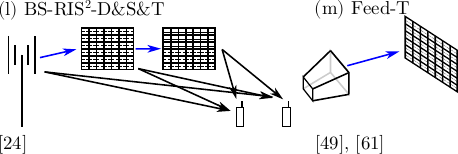}
	\caption[Models with Cascade and Transmissive RIS]{Cascade \gls{RIS} and Transmissive \gls{RIS} models, D\&S \& T: Direct and secondary and Ternary}
	\label{fig:models04-misc}
\end{figure*}

\begin{figure*}
	\centering
	\includegraphics{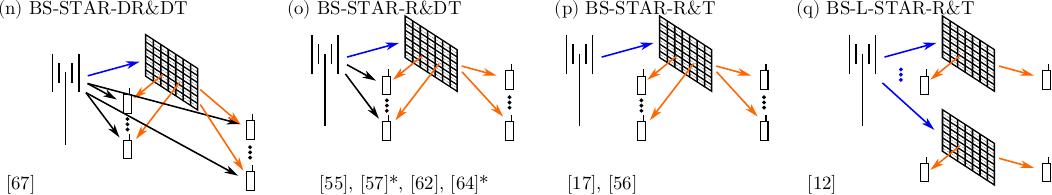}
	\caption[Models with STAR-RIS]{Models with \gls{STAR}-\gls{RIS}. R \& T: Reflect and Transmit.}
	\label{fig:models05-star}
\end{figure*}

\begin{figure*}
	\centering
	\includegraphics[width=\linewidth]{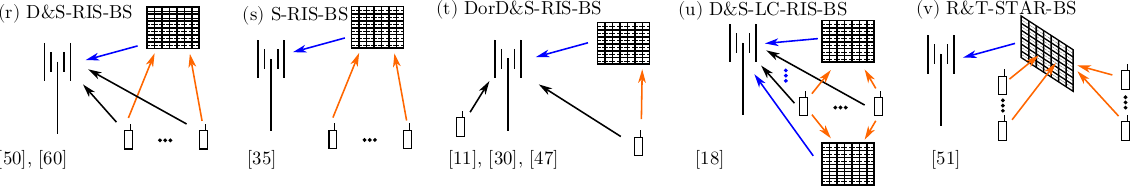}
	\caption[Uplink models.]{Uplink models. DorD\&S: Direct or (Direct and Secondary).}
	\label{fig:models06-uplink}
\end{figure*}

This section describes essentials for modelling different components of the \gls{RIS}-assisted \gls{RSMA} wireless communications link.
Most of the works focus on the \gls{DL} scenario \downlinkarticles, and a few focus on the \gls{UL} scenario \uplinkarticles. Figs. \ref{fig:models01-single-ris}, \ref{fig:models02-multi-ris}, \ref{fig:models03-multi-bs}, \ref{fig:models04-misc}, \ref{fig:models05-star} provide a general classification of the downlink models used by different authors, while Fig.~\ref{fig:models06-uplink} shows uplink models. 
Note that the classification is not fully detailed, which means that some authors might assume further details that are not depicted in these figures. 
Whenever special attention is required, an asterisk is used following the citation of the article in the figure. 
For example, two time slots are used in \cite{15Zhao2022}, and the \gls{RIS} is mounted on a \gls{UAV} in \cite{19Lima2022}. 
Throughout this paper, models will be referred to by their name without the figure number. Models 
(b) through (e) feature a single \gls{BS} and a single (reflecting) \gls{RIS}, 
(f) through (i) feature a single \gls{BS} and multiple \glspl{RIS}, 
(j) features multiple \glspl{BS} and multiple \glspl{RIS}, 
(k) features multiple \glspl{BS} and a single \gls{RIS}, 
(l) features a single \gls{BS} and two \glspl{RIS} in cascade, 
(m) features a feed antenna next to a transmissive \gls{RIS}, 
(n) through (p) feature a single \gls{BS} and a single \gls{STAR}-\gls{RIS}, 
(q) features a single \gls{BS} and multiple \gls{STAR}-\glspl{RIS}, 
and (r) through (v) features uplink models.

For the ease of exposition, the number of elements will be denoted as follows. 
\begin{itemize}
	\item The number of \glspl{BS} is $M$, and the number of antennas per \gls{BS} is $N_B$.
	\item The number of \glspl{RIS}  is $L$, and the number of elements per \gls{RIS} is $N_R$.
	\item The number of users is $K$, and the number of antennas per user is $N_U$.
\end{itemize}
The values for these parameters are used as headers in Table~\ref{tab:model-number-of-elements-across-papers} which summarizes the models keeping the notations of the authors. It also shows whether \gls{DL} or \gls{UL} channels were used, as well as \gls{CSIT} assumption.
The first subsection in this section covers the models, and the second subsection covers the performance evaluation metrics.

\begin{table*}
	\caption[Channel and number of elements assumed across the reviewed papers.]{Channel and number of elements assumed across the reviewed papers. Ch stands for Channel, DL for downlink, UL for uplink.}
	\label{tab:model-number-of-elements-across-papers}
	\centering

\begin{tabular}{|l|cc|l|cl|cl|cc|}
	\hline
	Ref                            & Ch &    Imperfect     &                        &   BS    &                               &    RIS     &                                        &       Users       &              \\
	                               &    &       CSIT       & Model Class            &   $M$   & $N_B$, UPA?                   &    $L$     & $N_R$, UPA?                            &        $K$        &    $N_U$     \\ \hline
	\cite{01Yang2020}              & DL &                  & f) BS-LC-RIS-D\&S      &   $1$   & $M$                           &    $L$     & $N_l$                                  &        $K$        &     $1$      \\
	\cite{02Bansal2021miso}        & DL &    \checkmark    & h) BS-L-RIS-DorS       &   $1$   & $N_s$                         &    $M$     & $N_r$             y                    & $K=K_0+\dots+K_M$ &     $1$      \\
	\cite{03Fu2021}                & DL &                  & b) BS-RIS-D\&S         &   $1$   & $N_t$                         &    $1$     & $L$                                    &        $K$        &     $1$      \\
	\cite{04Weinberger2021}        & DL &                  & k) M-BS-RIS-D\&S       &   $N$   & $L$                           &    $1$     & $R$                                    &        $K$        &     $1$      \\
	\cite{05Bansal2021}            & DL &    \checkmark    & h) BS-L-RIS-DorS*      &   $1$   & $1$                           &    $K$     & $N$                                    &    $K=K_1+K_2$    &     $1$      \\ \hline
	\cite{06Jolly2021}             & DL &    \checkmark    & h) BS-L-RIS-DorS*      &   $1$   & $M$                           &    $K$     & $N$                                    &       $2K$        &     $1$      \\
	\cite{07Fang2022}              & DL &                  & d) BS-RIS-S            &   $1$   & $M$                           &    $1$     & $N$                                    &        $K$        &     $1$      \\
	\cite{08Shambharkar2022}       & DL &    \checkmark    & g) BS-L-RIS-D\&S*      &   $1$   & $1$                           &    $K$     & $N$                                    &        $K$        &     $1$      \\
	\cite{09Weinberger2022csi}     & DL &    \checkmark    & b) BS-RIS-D\&S         &   $1$   & $L$                           &    $1$     & $N$                                    &       $K=2$       &     $1$      \\
	\cite{10Sena2022}              & DL &                  & f) BS-LC-RIS-D\&S*     &   $1$   & $4$                           &    $2$     & $50$                                   &        $2$        &     $1$      \\ \hline
	\cite{11LiuP2022}              & UL &                  & t) DorD\&S-RIS-BS      &   $1$   & $1$                           &            & $N$                                    &        $2$        &     $1$      \\
	\cite{12Dhok2022}              & DL &    \checkmark    & q) BS-L-STAR-R\&T      &   $1$   & $1$                           &    $K$     & $N_{kh} N_{kv}$     \checkmark             &       $2K$        &     $1$      \\
	\cite{13Li2022}                & DL &    \checkmark    & j) M-BS-LC-RIS-D\&S    & $N\ge1$ & $N_t$                         &    $L$     & $M$                                    &        $K$        &     $1$      \\
	\cite{14Weinberger2022}        & DL &    \checkmark    & k) M-BS-RIS-D\&S       &   $N$   & $L$                           &    $1$     & $R$                                    &        $K$        &     $1$      \\
	\cite{15Zhao2022}              & DL &                  & b) BS-RIS-D\&S*        &   $1$   & $N_t$                         &    $1$     & $N$                                    &        $2$        &     $1$      \\ \hline
	\cite{16Chen2022}              & DL &                  & d) BS-RIS-S            &   $1$   & $N_t$                         &    $1$     & $N_H N_V$      \checkmark              &       $K=2$       &     $1$      \\
	\cite{17Hashempour2022}        & DL & \checkmark (UER) & p) BS-STAR-R\&T        &   $1$   & $N_t$                         &    $1$     & $M$                                    &       $K+J$       &     $1$      \\
	\cite{18Katwe2022clustring}    & UL &                  & u) D\&S-LC-RIS-BS      &   $1$   & $N_r$                         &    $M$     & $N$                y                   &        $K$        &     $1$      \\
	\cite{19Lima2022}              & DL &                  & d) BS-RIS-S            &   $1$   & $N$                           &    $1$     & $L$                                    &       $K=2$       &     $1$      \\
	\cite{20Shambharkar2022edge}   & DL &                  & i) BS-L-RIS-S*         &   $1$   & $N_t$                         &    $K$     & $N$                                    &        $K$        &     $1$      \\ \hline
	\cite{21Gao2022}               & DL &    \checkmark    & b) BS-RIS-D\&S         &   $1$   & $1$                           &    $1$     & $N$                                   &        $K$        &     $1$      \\
	\cite{22Camana2022}            & DL &    \checkmark    & b) BS-RIS-D\&S*        &   $1$   & $N$                           &    $1$     & $M$                                    &        $K$        &     $1$      \\
	\cite{23Katwe2022shortPacket}  & DL &                  & e) BS-RIS-NUFU         &   $1$   & $M$                           &    $1$     & $N$                                    &        $K$        &     $1$      \\
	\cite{24Pang2022}              & DL &                  & l) BS-RIS2-D\&S\&T     &   $1$   & $N_t$                         &    $2$     & $M_1 \& M_2$                           &        $K$        &     $1$      \\
	\cite{25Singh2022}             & DL &    \checkmark    & d) BS-RIS-S*           &   $1$   & $1$                           &    $1$     & $N$                                    &        $K$        &     $1$      \\ \hline
	\cite{26Bansal2023}            & DL &    \checkmark    & g) BS-L-RIS-D\&S*      &   $1$   & $1$                           &    $K$     & $N$                &        $K$        &     $1$      \\
	\cite{27Li2023maxmin}          & DL &                  & b) BS-RIS-D\&S*        &   $1$   & $M$                           &    $1$     & $N$                                    &        $K$        &     $1$      \\
	\cite{28Gao2023}               & DL &                  & b) BS-RIS-D\&S         &   $1$   & $M$                           &    $1$     & $N$                                    &        $K$        &     $1$      \\
	\cite{29Soleymani2023workshop} & DL &                  & b) BS-RIS-D\&S         &   $1$   & $N_{BS}$                      &    $1$     & $N_{RIS}$                              &        $K$        &    $N_u$     \\
	\cite{30You2023}               & UL &                  & t) DorD\&S-RIS-BS      &   $1$   & $1$                           &    $1$     & $N$                                    &        $2$        &     $1$      \\ \hline
	\cite{31Aswini2023}            & DL &    \checkmark    & c) BS-RIS-DorS*        &   $1$   & $1$                           &    $1$     & $>1$                                    &       $1+1$       &     $1$      \\
	\cite{32Darabi2023}            & DL &                  & b) BS-RIS-D\&S*        &   $1$   & $M$                           &    $1$     & $N$                                    &        $K$        &     $1$      \\
	\cite{33Tian2023}              & DL &                  & b) BS-RIS-D\&S*        &   $1$   & $\ne 1$                       &    $1$     & $N$                                    &       $K+J$       &     $1$      \\
	\cite{34TianY2023}             & DL &                  & j) M-BS-LC-RIS-D\&S*     &   $F$   & $1$                           &    $N$     & $L$                                    &        $K$        &     $1$      \\
	\cite{35Katwe2023uplink}       & UL &                  & s) S-RIS-BS            &   $1$   & $1$                           &    $1$     & $N$                                    &        $K$        &     $1$      \\ \hline
	\cite{36Kim2023}               & DL &                  & d) BS-RIS-S            &   $1$   & $N_t$                         &    $1$     & $N_I$                                  &       $N_K$       &     $1$      \\
	\cite{37Soleymani2023signal}   & DL &                  & j) M-BS-LC-RIS-D\&S*   &   $L$   & $N_{BS}$                      &  $M\ge L$  & $N_{RIS}$                              &        $K$        &    $N_u$     \\
	\cite{38Sena2023}              & DL &                  & g) BS-L-RIS-D\&S       &   $1$   & $2\times M/2$                 &    $G$     & $D$                                    & $K=K_g \times G$  & $2 \times 1$ \\
	\cite{39Elganimi2023}          & DL &    \checkmark    & b) BS-RIS-D\&S         &   $1$   & $M$                           &    $1$     & $N_\text{REF}$         &       $K>M$       &     $1$      \\
	\cite{40Huang2023}             & DL &                  & b) BS-RIS-D\&S         &   $1$   & $N_t$                         &    $1$     & $L$                                    &        $N$        &     $1$      \\ \hline
	\cite{41Huang2023j}            & DL &                  & b) BS-RIS-D\&S         &   $1$   & $N_t$                         &    $1$     & $L$                                    &        $N$        &     $1$      \\
	\cite{42Li2023}                & DL &                  & b) BS-RIS-D\&S         &   $1$   & $N_t$                         &    $1$     & $L$                                    &        $K$        &     $1$      \\
	\cite{43Niu2023}               & DL &                  & b) BS-RIS-D\&S         &   $1$   & $N_s$                         &    $1$     & $N_R$                                  &        $L$        &     $1$      \\
	\cite{44Wu2023}                & DL &    \checkmark    & d) BS-RIS-S            &   $1$   & $K \times M_y M_z$ \checkmark &    $1$     & $K M_y M_z$         \checkmark         &        $K$        &     $1$      \\
	\cite{45Zhang2023}             & DL &                  & b) BS-RIS-D\&S         &   $1$   & $N_t$                         &    $1$     & $N$                                    &        $K$        &     $1$      \\ \hline
	\cite{46Karim2023}             & DL &    \checkmark    & b) BS-RIS-D\&S         &   $1$   & $1$                           &    $1$     & $N$                                    &        $K$        &     $1$      \\
	\cite{47Liu2023cognitive}      & UL &                  & t) DorD\&S-RIS-BS      &   $1$   & $1$                           &    $1$     & $N$                                    &        $2$        &     $1$      \\
	\cite{48Khisa2023}             & DL &                  & b) BS-RIS-D\&S*        &   $1$   & $N_t$                         &    $1$     & $M$                                    &        $2$        &     $1$      \\
	\cite{49Liu2023}               & DL &                  & m) Feed-T              &   $1$   & $1$                           &    $1$     & $M=M_r M_c$          y                 &       $N+K$       &     $1$      \\
	\cite{50Sun2023}               & UL &                  & r) D\&S-RIS-BS         &   $1$   & $M$                           &    $1$     & $N$                                    &        $2$        &     $1$      \\ \hline
	\cite{51Katwe2023}             & UL &                  & v) R\&T-STAR-BS        &   $1$   & $\ne 1$                       &    $1$     & $N$                                    &        $K$        &     $1$      \\
	\cite{52Yang2023}              & DL &                  & b) BS-RIS-D\&S         &   $1$   & $1$                           &    $1$     & $N$                                    &        $3$        &     $1$      \\
	\cite{53LiuP2023}              & DL &    \checkmark    & b) BS-RIS-D\&S*        &   $1$   & $N$                           &    $1$     & $M=M_x M_y$          y                 &        $K$        &     $1$      \\
	\cite{54Singh2023}             & DL &                  & d) BS-RIS-S            &   $1$   & $1$                           &    $1$     & $N$                                    &        $K$        &     $1$      \\
	\cite{55Xie2023}               & DL &                  & o) BS-STAR-R\&DT       &   $1$   & $1$                           &    $1$     & $L$                                    &       $2M$        &     $1$      \\ \hline
	\cite{56Mohamed2023}           & DL &                  & p) BS-STAR-R\&T        &   $1$   & $1$                           &    $1$     & $N$                                    &        $K$        &     $1$      \\
	\cite{57Wang2023multiF}        & DL &                  & o) BS-STAR-R\&DT*      &   $1$   & $N$                           &    $1$     & $M$                                    &       $K=2$       &     $1$      \\
	\cite{58Ge2023}                & DL &    \checkmark    & b) BS-RIS-D\&S         &   $1$   & $M$                           &    $1$     & $Q=Q_H Q_V$       \checkmark           &        $K$        &     $N$      \\
	\cite{59Lotfi2023}             & DL &                  & b) BS-RIS-D\&S         &   $1$   & $M$                           &    $1$     & $N$                                    &        $3$        &     $1$      \\
	\cite{60Hua2023}               & UL &    \checkmark    & r) D\&S-RIS-BS         &   $1$   & $M$                           &    $1$     & $N$             \checkmark             &        $K$        &     $1$      \\ \hline
	\cite{61LiB2023}               & DL &    \checkmark    & m) Feed-T              &   $1$   & $1$                           &    $1$     & $N N_e$                                &        $K$        &     $1$      \\
	\cite{62Zhao2023}              & DL &                  & o) BS-STAR-R\&DT       &   $1$   & $N_t$                         &    $1$     & $N$                                    &        $K$        &     $1$      \\
	\cite{63Huroon2023}            & DL &                  & k) M-BS-RIS-D\&S*      &   $G$   & $N$                           &    $1$     & $L$  &        $K$        &     $1$      \\
	\cite{64Xiao2023}              & DL &                  & o) BS-STAR-R\&DT*      &   $1$   & $1$                           &    $1$     & $K+L$                                  &      $2+Eve$      &     $1$      \\
	\cite{65Chen2024}              & DL &                  & b) BS-RIS-D\&S $^\ast$ &   $1$   & $M$                           &    $1$     & $N$                  &        $K$        &     $1$      \\ \hline
	\cite{66Tang2024}              & DL &                  & f) BS-LC-RIS-D\&S      & $L$     & $N_t$                         &    $R$     & $N$                                    &        $K$        &     $1$      \\
	\cite{67Meng2024}              & DL &                  & n) BS-STAR-DR\&DT      &   $1$   & $N$                           &    $1$     & $M$                                    &        $K$        &     $1$      \\
	\cite{68Pala2024}              & DL &    \checkmark    & f) BS-LC-RIS-D\&S      &   $1$   & $B$                           &    $M$     & $L$                                    &        $U$        &     $1$      \\ \hline
\end{tabular}
\end{table*}

\subsection{Models} \label{secModels}

\paragraph{BS-D}
There is no \gls{RIS} in this model, but it serves as a baseline in two ways: 1) using traditional beamforming techniques such as Weighted \gls{MMSE} \cite{Christensen2008} or 2) using \gls{RSMA} only without \gls{RIS} \cite{Mao2018}. 
This subsection describes the basic system model. 
Note that a matrix (denoted by capital bold letter) is used for the channel to the user, despite the prevailing assumption so far in the literature that a user is equipped with a single antenna, in which case the matrix is reduced to a vector. 
The $k$-th user receives \cite{Christensen2008}
\begin{equation}\label{eq:y_k_general}
	\mathbf{y}_k = \mathbf{H}_k \mathbf{x} + \mathbf{n}_k
\end{equation}

for the \gls{RIS}-free, \gls{RSMA}-free model, where $\mathbf{H}_k \in \mathbb{C}^{N_U\times N_B}$ is the direct channel from the \gls{BS} to the user $k$ which does not need to  be \gls{LoS} (e.g. it could be a Rayleigh fading link), $\mathbf{x} \in \mathbb{C}^{N_B}$ is the symbol transmitted by the \gls{BS}, and $\mathbf{n}_k \in \mathbb{C}^{N_Q}$ is the \gls{AWGN} at the $k$-th receiver. 
In addition, the transmitted symbol $\mathbf{x}$ is a combination of $K$ encoded symbols $\mathbf{s}_k \in \mathbb{C}^{N_U}$ that are intended for different users, and that combination is done linearly through a linear precoder represented by $\mathbf{P}_k \in \mathbb{C}^{N_B\times N_U}$.
That is $\mathbf{x} = \sum_{k=1}^{K} \mathbf{P}_k \mathbf{s}$.
Modelling and analysis of the  \gls{1L}-\gls{RS} will be discussed in Section \ref{downlink_problem_formulations}.

\paragraph{BS-RIS-D\&S}
This model is a simple downlink model and is useful to draw essential conclusions. It is the most studied model among all the models reviewed in this paper. 
This model assumes a single reflecting \gls{RIS} that can see all users, while the \gls{BS} still has a direct link to all users. 
Therefore, each user can receive a direct signal from the \gls{BS} and a reflected signal from the \gls{RIS}. 
This model is adopted in \cite{03Fu2021,09Weinberger2022csi,15Zhao2022,21Gao2022,22Camana2022,27Li2023maxmin,28Gao2023,29Soleymani2023workshop,32Darabi2023,33Tian2023,39Elganimi2023,40Huang2023,41Huang2023j,42Li2023,43Niu2023,45Zhang2023,46Karim2023,48Khisa2023,52Yang2023,53LiuP2023,58Ge2023,59Lotfi2023,65Chen2024}.
In \cite{03Fu2021}, discrete \gls{RIS} phase shifts are considered, and 2 to 5 users are considered in the simulation study.
A two-user system is considered in \cite{09Weinberger2022csi} and the \gls{CSI} is alternated between coherence blocks.
In \cite{21Gao2022}, a single-antenna \gls{BS} is assumed to serve multiple single-antenna users, and two users are considered in their simulation study.
Users in \cite{27Li2023maxmin} are grouped into multicast groups, where each group gets the same common and private messages.
\acrfull{IGS} is considered in \cite{29Soleymani2023workshop}, in addition to the possible I/Q imbalance at the receivers.
A single-antenna \gls{BS} serves $K$ single-antenna users in \cite{46Karim2023}, and an $N_R$-element \gls{RIS} assists the transmission.
Perfect \gls{CSI} is assumed in the analysis, and the  impact of imperfect \gls{CSI} is examined in the simulation study.

In \cite{40Huang2023} and the later journal article \cite{41Huang2023j}, a \gls{VR} streaming system is considered where the achievable rate of the $360\deg$ video is to be maximized.
They employ imitation learning and actor-critic in the solution of the optimization problem.

An active \gls{RIS} is considered in \cite{42Li2023}, and two users are assumed in the simulation study.
A \gls{UPA} active \gls{RIS} is also examined in \cite{53LiuP2023}, where the authors maximize the minimum rate and utilize \gls{2L}-\gls{RS}.
A hybrid antenna array is utilized at the \gls{BS} with digital and analog precoding.
In addition, users are grouped using coalition formation.

In \cite{32Darabi2023}, an active \gls{RIS} is considered for \gls{URLLC} transmission. Hence, short packets are considered.
This is motivated by the fact that the common rate of the \gls{RSMA} is limited by the rate support by the user having the lowest channel quality.
They consider an underloaded system where the number of users is less than the number of antennas.

The system proposed in \cite{15Zhao2022,48Khisa2023} builds upon the concept of cooperative-\acrshort{RSMA} \cite{Mao2020coopRS} to investigate the effect of incorporating the \gls{RIS}, and this combination proves useful. 
Two users are considered in the system model: a far user and a near user.
The system utilizes two time slots that are not necessarily equal.
In the first time slot, the \gls{BS} encodes the signal using the \gls{1L}-\gls{RS} protocol for both the near and the far user.
The near user adopts a \gls{NDF} relay mode to transmit the common stream once again to the far user at the second time slot.
Hence, the far user receives two copies of the common stream, one from the \gls{BS}, and the other from the near user.
A multi-user scenario is not considered.

In \cite{52Yang2023}, a transmitter, an \gls{RIS}, 2 (\gls{NOMA}) users, and 1 user (eavesdropper or warden)  are considered. 
Despite the two legitimate users being called \gls{NOMA} in \cite{52Yang2023}, their description fits \gls{1L}-\gls{RSMA}. 
They investigate \gls{DEP} (which is basically a hypothesis testing formulation), and covert communication rate. 
For \gls{DEP}, they use the gamma distribution to model the cascaded channel and then the \gls{DEP} expression is readily available. 
They discuss the effects of nulling one of the legitimate streams (that is, the effects of rate splitting), as well as the effect of the \gls{RIS} with different number of elements.
Direct channels from the transmitter to the users are assumed to experience Rayleigh fading, while channels through the \gls{RIS} experience Rician Fading.

In \cite{28Gao2023}, A system with \glspl{LU} and potential eavesdroppers is considered.
The authors study the two-legitimate-users system in addition to exploring the performance of three schemes for up to 6 \glspl{LU}.
The objective is to maximize the minimum \gls{SR}, and they show the advantage of \gls{RSMA} and \gls{RIS} over \gls{NOMA} and \gls{MU}-\gls{LP}.
In \cite{59Lotfi2023}, covert communications is considered where they maximize the covert rate to the covert user while maintaining a minimum rate for an ordinary user. 
The warden is the third user in the system. 

The authors in \cite{39Elganimi2023} focus on the overloaded case where they assume that the number of users is larger than the number of antennas at the \gls{BS}.
They consider two groups in the system model, the first group consists of $N_B$ users and is served through \gls{RSMA}, while the other group contains the extra users (more than the number of antennas at the \gls{BS}) is served through \gls{OMA}.
In fact, they employ \gls{TP}-\gls{RSMA} and \gls{PP}-\gls{RSMA}.
However, their model figure seems misleading as the direct link is considered in the analytical analysis. 
Hence, model (b) is used for classification.

The authors in \cite{43Niu2023} present an \gls{RSMA} network model with a single active \gls{RIS} and a single \gls{BS}. 
They assume an active \gls{RIS} which requires neither a \gls{DAC} nor an \gls{ADC}, and just utilizes power amplifiers and phase shifting circuits.
In the simulation study, the \gls{BS} is 10m high as well as the \gls{RIS} surface.
The users are assumed to be closer to the \gls{RIS}.

A \gls{SWIPT} system is considered in \cite{33Tian2023}, and the performance is compared to \gls{NOMA} and \gls{SDMA}.
A linear model is used for the energy harvester.
In addition, in the simulation study, the \glspl{ER} are located near the \gls{RIS} contrary to the \glspl{IR}.
Another correspondence considering the \gls{SWIPT} system model is \cite{22Camana2022}.
At each user, a power splitter delivers power to the \gls{EH} and information to the information decoder according to a splitting ratio.
A non-linear \gls{EH} model is adopted, and their objective is to minimize the transmit power while ensuring a minimum rate and a minimum harvested energy.
Imperfect \gls{CSI} is also considered.
The \gls{SWIPT} concept is also explored in \cite{45Zhang2023}. The transmitter employs \gls{1L}-\gls{RS}, and a single reflecting \gls{RIS} is used.
Each user has a power splitter after its receiving antenna that feeds two components: the information decoder and the energy harvester.
With \gls{RSMA} employed, the information decoder first decodes the common stream, removes it by \gls{SIC}, then decodes the private stream.
Furhtermore, they adopt a non-linear \gls{EH} model \cite{Boshkovska2015}. 
They use a \gls{PPO}-based \gls{DRL} method to maximize the \gls{EE}. 
The maximization of \gls{EE} will be discussed in Section \ref{maxEE}.

In \cite{58Ge2023}, users are assumed to have multiple antennas, which sets this article different. 
An \gls{MMSE} combiner is considered at the receiver, as well as the beamforming at the transmitter. 
However, the details of optimizing the common rate allocation and the \gls{RIS} phase shifts are not presented.
Furthermore, the simulation study does not consider single-antenna users for comparative performance study.

In \cite{65Chen2024}, an \gls{ISAC} system is considered, where the \gls{BS} does both communications and sensing of a target without using a special radar sequence. 
This study was motivated by previous studies on an \gls{ISAC} system either in an \gls{RIS}-assisted system, or an \gls{RSMA}-enabled system. 
In the \gls{RIS}-assisted \gls{ISAC} system, \gls{SDMA} is restricting the performance because it does not fully exploit the extra path created by the \gls{RIS}.
In addition, the gain of \gls{RSMA} can diminish in severe fading. 
Hence, the authors in \cite{65Chen2024} study the synergy of \gls{RSMA} and \gls{RIS} in a downlink \gls{MISO} system with a single target, and show the gains of both \gls{RSMA} and \gls{RIS} on the radar \gls{SNR}.

\paragraph{BS-RIS-DorS}
This model differs from the previous one \lq b) BS-RIS-D\&S\rq~ in the assumption that some users cannot receive the signal from the \gls{BS}, and hence can only be served through the \gls{RIS}. Meanwhile, the users served by the \gls{BS} do not receive any signal from the \gls{RIS}.
Only a single paper studies this model \cite{31Aswini2023}, where the author focuses on the capacity analysis for two users, one served by the \gls{BS} directly and the other served though the \gls{RIS} only.
Additional information can be found in Section \ref{DLcapacityanalysis}.

\paragraph{BS-RIS-S} \label{secTHz}
This model is a simple modification of \lq b) BS-RIS-D\&S\rq~ where the direct link from the \gls{BS} to the users is blocked. 
Hence, there is a single link from the \gls{BS} to the users which go through the \gls{RIS}, that is the secondary link.
This model is examined by \cite{07Fang2022,16Chen2022,19Lima2022,25Singh2022,36Kim2023,44Wu2023,54Singh2023}.
A fully-connected \gls{RIS} is considered in \cite{07Fang2022}.
A two-user system is considered in \cite{19Lima2022} where a \gls{UAV}-mounted \gls{RIS} provides the link to the blocked users.
The \gls{UAV} circulates a path of known radius with a constant velocity.
The authors in \cite{36Kim2023} present a grouping scheme for the impedance network of the \gls{RIS}.

In \cite{25Singh2022}, an \gls{RIS} is mounted on a \gls{UAV} along with a \gls{FD}-\gls{DF} relay to provide a channel link between a \gls{BS} and users.
The \gls{BS} and the users are equipped with a single-antenna each.
The relay decodes then re-encodes the signal using \gls{RSMA}.
A similar model is also studied by the same authors in \cite{54Singh2023} with more analyses.

References \cite{16Chen2022} and \cite{44Wu2023} consider terahertz models. 
In \gls{THz} networks, the channel matrix $\mathbf{H}$ can be modelled in a deterministic, or a statistical, or a hybrid way. 
The first one requires detailed knowledge of the specific installation site, while the second one can provide adequate model with some channel statistics \cite{Ning2023}. 
A popular model of the second type is the Saleh-Valenzuela (S-V) channel model \cite{Saleh1987}.
\gls{THz} channels show stronger directivity compared to \gls{mmWave} and \gls{cmWave}. 
Therefore, it is reasonable to use a rank-one matrix to model the channel, since more than 90\% of path gain is from the \gls{LoS} link \cite{Ning2023}. 
Consequently, shadowing will severely degrade the channel quality, and \gls{RIS}-assisted link may be valuable.
\gls{THz} signals, however suffer from degraded channel conditions due to molecular absorption loss and free-spreading loss\cite{Ning2023}. 
The former can generally be expressed as an exponential loss $e^{kd}$ where $d$ is the distance and $k$ is dependent on the frequency, temperature, and pressure and can be obtained from the HITRAN2012 database \cite{Rothman2013}. 
The latter is proportional to the squared frequency and squared distance\cite{Ning2023}. 
These significant losses support the use of \gls{UM}-\gls{MIMO} to provide a large gain to cover these losses.

\paragraph{BS-RIS-NUFU}
This model is similar to \lq c) BS-RIS-DorS\rq~ except that the near users can also receive a signal from the \gls{RIS}.
This model is studied by \cite{23Katwe2022shortPacket}. 
The main focus of \cite{23Katwe2022shortPacket} is on \gls{SPC}. 
The idea of \gls{SPC} is to reduce the packet size to achieve low latency at a low error probability, say, less than a millisecond with error probability less than $10^{-5}$.
The authors claim 80\% improvement in achievable rate for \gls{SPC} with \gls{RIS}.
In addition, motivated by the energy efficiency of \gls{RSMA} over \gls{NOMA} and \gls{SDMA} in general,
they propose that the \gls{RIS}-assisted \gls{RSMA} network can be suitable for \gls{SPC} maintaining low error probability as well as being energy efficient, especially for \glspl{FU} that are essentially blocked from any direct link to the \gls{BS}.
Their aim is optimizing the \gls{EE}. \gls{EE} maximization is discussed in Section \ref{maxEE}.

\paragraph{BS-LC-RIS-D\&S} \label{f}
This model is an extension of \lq b) BS-RIS-D\&S\rq~ where multiple \glspl{RIS} are possible. 
All \glspl{RIS} serve the same group of users. The users can also receive a signal directly from the \gls{BS}.
In this paper, $L$ stands for the number of \glspl{RIS}, and that is the \lq L\rq~ in the title of this model. 
In addition, $C$ stands for the common users. 
In other words, the same group of users can have $L+1$ links from the \gls{BS}.
This model is used by \cite{01Yang2020,10Sena2022,68Pala2024,66Tang2024}.
The authors in \cite{01Yang2020} consider a single \gls{BS}, along with multiple \glspl{RIS} and multiple users.
For the simulation study, they consider 3 \glspl{RIS}, 3 users, along with 3 transmit antennas at the \gls{BS}.

The article \cite{10Sena2022} is an overview of the potential advantages of \gls{RIS}-assisted \gls{RSMA} networks.
They start by noting the merits of \gls{SDMA} in providing the potential to reuse frequency over different spatial directions, 
yet this could be a source of interference to users in the same spatial direction.
Next, they motivate \gls{RSMA} by the robustness of \gls{CSI} imperfection compared to other \gls{OMA} techniques, as well as \gls{NOMA}.
In particular, they pose \gls{1L}-\gls{RS} as a middle strategy between precoding (\gls{ZF}) at the transmitter, and full \gls{SIC} at the receiver.
In addition, they also note that all private messages to the users are superimposed in the power domain to the combined message. 
Hence, each user does a single \gls{SIC} step.
They also mention the common dyadic channel model for communications links through the \gls{RIS}, and note possible construction materials for the \gls{RIS} like diodes and liquid crystals.
More importantly, they note three possible improvements of the \gls{RIS}-assisted \gls{RSMA} network, namely: rate enhancement, robustness to imperfect \gls{CSI}, as well as robustness to imperfect \gls{SIC}.
They illustrate each point with a figure based on a narrow-band, two-user, two-\gls{RIS} system model (where each user is served by a dedicated \gls{RIS}).
In the first figure, they show the advantage of the extra \gls{DoF} provided by the \gls{RIS} link, which enhance the optimization of the common rate of the \gls{RSMA} to meet the required common rate. Without the \gls{RIS}, this problem is somewhat difficult or infeasible.
In the second figure, they show that \gls{RSMA} and \gls{RIS}-\gls{RSMA} suffer less from the residual \gls{SIC} errors compared to \gls{NOMA} and \gls{RIS}-\gls{NOMA}. Comparison to \gls{TDMA} and \gls{RIS}-\gls{TDMA} is also presented.
In the third figure, the advantage of \gls{RIS}-\gls{RSMA} over \gls{RIS}-\gls{TDMA} and \gls{RIS}-\gls{NOMA} is shown in another realistic scenario, imperfect \gls{CSI}.
Finally, they note possible future advantages of \gls{RIS}-assisted \gls{RSMA} networks in the \gls{UAV}, high frequency, and \gls{LEO} satellites networks. This note is highly motivated by the robustness to \gls{CSI} imperfection, which happen in these systems due to various reasons.
In the focus of this paper, some articles considered \glspl{UAV} \cite{25Singh2022,26Bansal2023,54Singh2023,63Huroon2023,19Lima2022} and others considered \gls{THz} networks \cite{16Chen2022,44Wu2023}.
More on the \gls{THz} networks can be found in Section \ref{secTHz} (few paragraphs earlier).

In \cite{66Tang2024}, a mutli-cell multi-\gls{RIS} and multi user system is considered where the authors maximize the \gls{EE} for cell-edge users.
Users are divided intro groups and, within each group, users are assumed to request the same information from the \gls{BS}.
In the simulation study, three \glspl{BS}, three \glspl{RIS} and six users in three groups are considered.
In addition, the channel between the \gls{BS} and the users is assumed to follow Rayleigh distribution which implies non-dominant \gls{LoS} link in an urban micro cell environment with a path loss of $32.6+36.7\log_{10}(d)$ where $10\text{m} < d < 2\text{km}$. 
Furthermore, the channel between the \gls{BS} and the \gls{RIS} however, is assumed to follow Rician distribution suggesting a dominant \gls{LoS} path in an urban macro cell environment with path loss $35.6+22.0\log_{10}(d)$. 
Both path loss expressions assume a carrier frequency around 2.4GHz. \cite[Table B.1.2.1-1]{3GPP36.814}

A \gls{URLLC} system is studied in \cite{68Pala2024}. 
Slow fading channels and fixed \gls{RIS} deployment are considered.
It is assumed that the \gls{BS} has perfect \gls{CSI}, in addition to the \gls{RIS}.
However, the impact of channel estimation errors is investigated in the simulation study.
Each mobile broadband user has a target \gls{PEP} that must be satisfied in the throughput optimization.

\paragraph{BS-L-RIS-D\&S}
This model is a modification of the previous one \lq f) BS-LC-RIS-D\&S)\rq~ where the set of users served by each \gls{RIS} is different. 
Meanwhile all groups of users can have a direct link with the \gls{BS}.
Hence, the \lq C\rq~ is dropped from the title of this model.
This model is examined by \cite{08Shambharkar2022,26Bansal2023,38Sena2023}.
In \cite{08Shambharkar2022}, a single-antenna \gls{BS} serves multiple single-antenna cell-edge users.
Each user is associated to a dedicated \gls{RIS}, and can also have a direct link to the \gls{BS}.
Optimal and discrete phase shifts are considered, and channels are assumed to be Nakagami-m distributed.
A similar article with multi-antenna \gls{BS} is \cite{20Shambharkar2022edge} which assumes no direct paths to the users, hence classified in \lq i)BS-L-RIS-S\rq. 
A vehicular network is considered in \cite{26Bansal2023}, where a single-antenna \gls{UAV} serves vehicles.
The vehicles use the same spectrum and suffer from co-channel interference.
In the simulations study, they assume a two-lane road and a \gls{UAV} flying at a constant speed.
The operating frequency is assumed to be 240 MHz.
The authors in \cite{38Sena2023} utilize polarization multiplexing for the common and private rates to use \gls{RS} without \gls{SIC} at the receivers.

\paragraph{BS-L-RIS-DorS}
Following the same logic in separating users into groups as done from the model \lq f) BS-LC-RIS-D\&S\rq~ to the model \lq g) BS-L-RIS-D\&S\rq,
this model assumes that the \gls{BS} serves a group of users that is separate from the groups served by each \gls{RIS}.
Hence, each user receives a single message, either through the direct link from the \gls{BS}, or through the secondary link from the \gls{RIS}.
This model is used by \cite{02Bansal2021miso,05Bansal2021,06Jolly2021}.
The authors in \cite{02Bansal2021miso} utilize an on-off scheme for the \gls{RIS} phase shifts and study the \gls{2L}-\gls{HRS}.
They derive the outage probability for cell-edge and near users.
In \cite{05Bansal2021}, the closed-form expressions of the \gls{OP} for \gls{NU} and \gls{CEU} are derived, and the effect of the number of the \gls{RIS} elements on the outage behaviour for \gls{CEU} is explored, and they show performance gains over a \gls{DF}-\gls{RS} scheme and \gls{RIS}-\gls{NOMA} scheme.
In addition, in \cite{06Jolly2021}, the rate region for \gls{RIS}-assisted \gls{RSMA} is shown and compared to that of \gls{RIS}-assisted \gls{NOMA}.

\paragraph{BS-L-RIS-S}
This model is another simplification from the previous model \lq h) BS-L-RIS-DorS\rq, where the group of users served exclusively by the \gls{BS} is non-existent.
This model is examined by \cite{20Shambharkar2022edge} where a multi-antenna \gls{BS} serves multiple cell-edge users though a dedicated \gls{RIS} to each user.
Direct links between the \gls{BS} and the users are assumed to be weak and are ignored.
Elements of each \gls{RIS} are divided into groups equal to the number of transmission antennas at the \gls{BS}.
Channels are Rayleigh distributed, and discrete phase shifts are considered for the \gls{RIS}.

\paragraph{M-BS-LC-RIS-D\&S} \label{j}
This model assumes $M$ \glspl{BS} and $L$ \glspl{RIS}, where $M$ and $L$ are used in this paper to denote the number of \glspl{BS} and \glspl{RIS} respectively.
This model is similar to \lq f) BS-LC-RIS-D\&S)\rq~with the support of multiple \glspl{BS}.
This model is used in \cite{13Li2022,34TianY2023,37Soleymani2023signal}. 
The first article \cite{13Li2022} is described in the next paragraph.
The authors in \cite{34TianY2023} assume a \gls{CoMP} system.
In \cite{34TianY2023}, some users are blocked by an obstacle and do not get a direct link to any \gls{BS}.
Channels through the \glspl{RIS} are assumed to follow the Nakagami-$m$ distribution while channels without the \gls{RIS} follow the exponential distribution.
Few \glspl{RIS} are chosen to serve some users.
It is also assumed that the number of \glspl{BS} (or access points) are larger than the number of the users.
In the third article \cite{37Soleymani2023signal}, a multicell \gls{MIMO} model is used where users can receive signals from neighbouring cells. 
A motivation for this model is that it is one of the most practical scenarios.

The aim of \cite{13Li2022}, which is part of a three-article tutorial series on \gls{RSMA}, is to motivate the integration of \gls{RSMA} and \gls{RIS}. 
They briefly note some advantages of \gls{RSMA} including enhanced spectral efficiency, robustness to \gls{CSI} imperfection, and \gls{SIC} errors, as well as latency and mobility.
In addition, they also note the merits of the \gls{RIS} (e.g. passive, cheap, and configurable). 
Furthermore, they note that the \gls{RIS} can enhance the system by providing extra links, while making \gls{CSI} estimation more difficult, 
which is an opportunity to use \gls{RSMA} which is known for its robustness to \gls{CSI} errors, even with the simplest form of \gls{RSMA}: \gls{1L}-\gls{RS}.
In the paper, the models of the two most common schemes of \gls{RSMA}, namely, \gls{1L}-\gls{RS} and \gls{2L}-\gls{RS}, are  mentioned.
In addition, the authors mention the three architectures of \gls{RIS}, namely, single-connected, block connected, and fully-connected architectures.
They argue that this multi \gls{BS} model (termed as \lq j) M-BS-LC-RIS-D\&S\rq~in this paper) serves as a general framework for existing articles at the time, namely, 
\begin{itemize}
	\item b) BS-RIS-D\&S used by \cite{03Fu2021}.
	\item d) BS-RIS-S used by \cite{07Fang2022}.
	\item f) BS-LC-RIS-D\&S used by \cite{01Yang2020}.
	\item h) BS-L-RIS-DorS used by \cite{02Bansal2021miso,05Bansal2021,06Jolly2021}.
	\item k) M-BS-RIS-D\&S used by \cite{04Weinberger2021,14Weinberger2022}.
\end{itemize}

Furthermore, \cite{13Li2022} provides a rate region plot assuming a single \gls{BS} and a single \gls{RIS}. 
That is model \lq b) BS-RIS-D\&S\rq~ in this paper. 
The rate region provided by \cite{13Li2022} considers both perfect and imperfect \gls{CSI}, and shows consistent advantage of \gls{RSMA} over \gls{NOMA} and \gls{SDMA},
or at least similar level of performance as \gls{SDMA} for some rate pairs in the perfect \gls{CSI} condition.

\paragraph{M-BS-RIS-D\&S}
This model is a simplification of the previous one \lq j) M-BS-LC-RIS-D\&S\rq~where only a single \gls{RIS} remains in the model. 
This model is used by \cite{14Weinberger2022} and the earlier conference version \cite{04Weinberger2021}, in addition to \cite{63Huroon2023}. 
A \gls{C-RAN} network is used by \cite{04Weinberger2021,14Weinberger2022}, where multiple \glspl{BS} can serve the same user, and the \glspl{BS} are all connected to a central processor.
The \gls{C-RAN} is motivated by the cooperative interference reduction, which can ultimately improve the transmission rates at the same power.
The work in \cite{14Weinberger2022} adopts the  \gls{RS} approach of \cite{Gou2011}. 
In particular, the central processor splits the messages and then distributes the splitted messages to the desired \glspl{RRU}.
For the simulation study, they assume noise power spectral density of -169 dBm/Hz and the path loss model $148.1+37.6\log_{10}(d)$ for 2GHz \cite[Table A.2.1.1.2-3]{3GPP36.814}. 
On the other hand, \cite{63Huroon2023} is using multiple \glspl{UAV} as \glspl{BS}.

\paragraph{BS-RIS$^2$-D\&S\&T}
This model is  special  compared to the previous models.
In this model, two \glspl{RIS} are assumed in cascade. 
Meanwhile, the users can still receive a signal from the \gls{BS} directly, or from the first \gls{RIS}, in addition to the signal from the second \gls{RIS}.
There is only a single paper in this class \cite{24Pang2022}, where the authors aim to maintain a minimum power delivery and rate to multiple users through a double (cascade) \gls{RIS} system. 
Building on top of \eqref{eq:y_k_general}, an extra factor $\sqrt{\rho_k} \in [0,1], \forall k$ denoting the \gls{PS} ratio for the $k$th user is multiplied by the channel $\mathbf{H}_k$ which include 4 paths between the \gls{BS} and the $k$th user:
\begin{align}
	\mathbf{H}_k = &\mathbf{H}_{Bk} + \mathbf{H}_{R2 k} \mathbf{\Phi}_2 \mathbf{H}_{R2R1} \mathbf{\Phi}_1 \mathbf{H}_{BR1} +\\ \notag
	&\mathbf{H}_{R1 k} \mathbf{\Phi}_1 \mathbf{H}_{BR1} + \mathbf{H}_{R2 k} \mathbf{\Phi}_2 \mathbf{H}_{BR2}, 
\end{align}
where $\mathbf{H}_{Bk}$, $\mathbf{H}_{R1 k}$, and $\mathbf{H}_{R2 k}$ are the channels to the single antenna user and are $\in \mathbb{C}^{1\times N_B}$; $\mathbf{\Phi}_1 \in \mathbb{C}^{N_{R1}\times N_{R1}}$ and $\mathbf{\Phi}_2 \in \mathbb{C}^{N_{R2}\times N_{R2}}$ are both diagonal with unit-norm constraints; 
$\mathbf{H}_{R2R1} \in \mathbb{C}^{N_{R2}\times N_{R1}}$ is the channel from the first \gls{RIS} to the second \gls{RIS};
Finally, $\mathbf{H}_{BR1}$ and $\mathbf{H}_{BR1}$ are the channels from the \gls{BS} to the first and second \glspl{RIS}.
In summary, the four links to each user are: a direct link to the \gls{BS}, a link with double reflection through both \glspl{RIS}, and a link with a single reflection through either \gls{RIS}.

\paragraph{Feed-T}
It was pointed out \cite{DiRenzo2020} that the \gls{RIS} may be advantageous in the near field. 
This model assumes a \acrfull{TRIS} in the near field of a transmitting antenna at the \gls{BS},
and is used by \cite{49Liu2023,61LiB2023}.
This is quite an uncommon idea is to utilize a \gls{UPA} \gls{TRIS} in the near field (separation less than the Rayleigh distance) of a horn antenna to form a \gls{CBS} \cite{49Liu2023}. That is, a \gls{TRIS} is used as a  \gls{BS} in a cognitive radio network.  
The \gls{TRIS} is a \gls{UPA}, and \glspl{CSI} of \glspl{PU} and \glspl{CU} are assumed to be available at the transmitter. 

In \cite{61LiB2023}, a \gls{TRIS} is placed in front of an antenna at the transmitter in a \gls{MISO}-\gls{BC} channel with $K$ users. 
The \gls{RIS} has $N$ sub-arrays and each array has $N_e \ge K+1$ elements. 
With $N$ sub-arrays, the performance is equivalent to an $N$-antenna system. 
Yet, the transmissive \gls{RIS} system is cheaper and consumes less power.
The authors do not consider the channel between the transmit antenna and the \gls{RIS}, and they consider that each sub-array of the \gls{RIS} as a transmitter, therefore, the received signal at the $k$th user is
\begin{equation}
	y_k = \mathbf{h}_k^\mathsf{H} \mathbf{x} + n_k
\end{equation}
where $n_k \sim \mathcal{N}(0,\sigma_{n,k}^2)$, and it is assumed that  $\sigma_{n,k}^2 = \sigma_{n}^2 \forall k$, and $\mathbf{x}$ is a linear combination of messages weighted by the transmit power and \gls{RIS} coefficients according to the \gls{1L}-\gls{RS} protocol.

\paragraph{BS-STAR-DR\&DT}
This model and the next three models consider a \gls{STAR}-\gls{RIS}.
In this model, all users (before and behind the \gls{STAR}-\gls{RIS} surface) can get a direct link from the \gls{BS}.
That is, users in the transmit-domain of the \gls{STAR}-\gls{RIS} surface as well as users in the reflect-domain of the \gls{STAR}-\gls{RIS} surface can \lq see\rq~ the \gls{BS}.
This model is examined in \cite{67Meng2024} where the authors utilize a \gls{PPO}-based algorithm for sum-rate maximization.

\paragraph{BS-STAR-R\&DT}
This model is a simplification of the previous model \lq n) BS-STAR-DR\&DT\rq, where only the users in the reflect-domain of the \gls{STAR}-\gls{RIS} can get a direct link from the \gls{BS}.
While the users in the transmit-domain of the \gls{STAR}-\gls{RIS} are only served through the surface.
This model is considered in \cite{55Xie2023,57Wang2023multiF,62Zhao2023,64Xiao2023}.
An active \gls{STAR}-\gls{RIS} system is studied in \cite{55Xie2023}, and compared to the passive \gls{STAR}-\gls{RIS}.
All channels are assumed to be Rician channels.
A multi-functional \gls{RIS} is proposed in \cite{57Wang2023multiF} which retains the full space coverage like the \gls{STAR}-\gls{RIS}, in addition to resolving the problem of double-fading attenuation.
A \gls{STAR}-\gls{RIS} is assumed in \cite{62Zhao2023} along with cooperative \gls{RS} and the authors maximize the worst rate.
The authors in \cite{64Xiao2023} assume two \acrfullpl{LU}: Bob and Grace, one transmitter: Alice, one \gls{STAR}-\gls{RIS}, and one Eve.
Nakagami-$m$ channels are considered.
The \gls{RIS}'s phase shifts uncertainty \cite{Lv2022} as well as deliberate fading of the signal towards the eavesdropper can be useful in achieving covert communications.

\paragraph{BS-STAR-R\&T}
This model is a simplification of the previous one \lq o) BS-STAR-R\&DT\rq, where no users have a direct link from the \gls{BS},
and all users are served exclusively through the \gls{STAR}-\gls{RIS}.
This model is studied by \cite{17Hashempour2022,56Mohamed2023}.
In \cite{17Hashempour2022}, a \gls{BS} communicates with $K$ \glspl{IR} in the transmission half-space of the \gls{STAR}-\gls{RIS} as well as $J$ \glspl{UER} in its reflection half-space. 
All users are equipped with a single antenna, and no direct link from the \gls{BS} to any user. \glspl{UER} are considered potential eavesdroppers, and their \gls{CSIT} is unknown, unlike that of the \glspl{IR}. 
Therefore, the primary objective is to maximize the \gls{WCSSR} while maintaining a minimum sum energy for the \glspl{UER}.
In \cite{56Mohamed2023}, Stochastic geometry tools are employed to study the outage probability.
Both \gls{ES} and \gls{MS} modes are studied, along with Rician channels.
The users are assumed to be distributed according to a Poisson point process (PPP).

\paragraph{BS-L-STAR-R\&T}
This model assumes multiple ($L$) \gls{STAR}-\gls{RIS} surfaces, and that each surface serves two users: one in the reflect-domain, and the other in the transmit-domain.
This model is assumed in \cite{12Dhok2022}, with a pair of users served by each \gls{STAR}-\gls{RIS}.
The \gls{STAR}-\gls{RIS} is assumed to be a \gls{UPA}. 
The inter-element spacing is studied, along with \gls{ES} mode and \gls{MS} mode. 
In addition, the channels at the \gls{STAR}-\gls{RIS} are assumed correlated.

\paragraph{D\&S-RIS-BS}
Contrary to all previous models, this model and the next four models are uplink models.
This model is similar to \lq b) BS-RIS-D\&S\rq, except that the users are transmitting to the \gls{BS}.
That is, the \gls{BS} can receive a direct signal from each user in addition to the reflected signal through the \gls{RIS}.
This model is assumed in \cite{50Sun2023,60Hua2023}.
In \cite{50Sun2023}, two users are assumed where a single user does \gls{RS}.
In \cite{60Hua2023}, all users perform \gls{RS} and are assumed to be moving with a speed of $1.5$m/s in the system level simulation.
In addition, the \gls{BS} is assumed to have a \gls{ULA} of antennas and the \gls{RIS} is formed of a \gls{URA}.

\paragraph{S-RIS-BS}
This model is a simplification of the previous one, where the \gls{BS} only gets a single signal from each user through the \gls{RIS}.
This model is used in \cite{35Katwe2023uplink} where the \gls{RIS} is vital in supporting the users with no \gls{LoS}.

\paragraph{DorD\&S-RIS-BS}
This model is simple despite the complicated-looking title. 
In this model, the \gls{BS} either receives a direct signal from the user, or receives a direct and a reflected signal from the user.
This model is studied in \cite{11LiuP2022,30You2023,47Liu2023cognitive}.
The \gls{RIS} assists the secondary user which performs \gls{RS} in \cite{11LiuP2022}, which they term cognitive-\gls{RSMA}.
Similar cognitive-radio-inspired systems are presented in \cite{30You2023} and \cite{47Liu2023cognitive}.

\paragraph{D\&S-LC-RIS-BS}
This model is similar to \lq f) BS-LC-RIS-D\&S\rq, but in the uplink setting.
In this model, the \gls{BS} can have $L+1$ links from each user.
This model is examined in \cite{18Katwe2022clustring} where a \gls{mmWave} system is considered and users are dynamically clustered using $k$-means clustering before optimizing the resource allocation.

\paragraph{R\&T-STAR-BS}
This model is similar to \lq p) BS-STAR-R\&T\rq, but in the uplink setting.
That is, users communicate with the \gls{BS} exclusively through the \gls{STAR}-\gls{RIS}.
This model is used by \cite{51Katwe2023} where the authors focus on improving the spectral efficiency, fairness, and coverage probability for dead-zone users.
A phase-shift coupled \gls{STAR}-\gls{RIS} is assumed.

\subsection{Performance Metrics}
This section presents a few performance metrics that were used by different authors.
However, most authors utilize the main objective function such as \gls{WSR} for comparison and performance evaluation.
Different objective functions are discussed in the downlink section, in addition to \gls{WSR} for the uplink section, Section~\ref{uplink_resource_allocation}.

\subsubsection{\acrfull{JFI}}
When the users are divided into $L$ groups, 
and the rate of each group is $r_l$, 
the Jain's Fairness Index (JFI) is defined as \cite{Jain1998}
\[
\text{JFI} = \frac{\left( \sum_{l=1}^{L} r_l \right)^2}{L \sum_{l=1}^{L} r_l^2},
\]
where $0 \le \text{JFI} \le 1$. For example, the highest 
possible fairness can be achieved when all group rates $r_l$ are equal. When the JFI is 0.2, for example, then the system is unfair to 80\% of the users. This index is used by \cite{27Li2023maxmin}.
In addition, \cite{18Katwe2022clustring} also use this index for each decoding order, and they choose the decoding order that results in a higher \gls{JFI}.

\subsubsection{Level of Supportive Connectivity (LoSC)}
In an IRS-assisted C-RAN system, the \gls{LoSC}, introduced in \cite{14Weinberger2022} is  
the number of \acrshort{RRU}-to-user links divided by the number of \glspl{RRU}.
Whenever the fronthaul capacity of the \glspl{RRU} does not support all users, a minimum of one user can be imposed, and the number of dropped links will be indicated by \gls{LoSC}. 
On the other hand, when the \glspl{RRU} have enough capacity then they can serve all users, and the number of dropped links approaches zero.

\subsubsection{\acrlong{DEP}}
Consider the classical Alice-Bob-Willie covert communications model \cite{Chen2023covert}, where Alice want to transmit to Bob without Willie figuring out that the transmission took place. \acrfull{DEP} is
\begin{equation}
	\text{DEP} = \mathbb{P}_\text{FA} + \mathbb{P}_\text{MD},
\end{equation}
where \acrfull{FA} is the probability that Willie detects a non-existent transmission and \acrfull{MD} is the probability of Willie falsely thinking that Alice is silent. Willie would like the \gls{DEP} to be zero, and Alice would like the \gls{DEP} to be unity. The formulation of the \gls{DEP} is based on a binary detection problem that can be characterized by first finding the probability distributions of Willie's observations in both cases \cite{Chen2023covert}. Note that Willie has to set (or estimate) a detection threshold.  \gls{DEP} is considered in \cite{52Yang2023}.

\subsubsection{User Rate Satisfaction}
This measure can be useful to show the improvement of the \gls{RIS}-assisted system in terms of coverage where 
the user rate is plotted against the user distance from the \gls{BS}, and that rate will be reducing because of large-scale fading. 
However, the reduction may be slower in the case of \gls{RIS}-assisted system.
For instance, for a \gls{DL} \gls{MU}-\gls{MIMO} \gls{RSMA} system of 2 users and one \gls{RIS}, \gls{STAR}-\gls{RIS} system was shown to maintain satisfied rate (above 1bps/Hz) for the 2 users, compared to 1 user with the \gls{RO}-\gls{RIS}, and zero users without \gls{RIS} for distances between 30m and 180m.

\subsubsection{Constraint Satisfaction Ratio}
In \cite{67Meng2024}, the \gls{CSR} is defined 
to quantify the satisfaction of the minimum rate requirements of each user ($S_k$) and the power available at the transmitter ($S_p$). Denoting \lq Satisfaction\rq~by \lq S\rq, the \gls{CSR} satisfaction is a sum of two terms as follows: 
\[
\begin{aligned}
	S &= \frac{w_u}{K} \sum_{k=1}^K S_k + w_p S_p,
\end{aligned}
\]
where the sum of the weights $w_u$ and $w_p$ is unity, $S_p$ is zero if the power constraint is violated and unity otherwise, and $S_k$ is unity when the rate of the user is satisfied. More precisely,
\[
S_k = 
\begin{cases}
	1, & \text{if } (a_k + r_k) \ge Q_k \\
	e^{(a_k + r_k) - Q_k}, & \text{if } (a_k + r_k) < Q_k. \\
\end{cases}
\]

\subsubsection{Diversity Order}
In the high \gls{SNR} regime, the number of independent paths can be measured using the diversity order \cite{rsadve_mimo_lec}
\begin{equation}
	\Delta = \underset{\text{SNR} \rightarrow \infty}{\lim} - \frac{\log( \text{BER})}{\log( \text{SNR})},
\end{equation}
which can also be expressed in terms of the  outage probability \cite{Laneman2004}
\begin{equation}
	\Delta = \underset{\text{SNR} \rightarrow \infty}{\lim} - \frac{\log( \mathbb{P}_\text{outage})}{\log( \text{SNR})}.
\end{equation}
When $\Delta$ is small, the system is less robust to fading. 
On the contrary, when the outage probability decays faster with the increasing \gls{SNR}, the system is more robust to fading and $\Delta$ is high.
$\Delta$ is a more useful measure in the high \gls{SNR} regime.
In \cite{55Xie2023}, the diversity order is studied for active \gls{STAR}-\gls{RIS} with \gls{1L}-\gls{RS}.

\subsubsection{Latency}
One way to address latency is to assume a short packet in the problem formulation \cite{23Katwe2022shortPacket,25Singh2022,54Singh2023,68Pala2024}. 
Latency can also be addressed in the optimization problem by imposing conditions on the achievable rates, where these conditions relate to the error probability and packet size  \cite{23Katwe2022shortPacket}. 
Hence, if the transmission is successful, for given packet size and error probability, then the comparisons of the optimization objectives can be carried out to compare different schemes, e.g. \gls{NOMA} and \gls{RSMA}.

Some of the performance measures described above can be used with different design objectives. 
Optimization formulations will need to be carried out to allocate resources for the specified design objectives.
The next section examines \acrfull{DL} resource allocation while the following one examines \acrfull{UL} resource allocation.

\section{Downlink Resource Allocation}
\label{downlink_resource_allocation}
To support multiple users in a communications network, available resources must be allocated to serve the users. 
Limiting factors include the available resources such as power, time, frequency bands, \gls{RS} ratios, \gls{RIS} phase shifts, as well as the fairness or \gls{QoS} requirements. 
Ideally, the resources should be allocated optimally in the sense that we get the maximum possible efficiency. 
Often the best allocation strategy cannot be inferred directly in a simple closed-form formula. 
After defining the objective  function (e.g. for maximizing the sum-rate or maximizing the minimum-rate), an algorithm is devised to obtain the power allocations, \gls{RS} ratios, \gls{RIS} phase shifts, etc. 
This algorithm may utilize traditional optimization frameworks or a machine learning framework. 
In the next subsection, Table~\ref{tab:optimizations-across-papers} is introduced along with the abbreviations used in it.
Next, few notes are mentioned about terminology (Section~\ref{secOptTerminology}), capacity (Section~\ref{DLcapacityanalysis}) and outage analyses (Section~\ref{section-outage-analysis}).
Then, the main part of this section follows with a discussion on various optimization formulations along with selected results from the literature, in Section~\ref{downlink_problem_formulations}.
Before concluding with lessons learned in Section~\ref{DLlessons}, an example scenario is presented in Section~\ref{exampleScenario} which compares \gls{DPC}, \gls{MU}-\gls{LP}, and \gls{1L}-\gls{RS}, with and without an \gls{RIS}.

\subsection{Overview of Resource Allocation Studies} \label{secSummaryTable}
This subsection summarizes the main objectives of the surveyed papers as shown in Table~\ref{tab:optimizations-across-papers}, including those with uplink models.
The next subsections will expand more on capacity and outage analyses, as well as optimization problem formulations for the downlink models.
Table~\ref{tab:optimizations-across-papers} provides an overview of the optimization formulations by different authors. 
The table also includes uplink papers which are the subject of the next section. 
The table shows the optimization objective, the constraints, the general classification of the solution methodology, as well as the number of baseline schemes used in their simulations section. 
The \lq Objective\rq~column follows regular abbreviations in this article, except the new symbol $\otimes$ standing for cross-polar interference. 
For example, \acrshort{SR} is used for \acrlong{SR} in this article, hence, \acrshort{WSR} is used for sum-rate.
The `Constraints' column in Table~\ref{tab:optimizations-across-papers} follows the following notations:
\begin{itemize}
	\item P: Power constraint, that is the total power is less than or equal the available power budget.
	\item $\mathbb{P}$: The power of the active \gls{RIS} is limited by some maximum limit.
	\item D: Decodable common rate, that is the common rate of \gls{RS} is decodable by all users. This is the condition in \eqref{1lrsDecodable}. This condition is a requirement for \gls{SIC}. 
	\item C: Common rates of \gls{RS} are $\ge 0$ \eqref{1lrsNonNegative}. 
	\item $\mathbf{\Pi}$: decoding order permutations belong to the set $\mathbf{\Pi}$.
	\item $\approx$: user fairness. 
	\item :: prioritization of sub-messages or split proportions.
	\item U: Unit norm \gls{RIS} phase shifts.
	\item $\bigcap$: maximum norm for \gls{RIS} phase shifts is unity.
	\item $\mathbb{D}$ Dual polarized \gls{RIS}. Block matrices are diagonal and their magnitude is $\le1$, i.e. passive.
	\item $\tau$: The \gls{RIS} phase shifts belong to the range $[0,\tau)$, where $\tau=2\pi \approx 6.283185307179586$.
	Usually, this condition implies the unit norms condition (U) and vice-versa.
	\item $\Xi$: The \gls{RIS} phase shifts belong to a pre-determined set of values $\Xi$.
	\item b: The phase shifts of the \gls{RIS} are binary, that is 0 or 1 (on or off). 
	\item F: $\mathbf{\Theta} \in$ \acrshort{FC} means theta satisfies the two conditions on the scattering matrix of the fully connected RIS: $\theta=\theta^T$, $\theta^H$, $\theta=I$. Note that $\mathbf{\Theta}$ is different from the \gls{RIS} phase shifts matrix $\mathbf{\Phi}$.
	\item G: Group connected \gls{RIS} model, diagonal blocks.
	\item T: Transpose of the impedance matrix of RIS impedance model $\mathbf{X} = \mathbf{X}^T$.
	\item j: The definition for each group of impedance in the \gls{RIS} impedance model $(j\mathbf{X}+\mathbf{Z}_0\mathbf{I})^{-1}(j\mathbf{X}-\mathbf{Z}_0\mathbf{I})$
	\item $\rhd$: active \gls{RIS} conditions.
	\item *: \gls{STAR}-\gls{RIS} constraints.
	\item M: A minimum rate for each user.
	\item $M_1$: A minimum rate for a single user.
	\item w: receive beamforming 
	\item Q: Other \gls{QoS} constraints.
	\item $\alpha$: transmit power allocation factor between 0 and 1. 
	\item o: other constraints.
	\item $\rho$: In a \gls{SWIPT} system with the \gls{PS} mode, the power splitting ratio $\rho_k$ satisfies $0\le \rho_k\le1 \forall k$.
	\item E: Energy harvesting minimum requirement for user $k$ is $E_k\ge \gamma_k \forall k$.
	\item xyz: upper and lower bounds on the coordinates $x,y,$ and $z$ of the \gls{RIS}.
	\item $\mapsto$: Block length constraints.
\end{itemize}

Due to space limitations in Table~\ref{tab:optimizations-across-papers}, not all the details are shown.
In this table, the first listed method or objective is detailed in the subsequent columns.
For example, \cite{45Zhang2023} proposes a \gls{PPO} method, and an \gls{AO} method, but the last column describes only the \gls{PPO} method.
Another example is \cite{68Pala2024} where the maximization of the sum rate is considered, which is detailed in the next columns. But \cite{68Pala2024} also extends their formulation to another minimum latency formulation that seeks to reduce the block length. This extension, albeit mentioned in the \lq Objective\rq~column, is not detailed in the subsequent columns.

Also, due to space limitations, the `Baselines' column does not always list the total count and exact list of the baseline schemes. 
However, it lists the main acronyms used as baselines without detailing the exact combinations. The most combinations are however intuitive.
For instance, the table lists 3 keywords for \cite{66Tang2024}, which correspond to 5 baselines, namely, \lq \acrshort{SDMA}, \acrshort{SDMA}\textbackslash\acrshort{RIS}, \acrshort{NOMA},  \acrshort{NOMA}\textbackslash\acrshort{RIS}, \acrshort{RSMA}\textbackslash\acrshort{RIS}\rq.

\begin{figure*}[!t]
	\centering
	\includegraphics[width=\textwidth]{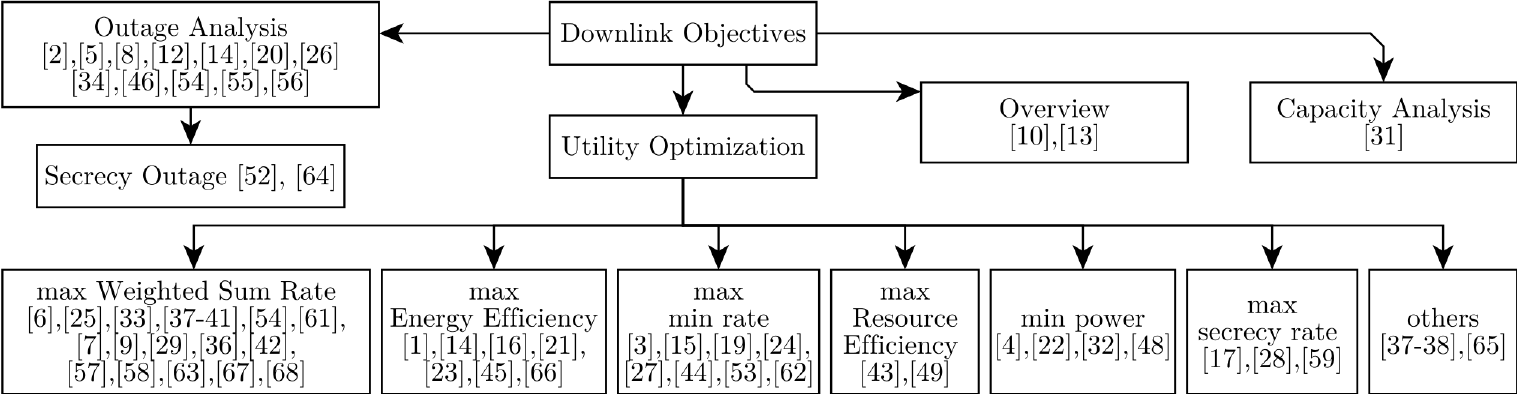}
	\caption{Classification of the downlink papers based on their focus.}
	\label{fig:downlink_objectives}
\end{figure*}

Another view of the first two columns of Table~\ref{tab:optimizations-across-papers} is depicted in Fig.~\ref{fig:downlink_objectives} and Fig.~\ref{fig:uplink_objectives} where the papers are classified according to their focus.
This section explores the resource allocation problem in the \gls{BC} channel, in addition to capacity analysis and outage analysis. Before that, few notes on the terminology are presented. The \gls{MAC} channel is the focus of the next section, but table~\ref{tab:optimizations-across-papers} also includes papers with uplink models.

\begin{table*}
	\caption{Signalling optimizations methods across the reviewed papers.}
	\label{tab:optimizations-across-papers}
	\centering

\begin{tabular}{ |c|c|l|l|c|c|l|}
	\hline
	Ref                            &      Class      & Objective                          & Constraints                             &                                                                      Baselines                                                                      &                Method                & Brief on Method                                                      \\ \hline
	\cite{01Yang2020}              &       opt       & max \acrshort{EE}                  & CDP$\tau$M                              &                                                       2: \acrshort{OFDMA} \& \acrshort{NOMA}                                                        &            \acrshort{AO}             & 1)$\mathbf{\Phi}$ 2)$\mathbf{P}$\&$\mathbf{c}$                       \\
	\cite{02Bansal2021miso}        &        -        & \hfill outage                      & -                                       &                3: \acrshort{MU}\acrshort{LP}, \acrshort{1L}-\acrshort{RS}, \acrshort{2L}-\acrshort{HRS}\textbackslash\acrshort{RIS}                 &                  -                   & -                                                                    \\
	\cite{03Fu2021}                &       opt       & max \acrshort{MR}                  & CDP$\Xi$    \hfill o                    & 4: \acrshort{MU}\acrshort{LP}?, \acrshort{1L}-\acrshort{RS}\textbackslash\acrshort{RIS}, \acrshort{MRT}, \acrshort{MRT}\textbackslash\acrshort{RIS} &            \acrshort{AO}             & 1)$\mathbf{P}$, $\mathbf{c}$ 2) $\mathbf{\Phi}$                      \\
	\cite{04Weinberger2021}        &       opt       & min power                          &                                         &                                                                                                                                                     &            \acrshort{AO}             &                                                                      \\
\hline
\end{tabular}

\end{table*}

\subsection{Notes on Optimization Terminology}\label{secOptTerminology}
The utility functions used by various authors (e.g. to maximize sum-rate or maximize energy efficiency) are non-convex problems. 
Solving these non-convex problems can be done using different methods that differ in complexity, accuracy, time, and convergence guarantees. Often surrogate functions, which are temporary (ideally convex) functions, are used to solve an approximated instance of the original problem near some initial point. 
Both \gls{SCA} and \gls{MM} approaches employ surrogate functions. 

The idea of \gls{SCA} is to start with a feasible solution and construct an approximation to the function in a block of variables (one or more variables) that is convex and easy to solve. 
Then, \gls{SCA} goes to the next block of variables, solve another approximate convex problem. 
\acrfull{SCA} does not necessarily converge to the global optimum, but often proofs for convergence to a stationary point can be established \cite{Scutari2018}. 

\acrfull{MM} has two steps \cite{Sun2017}. 
For a minimization problem, the first step is to find a surrogate function that always bounds the function of interest from above and intersects (or comes arbitrarily close) to the initial point. 
The next step is to minimize that surrogate function to get another initial point, and the cycle repeats. Constructing the surrogate function is an essential step of the algorithm. 
The desirable features of the surrogate function include convexity, fast convergence rate, and lower evaluation cost per iteration. 
More details on \gls{MM} along with techniques to construct surrogate functions can be found in \cite{Zhang2007,Sun2017}.

\acrfull{BCD} is an optimization framework where a block of variables are selected and the problem is optimized with respect to these variables. Next, another block is selected and optimized. 
Selection of variables affect convergence speed and can be cyclic, random, greedy, or according to other criteria \cite{Nutini2022}.

\subsection{Capacity Analysis}\label{DLcapacityanalysis}
The author in \cite{31Aswini2023} investigates the capacity for two users: one served by the \gls{BS} directly over a Rayleigh fading channel, and another user served through a reflective passive \gls{RIS}. 
The system includes a \gls{BS}, a direct user, and an indirect user.
Therefore, a common message and two private messages are sent by the \gls{BS}.
With \gls{RSMA}, and for \gls{SNR} values between 27dB and 40dB, they report higher capacity for the common message over the first private message, which in turn is higher than the second private message respectively. 
Before the 27dB to 40dB region, the rates are fairly close to each other. 
Furthermore, it is shown that, a reduced capacity results in the case of imperfect \gls{CSI}. The effect of varying the power splitting ratio of \gls{RSMA} on the rates of the three messages is also demonstrated.
The \gls{BS}-to-\gls{RIS} link uses Rayleigh fading. 
The placement of the \gls{RIS} is, however, not discussed.

\subsection{Outage Analysis} \label{section-outage-analysis}
The outage performance has been studied by many authors \cite{02Bansal2021miso,05Bansal2021,08Shambharkar2022,12Dhok2022,14Weinberger2022,20Shambharkar2022edge,26Bansal2023,34TianY2023,46Karim2023,54Singh2023,55Xie2023,56Mohamed2023}.
In addition, the secrecy outage was also examined by articles that study \gls{PLS} aspects \cite{52Yang2023,64Xiao2023}.
This classification is depicted in Fig.~\ref{fig:downlink_objectives}.
Another classification is based on the type of the \gls{RIS}, where two classes exist: reflective \gls{RIS} and \gls{STAR}-\gls{RIS}.
These two classes are examined next.

\subsubsection{Reflective RIS}
Many articles have examined the outage performance of \gls{RSMA} systems that are assisted by a reflective \gls{RIS} \cite{02Bansal2021miso,05Bansal2021,08Shambharkar2022,14Weinberger2022,20Shambharkar2022edge,26Bansal2023,34TianY2023,46Karim2023,52Yang2023,54Singh2023}.
The authors in \cite{02Bansal2021miso} analyzed the outage performance of cell-edge users using \gls{2L}-\gls{HRS} and found the performance better compared to the situations of no \gls{RIS} with \gls{2L}-\gls{HRS}, \gls{MU}-\gls{LP}, and \gls{1L}-\gls{RS} with \gls{RIS}.
Another article by mostly the same authors \cite{05Bansal2021} analyzed the outage performance of cell-edge users and near users using 1L-RS and derived closed-form expressions. 
They also show that the proposed system outperforms \gls{RIS}-assisted \gls{NOMA} and \gls{DF}-\gls{RS}.
Both articles utilize an on-off scheme to control the \gls{RIS} phase shifts.

The authors in \cite{08Shambharkar2022,20Shambharkar2022edge} consider discrete \gls{RIS} phase shifts. 
In \cite{08Shambharkar2022}, the authors derive the \gls{OP} for the discrete as well as the optimal phase shifts.
They observe that equal power allocation provides the best system performance. 
For instance, they use 4 users in the simulations, and that corresponds to a 0.2 power allocation to each stream, that is, 0.2 to the common stream and 0.2 to each of the other 4 streams.
The work in \cite{20Shambharkar2022edge} considers other details like asymptotic analysis as well as comparison to other multiple access techniques.
It also assumes Rayleigh channels instead of Nakagami-$m$ channels, and assumes \gls{MISO} instead of \gls{SISO}.
With \gls{RSMA}, \gls{ZFBF} provides lower outage probability compared to \gls{MRT}.
In addition, \gls{RSMA} outperforms \gls{NOMA}.

In \cite{14Weinberger2022}, a \gls{C-RAN} network is employed and the outage performance of two approaches for phase-shifts optimizations, namely, \gls{SCA} and \gls{SDP}, are considered.
It is noted that the \gls{SDP} approach has a better outage performance compared to the \gls{SCA} approach.
A \gls{CoMP} system is considered in \cite{34TianY2023} with on-off control on the \glspl{RIS}.
They assume some users are only served through one or more \glspl{RIS} because of blocked path to any access point.
They discuss the selection of the \glspl{RIS} along with the design of signalling.
They show improved outage performance of the opportunistic \gls{RIS} scheme in \gls{RSMA} and \gls{NOMA} when some users are blocked from directly communicating with any of the access points.
Comparisons of simulations and analytical expressions are also provided.

A vehicular network served by a \gls{UAV} is considered in \cite{26Bansal2023}.
Over the flying time of the \gls{UAV}, they derive the \gls{AOP} for the vehicle while considering co-channel interference.
Increasing the number of interfering vehicles significantly affects the performance.
They also propose an algorithm to minimize the sum-\gls{AOP} for selected vehicles.
The impact of channel estimation error on the outage probability is also examined.
Moreover, the \gls{RSMA} power allocation coefficients are optimized for the flight time.
Results show the advantage of the \gls{RIS} as well as the \gls{RSMA} over \gls{NOMA}.

In \cite{46Karim2023}, an approximate outage probability expression is derived for \gls{RSMA} and \gls{NOMA} after finding the \gls{PDF} of the \gls{SINR} for each user.
The derived \gls{PDF} is verified with Monte-Carlo simulations. 
Increasing the number of elements of the \gls{RIS} or using \gls{RSMA} instead of \gls{NOMA} are found to provide better performance.
In addition, they show better outage performance for \gls{RSMA} when the \gls{SNR} is relatively high, and a similar trend when the number of \gls{RIS} elements are increased for fixed transmit power. 
Moreover, they show that increasing the \gls{RIS} elements can compensate for small 
\gls{CSI} imperfection.

The authors in \cite{54Singh2023} examine the outage probability, among other performance measures, for a system with a relay and an \gls{RIS} mounted on a \gls{UAV}.
\gls{FBL} and \gls{IBL} in addition to imperfect \gls{SIC} at the users as well as \gls{RSI} at the \gls{UAV} are considered.
\gls{SIC} errors are found to negatively affect the outage performance at the users.
In addition, the \gls{DF} relaying link is better than the \gls{RIS} link for low transmission power.
\gls{BLER} expressions are also derived in \cite{54Singh2023}.

In \cite{52Yang2023}, the covert communication rate and \gls{DEP} are evaluated in a system of three users and a single transmitter.
\gls{RS} is conducted at the \gls{BS} to serve two \acrfullpl{LU}.
Moreover, an approximate expression for the \gls{SOP} is derived, and the asymptotic \gls{SOP} is analyzed.
Comparisons of simulation and analytical results are provided, with a small gap because they approximated the channel distribution by a gamma distribution.
The use of \gls{RIS} is found to lower the \gls{DEP} for the covert user, as well as increase the secrecy rate.
The \gls{DEP} shows a convex behaviour for increasing total transmission power, and an optimal transmission power should be used to get the lowest \gls{DEP}.
When the total power is low, the probability of mis-detection is the dominant term and it decreases with increasing power, resulting in a decrease of \gls{DEP}.
Then, the probability of false alarm becomes the dominant term and increases with increased total power, which results in an increase in \gls{DEP} again.

\subsubsection{STAR-RIS}
Several works have considered a \gls{STAR}-\gls{RIS} for the outage analysis \cite{12Dhok2022,55Xie2023,56Mohamed2023,64Xiao2023}.
In \cite{12Dhok2022}, the authors consider the \gls{ES} and \gls{MS} modes with continuous phase for their analysis where two users are served by each \gls{STAR}-\gls{RIS}.
They also consider that Rician fading are spatially correlated due to the proximity of cells of the \gls{STAR}-\gls{RIS}.
They derive the outage probability for the \gls{IBL} transmission.
In addition, they derive the \gls{BLER} and goodput for the \gls{FBL} system.
Capacity analysis and asymptotic outage probability are also presented.
Conclusions include a preference for alternating elements in the \gls{MS} mode.
Furthermore, the outage probability follows a convex-like shape, depending on whether the private message or the common message are in outage.
The invalid \gls{SINR} thresholds for the common message are highlighted.

An active \gls{STAR}-\gls{RIS} is considered in \cite{55Xie2023} with perfect and imperfect \gls{SIC}.
They derive the outage expressions for users and find the diversity order.
The diversity order is zero with imperfect \gls{SIC} due to residual interference, and it is related to the number of elements of the \gls{STAR}-\gls{RIS} when the \gls{SIC} is perfect.
In \cite{55Xie2023}, the \gls{EE} of the system with active \gls{STAR}-\gls{RIS} is compared to that with passive \gls{STAR}-\gls{RIS} in the delay-limited mode.
The system throughput for a fixed data rate can be related to the outage probability in that delay-limited mode.
They report that, despite the extra power requirement of the active \gls{STAR}-\gls{RIS}, it can still achieve overall better \gls{EE} for the system.

In \cite{56Mohamed2023}, the users are assumed to be distributed according to a Poisson point process (PPP), and stochastic geometry is utilized to study the outage probability of users in the transmit zone as well as the reflect zone of the \gls{STAR}-\gls{RIS}.
Closed-form expressions are found for the outage probability in terms of the Meijer G-functions.
Increasing the  density of the devices, the \gls{RS} power coefficient, or the number of elements in the \gls{RIS} are all found to improve the outage probability.

The authors in \cite{64Xiao2023} study the system secrecy performance.
They find a closed-form expression for the \gls{SOP} with Nakagami-$m$ fading.
Moreover, the asymptotic performance at high \gls{SNR} is also studied.
The effects of the number of elements of the surface, transmit power, and target secrecy rate on the system are examined.
In addition, comparisons to \gls{NOMA}, \gls{OMA}, and no \gls{RIS} cases are provided.
The choice of the power allocation factor is found to be a critical parameter.
If it is too low or too high, the \gls{SOP} is high.
Only for moderate values is the \gls{SOP} low. 
Ideally, the optimum value should be chosen which results in the lowest \gls{SOP}.

\subsection{Problem Formulations} \label{downlink_problem_formulations}
For the downlink system with \gls{1L}-\gls{RS}, assume that the \gls{BS} is sending a message $m_k$ to the $k$th user. 
The \gls{BS} splits each message $m_k$ into a private part $m_{k,p}$ and a common part $m_{k,c}$, as depicted earlier in Fig.~\ref{fig:rs1l_draw}.
Hence, there is a total of $2K$ messages. 
The \gls{BS} then combines all common messages into a single common message $m_c$. 
Hence, the total number of messages is reduced to $K+1$.
The single common message is encoded into a common stream $s_c$, and each private message is encoded into a private stream $s_k$.
Subsequently, all $K+1$ messages are superimposed in the power domain by the precoder to form the transmitted signal
\begin{equation}
	\mathbf{x} = \mathbf{p}_c s_c + \sum_{k=1}^{K} \mathbf{p}_k s_k
\end{equation}
where $\mathbf{p}_c$ is the precoding vector applied to the common message, and $\mathbf{p}_k, \forall k \ne 0$ is the precoder vector applied to the private message.
Note that $\mathbf{p} \in \mathbb{C}^{N_B}$.
In other words, all antennas are used to transmits all $K+1$ messages.
Furthermore, a common assumption is to assume that the symbols are uncorrelated, that is $\mathbb{E}[|s_k|^2]=1$.
In addition, the power is limited by the power budget $\left|\left| \mathbf{p}_c \right|\right| + \sum_{k=1}^{K} \left|\left| \mathbf{p}_k \right|\right| \le P$

In each time slot of a discrete-time system, the received signal at the $k$th user is 
\begin{equation}
	y_k = \left( \mathbf{g}_k^H + \mathbf{h}_k^H \mathbf{\Phi} \mathbf{H} \right) \mathbf{x} + n_k \quad \forall k = 1,2,\dots,K
\end{equation}
Here the received signal is based on the model \lq (c) BS-RIS-DorS\rq, where each user gets a direct link and a reflected link through the \gls{RIS}.
The direct link is represented by $\mathbf{g}_k \in \mathbb{C}^{N_B}$, and the reflected link is represented by $\mathbf{h}_k^H \mathbf{\Phi} \mathbf{H}$,
where $\mathbf{H} \in \mathbb{C}^{N_R \times N_B}$ is the channel from the \gls{BS} to the \gls{RIS},
$\mathbf{\Phi} \in \mathbb{C}^{N_R \times N_R}$ is the matrix of phase shifts introduced by the \gls{RIS},
$\mathbf{h}_k \in \mathbb{C}^{N_R}$ is the channel from the \gls{RIS} to the $k$th user.
In addition, $n_k$ is the additive noise at the $k$th user.

Other formulations follow in a similar way. Then, relevant conditions are imposed based on the objectives and assumptions in each paper.
For instance, the phase shift matrix of the \gls{RIS} can be assumed to be diagonal, block diagonal, or full.
In addition, the structure of $\mathbf{\Phi}$ depends on the assumption of the array type, e.g. a \gls{ULA} or a \gls{UPA}.
The next step is to optimize some utility based on a specific model. This section discusses multiple optimization objectives that exist in the reviewed literature.
A classification of the articles based on the optimization objective can be found in Fig.~\ref{fig:downlink_objectives}.
In addition, the same information can be found in the \lq Objectives\rq~ column of Table~\ref{tab:optimizations-across-papers}.
In particular, seven subsections are now presented: 
\acrfull{WSR} maximization,
\acrfull{EE} maximization,
\acrfull{MR} maximization,
\acrfull{RE} maximization, 
transmit power minimization,
\acrfull{SR} maximization, 
as well as other optimization objectives.

\subsubsection{Maximize \acrlong{WSR}}
The \acrfull{WSR} maximization problem is the most widely studied objective \cite{06Jolly2021,25Singh2022,33Tian2023,37Soleymani2023signal,38Sena2023,39Elganimi2023,54Singh2023,61LiB2023,07Fang2022,09Weinberger2022csi,29Soleymani2023workshop,36Kim2023,42Li2023,57Wang2023multiF,63Huroon2023,67Meng2024,68Pala2024}.
In \cite{25Singh2022}, a \gls{UAV} carries an \gls{RIS} and a relay.
Three modes are considered, \gls{RIS} mode, relay mode, and hybrid mode.
For the \gls{RIS} mode, they use \gls{AO} to optimize the \gls{UAV} position and the \gls{RIS} phase shifts.
The former is done through a one-dimensional search, and the latter through a modified \gls{RCG} method.
The authors investigate the effect of \gls{SIC} errors and show an advantage of \gls{RSMA} over \gls{NOMA} by a maximum of 16\%.
They also conclude that the hybrid mode of the \gls{UAV} is better than either the \gls{RIS} or the \gls{DF} relay alone.
The same authors provide a more detailed study in \cite{54Singh2023} where they consider \gls{BCD} method to decompose the problem into three blocks: power allocation, phase shift optimization, and \gls{UAV} position optimization.
The other variables are fixed in each block.
Power allocation follows a heuristic approach, while phase shifts optimization follows a method based on \gls{RCG}.
\gls{IBL} and \gls{FBL} are also considered.
Since selection combining is used, the hybrid mode is found to surpass either the relay or the \gls{RIS} modes with \gls{IBL} or \gls{FBL} in terms of \gls{WSR}.
Nevertheless, for large number of \gls{RIS} elements, the \gls{RIS}-aided mode provides similar or better performance to the relay-aided mode.
The effect of imperfect \gls{SIC} on reducing the achievable \gls{WSR} is also examined.

The authors in \cite{61LiB2023} use a \gls{SAA} and \gls{WMMSE} methods to obtain the average rate which they relate to the long-term \gls{WESR}. 
Then a \gls{BCD} method is used to solve the resulting problem which is convex in one variable if others are fixed. 
Specifically, the algorithm alternates between the transmit power, the \gls{RIS} phase shifts, and the common rates of users. 
The algorithm has a finite lower bound, and thus is guaranteed to converge. 
Simulation results show convergence within 10 iterations. 
In addition, considering inaccurate estimation of \gls{CSI}, they show a consistent advantage of \gls{RSMA} over \gls{NOMA} and \gls{SDMA}.

A multi-antenna system is considered in \cite{58Ge2023}, where the users have more than one antenna.
They utilize \gls{MMSE} for precoder design and use \gls{PGAM} for \gls{RIS} phase shift optimization.
Furthermore, they study the perfect and statistical \gls{CSI} scenarios.
A maximum sum-rate deterioration of 34\% was observed with imperfect \gls{CSI}.
Furthermore, using 4 or 6 antennas at the users are recommended due to reduced complexity, while using 8 antennas reduces the sum rate due to interference among the antennas.

The authors in \cite{40Huang2023} and the later journal article \cite{41Huang2023j} consider maximizing the rate of video streaming.
Specifically, they consider that multiple users are requesting tiles of a 360-degree video stream.
Different users can request different tiles of the video.
The bitrate of these tiles, the \gls{RS} power allocation coefficients, \gls{RIS} phase shifts, and beamforming vectors are jointly optimized.
They exploit \gls{DDPG}, actor-critic, and imitation learning to propose the `Deep-GRAIL algorithm'.
They also propose a \gls{DNN} module for policy learning in their algorithm, which includes a \gls{DCO} layer.
The report better performance of their algorithm compared to the cases of no \gls{RS}, \gls{RS} with \gls{AO} algorithm, as well as \gls{RS}-\gls{NOUM} with \gls{AO}.
They also compare to an algorithm that uses \gls{SL} to train the \gls{DNN} module.
Despite the superiority of their algorithm, the \gls{RIS}-aided \gls{RSMA} system with \gls{AO} can attain performance `close-enough' to that of their proposed algorithm, surpassing the other \gls{SL} benchmark approach as well.

A less popular way of solving the optimization problem of \gls{DL} \gls{MU}-\gls{MISO} system with \gls{RSMA} and \gls{STAR}-\gls{RIS} is to utilize the fairly recent \gls{PPO} family of \gls{DRL} \cite{67Meng2024}.
By properly designing the reward function, state space, and action space, a \gls{PPO}-based algorithm was shown not only to have good convergence behaviour, but also enhance the sum rate compared to \gls{SAC}, \gls{DDPG}, \gls{GA}, 
\gls{MRT}, \gls{ZF}, and the random solution; for any maximum available power between 1W and 30W \cite{67Meng2024}.  
Moreover, after studying the learning rates of all the 36 combinations of $K\in\{2,3,4\}$ users, $M\in\{10,20,30,40\}$ elements, and $P\in \{40\mbox{ dBm}, 43\mbox{ dBm}, 45\mbox{ dBm}\}$, the authors show that increasing the available power causes a slow improvement in the training reward which is attributed to the simultaneous increase in interference \cite{67Meng2024}. 
Although the convergence behaviour is consistent among all these scenarios, it is interesting to note that the case with $K=3$ users generally gets higher reward for a low number of training episodes. Then the case of $K=4$ users surpasses the former case anywhere between 6000 and 8000 training episodes.

In \cite{68Pala2024}, they aim to maximize the throughput for a specified \gls{PEP}.
The study is primarily an extension of \cite{23Katwe2022shortPacket} to consider multiple \glspl{RIS}.
The authors propose a three-stage \gls{AO} algorithm to optimize beamforming, block-length, and \gls{RIS} phase shifts.
In addition, another optimization formulation is presented to minimize the maximum block length.
This formulation is an extension to the main formulation and is targeted at minimizing the worst-case latency while maintaining the rate and power requirements.
The impact of channel estimation errors is also studied.
Using a few \glspl{RIS} with low number of elements is better than a single \gls{RIS} with large number of elements due to the additional paths.
Moreover, relaxing the \gls{PEP} requirement enhances the rate and it gets closer to the \gls{IFBL} case.
In other words, reliability can be traded for throughput.
Furthermore, \gls{RSMA} with \gls{FBL} can achieve rates similar to those of \gls{NOMA} and \gls{SDMA} with \gls{IFBL}.

The authors of \cite{39Elganimi2023} consider the overloaded system and show that the same sum-rate can be achieved at 5dB-less (depending on the number of \gls{RIS} elements) when the \gls{RIS}-assisted \gls{RSMA} system is used compared to \gls{RSMA} alone.
They include comparisons of the \gls{TP}-\gls{RSMA} and \gls{PP}-\gls{RSMA} for the two groups of users in their system model.
One group is served by \gls{OMA} while the other is served by \gls{RSMA}.

\subsubsection{Maximize \acrlong{EE}} \label{maxEE}
The \acrfull{EE} maximization problem is similar to the \gls{WSR} maximization problem, but with a term in the denominator that accounts for the power consumption.
Several works studied this objective \cite{01Yang2020,14Weinberger2022,16Chen2022,21Gao2022,23Katwe2022shortPacket,45Zhang2023,66Tang2024}.
The authors in \cite{01Yang2020} aim to maximize the \gls{EE} under a minimum rate constraint.
They employ \gls{AO} to solve the subproblems of beamforming and \gls{RIS} phase shifts' optimizations, and the subproblems are solved using \gls{SCA}.
The authors show that \gls{RSMA} provides a higher \gls{EE} compared to \gls{OFDMA} or \gls{NOMA}. 
However, the difference of performance between \gls{NOMA} and \gls{RSMA} is not huge (up to 12\%).
Three \glspl{BS}, three \glspl{RIS} with $4$ elements each, and three users are used in their simulation study.

The \gls{C-RAN} network is considered in \cite{14Weinberger2022}. The authors note that they aim to dynamically allocate user clusters and sets of common messages. 
Static sets are assumed for the latter in \cite{01Yang2020} which do not fully exploit the integration of the \gls{RIS} and \gls{RSMA} \cite{14Weinberger2022}.
In the \gls{EE} maximization in \cite{14Weinberger2022}, they consider the fronthaul capacity and power constraints of each \gls{RRU}, which clusters of \glspl{RRU} serve which users, passive \gls{RIS} phase shifts, as well as \gls{QoS} constraints for users.
An \gls{AO} algorithm is employed to solve the problem.

In \cite{16Chen2022}, a \gls{THz} model is considered; the \gls{RIS} is assumed to be a \gls{UPA} and its phase shifts are uniformly quantized.
They utilize \gls{SSA} algorithm and show that it provides a better \gls{EE} in less time compared to a \gls{SCA} algorithm.
The results generally show an advantage for \gls{RSMA} over \gls{SDMA} or \gls{NOMA}, but not in all cases.

An \gls{AO} algorithm is presented in \cite{21Gao2022} for the \gls{EE} maximization, and it shows better performance compared to their benchmarks. 
Fractional programming is first employed to get rid of the power terms in the denominator.
They compare the performance to the benchmark cases of no \gls{RIS}, random \gls{RIS} phases, and fixed \gls{RIS} phases.
When increasing the number of \gls{RIS} elements, the \gls{EE} first increases and then decreases. Hence, increasing the number of \gls{RIS} elements does not always increase the \gls{EE}.
Moreover, the \gls{EE} of the system is the lowest when the \gls{RIS} is not close to either the transmitter or receivers.
Furthermore, the \gls{EE} increases logarithmically for increasing power in decibels.

In \cite{23Katwe2022shortPacket}, the objective is to maximize the \gls{EE} with \acrfullpl{FU} only reachable via the \gls{RIS}, while ensuring that the rates satisfy a given error probability. They proceed with an \gls{AO} algorithm for both \gls{SPC} and \gls{LPC} settings. 
The algorithm performs better than random phase shifts,
and also validate the convergence.
Next, they show that \gls{RIS}-aided \gls{RSMA} \gls{SPC} not only outperforms its \gls{NOMA} and \gls{SDMA} counterparts in terms of \gls{EE}, 
but also can surpass  \gls{RIS}-aided \gls{NOMA} \gls{LPC} for packet sizes above 1500 bits, as well as surpass \gls{RIS}-aided \gls{SDMA} \gls{LPC} for packet sizes above 2500 bits.
This implies that very short packet length are less energy efficient compared to \gls{LPC}, which is true for all of \gls{SDMA}, \gls{NOMA}, and \gls{RSMA}.
They also illustrate the benefit of the passive beamforming at the \gls{RIS} to enhance the gain for both \glspl{NU} and \glspl{FU}.

A \gls{PPO}-based \gls{DRL} method is the main focus of \cite{45Zhang2023} in a \gls{SWIPT} network. 
It is a non-alternating algorithm, that is, it does not alternate between phase shift and beamforming optimizations.
In the optimization, they consider discrete phase shifts, the non-linear \gls{EH} model, power budget, and \gls{QoS} constraints for information and energy transfer.
The beamforming vectors, \gls{RIS} phase shifts, common rates of \gls{RSMA}, and \gls{PS} ratios at the receivers are jointly optimized.
They also present another solution based on \gls{SCA} as a benchmark, where the constraint on discrete phase shifts is relaxed.
\gls{RSMA} is found to enhance the \gls{EE} of the system. 
In addition, the \gls{PPO}-based approach only loses small \gls{EE} in exchange of large time savings. This can be very important to quickly arrive at the near-optimum result in time-varying channels.
For the \gls{RIS} position, they show that it is not always best to keep the \gls{RIS} near the transmitter or the receiver, rather, in a specific case it is optimal to keep the \gls{RIS} in the middle.

A multi-cell system is considered in \cite{66Tang2024} where they focus on enhancing the performance of cell-edge users.
An \gls{AO} framework is utilized where an \gls{SCA} method is employed for beamforming optimization and \gls{SDR} along with a penalty function are used for phase shifts optimization.
Increasing the minimum rate demand is found to reduce the \gls{EE}.
In addition, increasing the transmit power reaches a saturation point where further increase does not result in additional \gls{EE}.
Moreover, the \gls{EE} increases when the number of elements in the \gls{RIS} increase.
Nonetheless, it seems that further increase would result in reduction in the \gls{EE} as observed by \cite{21Gao2022}.

\subsubsection{Maximize \acrlong{MR}}
\gls{MR} maximization is another optimization objective that aims to maximize the rate of the user having the lowest rate.
This objective embeds some fairness in the main objective.
It can also be coupled with other constraints like minimum energy.
A few works have considered this objective \cite{03Fu2021,15Zhao2022,19Lima2022,24Pang2022,27Li2023maxmin,44Wu2023,53LiuP2023,62Zhao2023}.

In \cite{03Fu2021}, the authors study minimum rate maximization and consider discrete \gls{RIS} phase shifts.
They avoid the traditional \gls{SCA} algorithm to avoid the challenging task of finding the initial feasible point, and use a penalized \gls{SCA} which can start form an arbitrary point.
Comparison with \gls{NOMA} is not provided, but other benchmarks are used which show the advantage of their algorithm.

The authors in \cite{24Pang2022} propose an \gls{AO} algorithm for the double-\gls{RIS} system. As baselines, they compare to \gls{SDMA} and \gls{NOMA} and single \gls{RIS}, and they show that using double-\gls{RIS} achieves better max-min rate compared to all previous baselines.
The optimization consists of 4 subproblems:
\begin{enumerate}
	\item Optimization of precoders and the common rate given the \gls{PS} ratio $\rho_k$ and the \gls{RIS} matrices $\mathbf{\Phi}_1$ and $\mathbf{\Phi}_1$ which is done through \gls{SDR} and \gls{SCA}.
	\item Optimization of \gls{PS} ratio $\rho_k$ given the precoders, common rates, and \gls{RIS} matrices, via a closed-form solution.
	\item Optimization of the phases of the first \gls{RIS} utilizing \gls{SDR} and \gls{SCA}.
	\item Optimization of the phases of the second \gls{RIS} utilizing \gls{SDR} and \gls{SCA}.
\end{enumerate}
The authors explain that iterations are non-decreasing and capped by the maximum available power which guarantees convergence.

The authors of \cite{44Wu2023} utilize the achievable rate of the worst \gls{UE} as the loss function in the end-to-end training of their proposed \gls{DFAPN} and \gls{RRN}.
Their \gls{DFAPN} is model-driven, and is utilized to get key parameters in the iteration of their \gls{AWMMSE} precoding scheme.
The \gls{RRN} is utilized to design a frequency-flat \gls{RIS} matrix despite the frequency-selective channel.
The \gls{RRN} is based on the emerging transformer structure, and it extracts the correlation features of the \gls{OFDM} channel.
Furthermore, the emerging transformer structure is also utilized for \gls{CSI} acquisition.
Correlation features are also extracted from the multiple subcarriers of \gls{CSI} to reduce feedback and pilot overhead.

\subsubsection{Maximize \acrlong{RE}}
The main objective of this \gls{RE} maximization formulation is to consider the tradeoff between \gls{EE} and \gls{SE}. 
This formulation is used in \cite{43Niu2023}. 
Furthermore, a similar formulation that considers both \gls{EE} and \gls{SE} is proposed in \cite{49Liu2023} where a multi-objective problem is formulated.
In this paper, we expand the meaning of \gls{RE} to also include the formulation of \cite{49Liu2023} under one header.
In \cite{43Niu2023}, the objective function is a weighted linear combination of the \gls{EE} and the \gls{SE}.
Their algorithm has two-stages. The problem is prepared and simplified in the first stage and a closed-form solution is found.
Next the second stage carries out the \gls{AO} over the beamforming and the phase shifts and gains of the active \gls{RIS}.
In addition, they extend the optimization to the \gls{2L}-\gls{RS} scheme.
The gains and phase shifts of the \gls{RIS} are considered to be quantized.
In the simulation study, different schemes are compared including \gls{NOMA}, \gls{SDMA}, \gls{FD}-\gls{RIS}, group-connected \gls{RIS}, as well as passive \gls{RIS}.
They also show that the performance of the quantized scheme is close to that of the continuous scheme.

The authors in \cite{49Liu2023} propose a joint optimization scheme for the non-convex fractional problem of optimizing \gls{EE} and \gls{SE} with \gls{RSMA} and other constraints like \gls{QoS} guarantees for \glspl{PU}; based on \gls{DC} and \gls{SCA} techniques. 
They show a consistent advantage of their scheme over \gls{NOMA}, \gls{SDMA}, random and constant precoding. 
The performance of \gls{NOMA} is sometimes close to their proposed scheme. 
What is interesting is that using random or consistent precoding offers an advantage over the case without the \gls{RIS} surface. 
They also show the trade-off between \gls{EE} and \gls{SE} where the proposed scheme maintains better \gls{EE} for values of \gls{SE} between 10 and 20 bps/Hz compared to other schemes.

\subsubsection{Minimize Transmit Power}
Instead of maximizing the \gls{SE}, \gls{EE} or \gls{MR}, the formulation could address the transmit power directly.
Four articles consider this formulation \cite{04Weinberger2021,22Camana2022,32Darabi2023,48Khisa2023}.
The authors in \cite{04Weinberger2021} minimize the transmit power under \gls{QoS} constraints and find that the use of \gls{RS} with \gls{RIS} yields more power savings than the individual power savings by each technique.

A \gls{SWIPT} system is examined by \cite{22Camana2022} where they aim to reduce the transmission power subject to \gls{EH} and rate constraints.
They decompose the non-convex problem into a bi-level problem. The outer level is solved with \gls{GA} and the inner level is solved via \gls{SDR} with a penalty method.
In addition, an \gls{AO} algorithm is implemented to show its inferior performance and extra complexity compared to the \gls{GA}-based algorithm.
\gls{SDMA} and \gls{NOMA} are considered for comparison in perfect and imperfect \gls{CSI}.
It is observed that as the \gls{EH} requirements increase, the gain of \gls{RSMA} over \gls{SDMA} increases.
Moreover, the \gls{RIS}-aided \gls{RSMA} system is found to achieve the lowest power consumption compared to the \gls{SDMA} or \gls{NOMA} systems.
Furthermore, \gls{RIS}-assisted \gls{NOMA} system becomes impractical as the rate requirements increase, because \gls{NOMA} requires multiple layers of \gls{SIC}, which in turn requires the \gls{BS} to use more power.

An active \gls{RIS} is considered in \cite{32Darabi2023} for \gls{URLLC} transmission.
They aim to minimize the power consumption at the \gls{BS} as well as the active \gls{RIS}.
An \gls{AO} algorithm is presented where the \gls{RSMA} optimization employs Taylor approximation, fractional programming, and \gls{SCA}.
Moreover the \gls{RIS} optimization employs the big-M formulation, \gls{DC}, and fractional programming. 
Overall, the authors found that the active \gls{RIS}-assisted system consumes less power and requires less number of elements compared to its passive counterpart.
In addition, the \gls{RSMA} system consumes less power compared to the \gls{SDMA} system.

A cooperative system is considered in \cite{48Khisa2023} where the near user re-transmits the common message to the far user.
This introduces two additional design parameters compared to the common formulation: the time split ratio, and the transmission power by the near user.
In addition, the first and second time slots could have different phase shifts at the \gls{RIS}.
The objective is to minimize the transmit power of the \gls{BS} in addition to minimizing that of the near user.
The authors focus on two users and show lower energy consumption of the \gls{RIS}-assisted cooperative-\acrshort{RSMA} compared to cooperative-\acrshort{NOMA} (with and without \gls{RIS}), regular \gls{RIS}-assisted \gls{1L}-\gls{RSMA}, \gls{RIS}-assisted \gls{NOMA} and cooperative-\acrshort{RSMA} without \gls{RIS}.

\subsubsection{Maximize \acrlong{SR}}
Three papers consider this objectives where the secrecy rate is to be maximized \cite{17Hashempour2022,28Gao2023,59Lotfi2023}.
In \cite{17Hashempour2022,28Gao2023}, the objective is to maximize the minimum secrecy rate, and in \cite{59Lotfi2023}, it is maximizing the rate of the covert user.
Furthermore, in \cite{28Gao2023}, \gls{AO} is used, where in the first alternation, the \gls{BS}'s beamformers are optimized (along with the noise that is assumed unknown). 
The noise here refers to the exploitation of the common message to create artificial noise at the eavesdropper. 
In the other alternation, the \gls{RIS} phase shifts are optimized.
They also show that the proposed \gls{RIS}-assisted \gls{RSMA} system attains the same or higher \gls{MMF} multiplexing gain compared to both \gls{NOMA} and \gls{MU}-\gls{LP}.

\subsubsection{Other Optimization Objectives}
This category includes three papers as depicted in Fig.~\ref{fig:downlink_objectives}, namely \cite{37Soleymani2023signal,38Sena2023,65Chen2024}.
The authors in \cite{37Soleymani2023signal} proposed a general framework that considers different objectives including \gls{EE} or \gls{WSR} maximization.
Although the authors focus on a downlink system, the framework can be extended to other scenarios.
They consider realistic assumptions like devices suffering from \gls{IQI}, and they employ \gls{IGS} as an additional interference management technique along with \gls{RS}.
In addition, they consider the reflective \gls{RIS} and also present an extension to the \gls{STAR}-\gls{RIS}.
An \gls{AO} formulation is presented where the precoder optimization is done through \gls{MM} and they utilize lower bound surrogate functions for the \gls{RIS} optimization where three feasibility sets are considered.
They compare the performance with \gls{PGS}, \gls{TIN}, and the case without \gls{RIS}.
They show the advantage of \gls{RS} with \gls{IGS} in improving the \gls{EE} and \gls{SE} of the overloaded \gls{DL} system. 

The authors in \cite{38Sena2023} follow a non-traditional approach where they utilize a modified variant of \gls{RSMA} that encodes the common stream in a polarization orthogonal to that of the private streams.
The benefit is that the users can decode the private symbols without \gls{SIC}.
However, since the channel can cause cross-polar interference, the authors develop a \gls{MOOP} and solve it using a Frank-Wolfe (FW)-based method.
The precoding design is explained with three main matrices. 
The first two are for cancelling the inter-group interference and directing the group transmission to the intended group, while the third is for reducing intra-group interference.
Next the optimization of the dual-polarized \gls{RIS} is presented through a FW-based method.

The \gls{RIS}-assisted \gls{RSMA} \gls{ISAC} system is considered in \cite{65Chen2024}, where the authors aim to maximize the radar \gls{SNR}. 
Typical constraints for \gls{1L}-\gls{RS} are considered, namely, decodable non-negative common rate and a \gls{QoS} constraint, where each user is guaranteed a total rate above some threshold.
In addition, power limitation at the \gls{BS} is considered in addition to unit-norm constraint on the \gls{RIS}'s phase shifts matrix.
An \gls{AO} framework is presented which utilize \gls{MM} and \gls{DC} approximation.
The authors show that the \gls{RSMA} \gls{ISAC} system almost matches the performance of the radar-only system regardless of the use of the \gls{RIS}. 
Of course, the \gls{RIS} provides an independent gain in both systems, and that gain is linear with increasing number of \gls{RIS} elements.
This is in contrast to the \gls{SDMA} \gls{ISAC} system, which achieves lower radar \gls{SNR}.

Multiple optimization objectives have been studied in the literature as presented in this section.
The objectives include \gls{WSR} maximization, \gls{EE} maximization, \gls{MR} maximization, \gls{RE} maximization, \gls{SR} maximization, and transmit power minimization.
The next subsection provides an example of \gls{WSR} maximization, followed by a subsection that draws a few lessons to conclude this downlink section, then the next section (Section~\ref{uplink_resource_allocation}) provides an overview of works studying uplink channel models.

\subsection{Example of \acrlong{WSR} Maximization}\label{exampleScenario}
\begin{figure}
	\centering
	\includegraphics[width=\columnwidth]{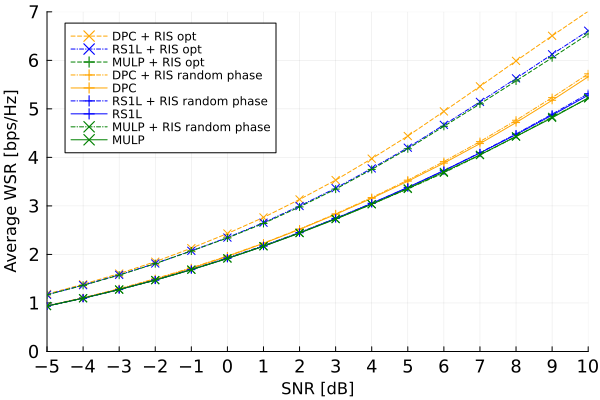}
	\caption{Performance of \gls{MU}-\gls{LP} and \gls{1L}-\gls{RS} with and without \gls{RIS} with varying \gls{SNR}. \gls{DPC} is added for comparison.}
	\label{fig:WSR_vs_SNR}
\end{figure}

\begin{figure}
	\centering
	\includegraphics[width=\columnwidth]{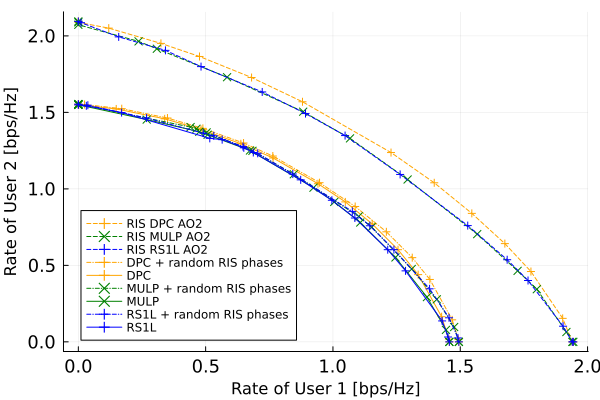}
	\caption{Rate region of \gls{MU}-\gls{LP} and \gls{1L}-\gls{RS} with and without \gls{RIS} at 0dB \gls{SNR}. \gls{DPC} is added for comparison.}
	\label{fig:WSR_vs_weights}
\end{figure}

The previous section described multiple optimization objectives including \gls{WSR} maximization, \gls{EE} maximization, transmit power minimization, etc.
This section provides an example of \gls{WSR} maximization for an \gls{RIS}-assisted \gls{RSMA} downlink \gls{MISO} system that employs \gls{1L}-\gls{RS}.
A common type of \gls{RS}, known as \gls{1L}-\gls{RS} will be explored in this section, along with comparisons to \gls{SDMA} and \gls{NOMA}. 
In addition, two benchmarks are briefly explained, namely, \acrfull{DPC} and \gls{MU}-\gls{LP}. 
In Fig.~\ref{fig:WSR_vs_SNR} and Fig.~\ref{fig:WSR_vs_weights}, we provide sample performance results for the optimization algorithms when 
perfect \gls{CSIT} is assumed and model \lq b) BS-RIS-D\&S\rq~is adopted. 
The model is very similar to that of \cite{Guo2020} but with two users to draw the rate region in Fig.~\ref{fig:WSR_vs_weights}. 
We utilize a nested alternating optimization formulation to solve the resource allocation problem. 
In particular, the \gls{RIS} phase shifts optimization is almost the same as presented in \cite{Guo2020}, while the \gls{MU}-\gls{LP} and \gls{1L}-\gls{RS} follow the transformation of the sum-rate maximization problem into an equivalent \gls{WMMSE} minimization problem as described in \cite{Christensen2008} and \cite{Mao2018} respectively.
A modification of the codes published in \cite{Guo2020} and \cite{Zhao2019} is utilized in these simulations.
The simulation parameters are listed in Table~\ref{tab:optimizations-parameters}.
Fig.~\ref{fig:WSR_vs_SNR} shows two sets of traces, the upper set corresponds to the nested \gls{AO}, while the lower set is without \gls{RIS} optimizations.
In particular, the lower set shows the cases of random \gls{RIS} phase shifts and the case of no \gls{RIS}.
It is evident that the \gls{RIS} can provide a decent gain.
In addition, since perfect \gls{CSIT} is assumed, \gls{MU}-\gls{LP} is already providing a very good performance.

\begin{table}
	\caption{Parameters for the simulations in this article}
	\label{tab:optimizations-parameters}
	\centering
	\begin{tabular}{|l|l|}
		\hline
		Parameter                                & Value                   \\ \hline
		Number of \acrshortpl{BS}, $M$           & $1$                     \\
		Antennas per \acrshortpl{BS},  $N_B$     & $2$                     \\
		Location of the \acrshortpl{BS}          & $(0,0)$                 \\ \hline
		Number of \acrshortpl{RIS}, $L$          & $1$                     \\
		\acrshortpl{RIS} elements ,  $N_R$       & $100$                   \\
		Location of the \acrshortpl{RIS}         & $(200,0)$               \\ \hline
		Number of users ,  $K$                   & $2$                     \\
		Antennas per user , $N_U$                & $1$                     \\
		Location of user 1                       & $(205.65,34.48)$        \\
		Location of user 2                       & $(193.47,30.24)$        \\ \hline
		\acrshortpl{BS}-\acrshortpl{RIS} channel & Rician                  \\
		\acrshortpl{RIS}-user channel            & Rician                  \\
		\acrshortpl{BS}-user channel             & Rayleigh                \\
		Rician Factor ,  $\kappa$                & 10                      \\
		Random Channels                          & 100                     \\
		\acrshort{LoS} Path Loss ,  $L(d)$       & $35.6+22\log_{10}(d)$   \\
		\acrshort{NLoS} Path Loss ,  $L(d)$      & $32.6+36.7\log_{10}(d)$ \\
		Noise                                    & $-170+10\log_{10}(B)$   \\
		Bandwidth,   $B$                         & $180$kHz                \\ \hline
	\end{tabular}
\end{table}

\subsubsection{Dirty Paper Coding}
In a \gls{MU}-\gls{MIMO} system, when the \gls{CSI} is known at the transmitter, \gls{DPC} achieves the capacity region of the \gls{MIMO}-\gls{BC} channel \cite{Weingarten2006,Caire2003}. 
Nevertheless, implementing \gls{DPC} is computationally complex and not practical.
As such, it can be considered as a benchmark for other multiple access schemes in \gls{MU} networks.
This section briefly describes the procedure of precoder optimization with \gls{DPC} \cite{Viswanathan2003}.
Users are encoded sequentially in \gls{DPC} ensuring that the latter users are not interfered by the previous users.
Hence, all possible permutations of the users should be considered, and the rate region is the union of all rate regions with each permutation.
The rate region of the \gls{BC} channel can be characterized through characterization of the dual \gls{MAC} channel.
Building on top of \eqref{eq:y_k_general}, the dual \gls{MAC} channel for the $k$th user is $\mathbf{H}_k^\mathsf{H}$.
Let $\mathbf{Q}_k$ denote the transmit covariance matrix of the $k$th user, and $P_k$ denote the power budget of the $k$th user
Assuming the user weights vector $\mathbf{u}$ is sorted from largest to lowest weight, the \gls{WSR} can be expressed as \cite{Viswanathan2003}
\begin{align}  \label{dpcWSR}
	\text{WSR} 	&=  \underset{\mathbf{Q}_{k}}{\max}\quad f(\mathbf{Q}_k) \\
	& \text{s.t.}\quad  \sum_{k=1}^{K} \text{tr}\left( \mathbf{Q}_{k} \right) = P
\end{align}
where 
\begin{equation} 
	f(\mathbf{Q}_k) = \sum_{i=1}^{K} (\mathbf{u}_i - \mathbf{u}_{i+1}) \log \det \left\{ \mathbf{I} + \sum_{l=1}^{i} \mathbf{H}_l^\mathsf{H} \mathbf{Q}_l \mathbf{H}_l \right\}
\end{equation}
and the derivative with respect to the $j$th covariance matrix $\mathbf{Q}_j, 1\le j \le K$ is
\begin{equation} 
	\nabla f_j(\mathbf{Q}_k) = \sum_{i=j}^{K} (\mathbf{u}_i - \mathbf{u}_{i+1}) \left[ \mathbf{H}_j \left\{ \mathbf{I} + \sum_{l=1}^{i} \mathbf{H}_l^\mathsf{H} \mathbf{Q}_l \mathbf{H}_l \right\}^{-1} \mathbf{H}_j^\mathsf{H}\right].
\end{equation}
To simplify the expression, with an abuse of the notations, we assume that $\mathbf{u}_{K+1} = 0 $.
The algorithm in \cite{Viswanathan2003} updates the $\mathbf{Q}_k$'s iteratively. 
In the next two equations, the left-hand side represents the value at the new iteration:
\begin{align}
	\mathbf{Q}_{j^*} &= t^* \mathbf{Q}_{j^*} + (1-t^*) P \mathbf{v}_{j^*} \mathbf{v}_{j^*}^\mathsf{H} \\
	\mathbf{Q}_{i} &= t^* \mathbf{Q}_{i}, \qquad \forall i \ne j^*,
\end{align}
where $j^* = \arg \max(\lambda_1, \dots, \lambda_K)$, $\lambda_j$ is the principal eigenvalue of the gradient $\nabla f_j$, and $\mathbf{v}_{j}$ is the corresponding principal eigen vector.
In addition, $t^*$ is found by the line search:
\begin{equation}
	t^* =  \underset{0 \le t \le 1}{\arg \max} f\left(t \mathbf{Q}_1 \dots, \mathbf{Q}_{j^*} + (1-t) P \mathbf{v}_{j^*} \mathbf{v}_{j^*}^\mathsf{H}, \dots, t \mathbf{Q}_K\right)
\end{equation}

\subsubsection{Multi User Linear Precoding}
The objective of this formulation is to maximize the \gls{WSR} in \eqref{mulpWSR}. The achievable rate for the $k$th user is
\begin{equation} 
	r_k = \log \det \left( \mathbf{I}_k + \mathbf{P}_{k}^\mathsf{H} \mathbf{H}_{k}^\mathsf{H} \mathbf{R}_{n_k n_k}^{-1} \mathbf{H}_{k} \mathbf{P}_{k} \right),
\end{equation}
assuming Gaussian signalling. With the user weights vector $\mathbf{u}$ and power budget $P$, the \gls{WSR} maximization problem can be written as

\begin{align}  \label{mulpWSR}
	\left[ \mathbf{P}_{1}, \dots, \mathbf{P}_{K}  \right] &=  \underset{\mathbf{P}_{1}, \dots, \mathbf{P}_{K}}{\arg \max} \sum_{k=1}^{K} \mathbf{u}_k r_k \\
	& \text{s.t.}\quad  \sum_{k=1}^{K} \text{tr}\left( \mathbf{P}_{k} \mathbf{P}_{k}^\textsf{H} \right) = P.
\end{align}

The constrained problem in \eqref{mulpWSR} can be solved by introducing two auxiliary variables $\mathbf{A}_k$ and $\mathbf{W}_k$ to relate a \gls{WMMSE} minimization problem to this \gls{WSR} maximization problem.
Note that these two matrices can be used directly to solve the \gls{WSR} maximization problem in the following alternating fashion:
\begin{equation}
	\mathbf{P}_k \Rightarrow \left[\begin{array}{c}
		\mathbf{A}_{k}\\
		\mathbf{W}_{k}
	\end{array}\right]
	\Rightarrow
	\mathbf{P}_k 
	\qquad \forall k.
\end{equation}
That is, once the first $\mathbf{P}_k$ is initialized, $\mathbf{A}_{k}$ and $\mathbf{W}_{k}$ can be found by equations \eqref{mulpA} and \eqref{mulpW}, respectively. 
After that, a new $\mathbf{P}_k$ can be found using equation \eqref{mulpP}.
The algorithm stops when the relative rate error among the last two iterations is small, say $1\times 10^{-6}$.
The \gls{WMMSE} matrices $\mathbf{A}_{k}$ and $\mathbf{W}_{k}$ are detailed in \cite{Christensen2008}.
The expressions for $\mathbf{A}_{k}$, $\mathbf{W}_{k}$, and $\mathbf{P}_k$ are now listed.
The \gls{MMSE} receive filter is
\begin{equation} \label{mulpA}
	\mathbf{A}_k = \mathbf{P}_{k}^\mathsf{H} \mathbf{H}_{k}^\mathsf{H} \left( \mathbf{H}_{k} \mathbf{P}_{k} \mathbf{P}_{k}^\mathsf{H} \mathbf{H}_{k}^\mathsf{H} + \mathbf{R}_{n_k n_k} \right)^{-1},
\end{equation}
and the \gls{MMSE} matrix is
\begin{equation}
	\mathbf{E}_{k} = \left( \mathbf{I}_k + \mathbf{P}_{k}^\mathsf{H} \mathbf{H}_{k}^\mathsf{H} \mathbf{R}_{n_k n_k}^{-1} \mathbf{H}_{k} \mathbf{P}_{k}  \right)^{-1},
\end{equation}
where the noise covariance matrix is 
\begin{equation}
	\mathbf{R}_{n_k n_k} = \mathbf{I}_k + \sum_{i=1,i\ne k}^{K} \mathbf{H}_{k} \mathbf{P}_{i} \mathbf{P}_{i}^\mathsf{H} \mathbf{H}_{k}^\mathsf{H}.
\end{equation}
Furthermore, the \gls{MMSE} weight matrix is
\begin{equation} \label{mulpW}
	\mathbf{W}_{k} = \mathbf{u}_k \mathbf{E}_{k}^{-1}.
\end{equation}
Finally, defining $\mathbf{P} = \left[ \mathbf{P}_{1}, \dots, \mathbf{P}_{K}  \right] \in \mathbb{C}^{N_B \times N_U K}$, 
$\mathbf{H} = \left[ \mathbf{H}_{1}^\top, \dots, \mathbf{P}_{H}^\top  \right]^\top \in \mathbb{C}^{N_U K \times N_B}$,
$\mathbf{A} = \text{diag}\left( \mathbf{A}_{1}, \dots, \mathbf{A}_{K}  \right) \in \mathbb{C}^{N_U K \times N_U K}$,
$\mathbf{W} = \text{diag}\left( \mathbf{W}_{1}, \dots, \mathbf{W}_{K}  \right) \in \mathbb{C}^{N_U K \times N_U K}$
the precoder can be found by
\begin{equation} \label{mulpP}
	\mathbf{P} = b \overline{\mathbf{P}},
\end{equation}
where
\begin{equation}
	b = \sqrt{\frac{P}{\text{tr}\left( \overline{\mathbf{P}} \overline{\mathbf{P}}^\mathsf{H}\right)}},
\end{equation}
is a gain scaling factor and 
\begin{equation}
	\overline{\mathbf{P}} = \left( \mathbf{H}^\mathsf{H} \mathbf{A}^\mathsf{H}  \mathbf{W} \mathbf{A} \mathbf{H} + \frac{\text{tr} \left( \mathbf{W} \mathbf{A} \mathbf{A}^\mathsf{H} \right)}{P} \mathbf{I}_{N_B} \right)^{-1} \mathbf{H}^\mathsf{H} \mathbf{A}^\mathsf{H}  \mathbf{W}.
\end{equation}

\subsubsection{1-Layer Rate Splitting}
With \gls{RS}, just like \gls{MU}-\gls{LP}, we utilize the conversion of the sum-rate maximization problem into an equivalent \gls{WMMSE} problem \cite{Mao2018}.
Note that a globally-optimum algorithm for a 2-user \gls{RS} \gls{MISO}-\gls{BC} was proposed \cite{Matthiesen2022}, 
but the difference to the \gls{WMMSE} formulation was reported to be small.
Compared to the \gls{MU}-\gls{LP} \gls{WSR} maximization problem in \eqref{mulpWSR}, the optimization formulation gets two new conditions. 
The first is that the sum of the allocated common rates must not exceed the achievable common rate at the worst user, that is
\begin{equation} \label{1lrsDecodable}
	\sum_{k=1}^{K} c_k \le r_c,
\end{equation}
where $r_c = \min\{r_{1,c}, \dots, r_{K,c}\}$, and $c_k$ is the common rate allocated to user $k$.
The second condition is 
\begin{equation} \label{1lrsNonNegative}
	\mathbf{c} \ge 0,
\end{equation}
which may be dropped if the algorithm is guaranteed to be restricted to non-negative values.
The optimization can also be solved in an alternating fashion after relating the \gls{WSR} maximization problem to a \gls{WMMSE} minimization problem as described in \cite{Mao2018}.

\subsection{Lessons Learned}\label{DLlessons}

The following points summarize the lessons learned for downlink resource allocation in RIS-aided RSMA systems.
\begin{itemize}
	\item When the common message is in outage, both messages are in outage. But the converse is not true.
	This leads to convex-like shape of the outage probability for both the transmit user and the reflect user in the \gls{STAR}-\gls{RIS}-aided system, where there is an optimum value of the common rate split ratio \cite{12Dhok2022,08Shambharkar2022,20Shambharkar2022edge}.
	\item A similar convex-like shape is also observed for the secrecy outage probability in \cite{64Xiao2023} because either of the legitimate users will be negatively affected at very high or very low values of the common rate splitting ratio.
	\item The use of fully-connected \gls{RIS} with \gls{RSMA} can improve the sum rate compared to the single-connected \gls{RIS} \cite{07Fang2022,43Niu2023}.  
	\item Using \gls{RIS} with quantized bits in the \gls{RSMA} system is more energy efficient even though the \gls{SE} is slightly reduced due to quantization \cite{43Niu2023}.
	\item The use of double-\gls{RIS} (in cascade) can achieve better max-min rate compared to \gls{SDMA}, \gls{NOMA}, and a single \gls{RIS} with \gls{RSMA}. \cite{24Pang2022}
	\item The \gls{EE} increases with increasing number of \gls{RIS} elements up to a certain point. After that, further increase in the number of \gls{RIS} elements results in a decrease of the \gls{EE} \cite{21Gao2022}.
	\item The \gls{EE} is the lowest when the \gls{RIS} is in the middle between the transmitter and users, i.e. when the \gls{RIS} is not close to either of them.
	\item In downlink \gls{RIS}-assisted \gls{HD} \gls{RSMA} system with two users, if the user with the better channel cooperates to deliver the common stream to the other user with less-favourable channel conditions, the overall energy consumption of the system will be lower compared to traditional \gls{RIS}-assisted \gls{1L}-\gls{RSMA}, \gls{RIS}-assisted cooperative-\gls{NOMA},  as well as \gls{HD} \gls{RSMA} without \gls{RIS}. \cite{48Khisa2023}
	\item In a downlink system with a single transmit antenna with transmissive \gls{RIS}  (TRIS) of $N$ sub-arrays, the performance is similar to that of a conventional $N$-antenna system but at a cheaper cost and reduced complexity. \gls{RSMA} consistently outperforms \gls{NOMA} and \gls{SDMA} in such a system \cite{61LiB2023}.
	\item \gls{RSMA} with \gls{FBL} can achieve rates similar to those of \gls{NOMA} and \gls{SDMA} with \gls{IFBL} \cite{68Pala2024}.
	\item \gls{SPC} can benefit from \gls{RIS}-assisted \gls{RSMA} networks compared to those with \gls{NOMA} and \gls{SDMA} in terms of \gls{EE}. In fact, \gls{RIS}-assisted \gls{RSMA} \gls{SPC} can surpass \gls{NOMA} and \gls{SDMA} \gls{LPC} for some packet sizes \cite{23Katwe2022shortPacket}.
	\item The \gls{ISAC} system can benefit from \gls{RSMA} where both communications and sensing can be done simultaneously and achieve a radar \gls{SNR} similar to that achievable by the radar alone, as illustrated in \cite{65Chen2024}. 
	Furthermore, the \gls{RIS} provides an extra independent gain.
	\item An active \gls{RIS} could achieve a lower overall system power consumption and require less number of elements compared to its passive counterpart \cite{32Darabi2023}.
	\item The use of an active \gls{STAR}-\gls{RIS} may be more energy efficient for the full system \cite{55Xie2023}. 
	Despite the fact that the active \gls{STAR}-\gls{RIS} consumes more power than its passive counterpart.
	\item Increasing the number of \gls{STAR}-\gls{RIS} elements, or the \gls{RS} power splitting factor, or the density of users are all found to reduce the outage probability \cite{56Mohamed2023}.
	\item For a system with two legitimate users and an eavesdropper (or warden), the detection error probability first decreases then increases as the total transmit power is increased at the transmitter.
	This can be attributed to the initial decrease of the probability of missed detection, followed by the increase of the probability of false alarm \cite{52Yang2023}.
	\item \gls{SIC} errors cause noticeable degradation in the outage performance of the \gls{RIS} assisted or relay-assisted channels from the \gls{BS} to the users \cite{54Singh2023}.
	The \gls{SIC} errors also reduce the achievable \gls{WSR} \cite{25Singh2022,54Singh2023}.
	\item Using selection combining, a hybrid \gls{RIS}-assisted and \gls{DF}-relay-assisted link outperforms either of the individual systems \cite{54Singh2023,25Singh2022}.
	\item In an \gls{RSMA} system with a dedicated \gls{RIS} to each cell-edge user, \gls{ZFBF} is found to outperform \gls{MRT} \cite{20Shambharkar2022edge}.
	\item In \gls{SWIPT} systems, using \gls{RSMA} is a good idea to deliver power to the \gls{IoT} devices, since the common stream is delivered to all devices.
	In addition, the \gls{RIS} can establish a virtual \gls{LoS} link between the transmitter and the device. Improved \gls{EE} in \gls{RIS}-assisted \gls{RSMA} \gls{SWIPT} systems has been reported \cite{45Zhang2023}.
	\item In a \gls{SWIPT} system with \gls{EH} and rate requirements, increasing the rate requirements renders the \gls{RIS}-assisted \gls{NOMA} system impractical compared to the \gls{SDMA} and \gls{RSMA} counterparts in terms of transmission power \cite{22Camana2022}. 
	The reason is that \gls{NOMA} requires multiple layers of \gls{SIC} which require more transmission power from the \gls{BS}.
	\item The \gls{RIS}-assisted \gls{RSMA} system maintains the best performance in underloaded and overloaded scenarios when \gls{SR} maximization is considered \cite{28Gao2023}. 
\end{itemize}

\section{Uplink Resource Allocation}
\label{uplink_resource_allocation}

\begin{figure}[!t]
	\centering
	\includegraphics[width=0.45\textwidth]{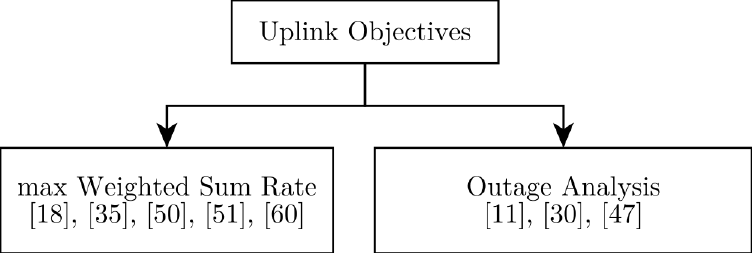}
	\caption{Classification of the uplink papers based on their focus.}
	\label{fig:uplink_objectives}
\end{figure}

A few papers \uplinkarticles~considered the uplink channel models. 
These papers are included in all tables in this document.
Details of the models were presented earlier in Fig.~\ref{fig:models06-uplink}. 
For Table~\ref{tab:optimizations-across-papers}, the abbreviations for the `Constraints' column follow the same definitions as presented at the beginning of Section \ref{downlink_resource_allocation}.
Fig. \ref{fig:uplink_objectives} shows a classification of the objectives of the uplink papers.
In this section, an overview of these uplink articles is presented. 
Then, some open challenges are discussed in the next section.

\subsection{Maximize \acrlong{WSR}}
Five articles considered \gls{WSR} maximization in the uplink setting \cite{18Katwe2022clustring,35Katwe2023uplink,50Sun2023,51Katwe2023,60Hua2023}.
A group of four authors focused on the uplink system in three papers \cite{18Katwe2022clustring,35Katwe2023uplink,51Katwe2023}.
A \gls{mmWave} model is considered in \cite{18Katwe2022clustring}, where \gls{RSMA} was shown to provide up to 40\% higher sum-rate compared to \gls{NOMA}, and up to 80\% compared to \gls{OMA}. 
To reduce interference among users, they consider clusters of users where \gls{RSMA} is only used within the cluster. 
Clusters with higher user density are allocated larger bandwidth to ensure fairness. 
In addition, a multi-layer \gls{RS} model is assumed, where each user can split its message into multiple streams.
In \cite{35Katwe2023uplink}, clustering is not considered, but they derive a near-optimal decoding order. 
The authors note that the ascending order of the users' split proportions, as well as the descending order of channel conditions can be used to form a decoding order that attains better sum-rate.
A multi-layer \gls{RS} was considered in \cite{51Katwe2023}, that is, each user can split its message into more than two parts. 
Then, the decoding order is decided as the first step in the optimization process.

In \cite{50Sun2023}, two users are assumed and only one of them does \gls{RS} while the other does not. 
This is because, for $K$ users, only $K+1$ streams are needed to achieve the capacity of the \gls{MAC} channel \cite{Rimoldi1996}. 
The objective is to maximize the rate of the second user which does \gls{RS}, while maintaining a minimum rate for the first user.
Note that this objective can still be considered a \gls{WSR} maximization since the rate of user 1 is a constraint.
In addition, the rate of the first user is optimized first to get the initial values of beamforming and \gls{RIS} phase shifts.
The optimization is carried out using \gls{AO} in three steps to allocate the power at the transmitter, the beamforming at the receiver, and the phase shifts of the \gls{RIS}.
The authors show that the \gls{RIS}-assisted \gls{RSMA} system achieves a better rate compared to other schemes with \gls{NOMA}, without \gls{RIS}, or with a suboptimal \gls{AO} approach.
However, the performance of the second best case, the \gls{RIS}-assisted \gls{NOMA} system, closes the gap and approaches the performance of \gls{RIS}-assisted \gls{RSMA} system as the number of antennas at the \gls{BS} or the number of elements at the \gls{RIS} is increased.
In \cite{60Hua2023}, each user splits its message into two streams, and they propose a \gls{DDPG} approach to jointly optimize the beamforming, power allocation, and \gls{RIS} phase shifts to achieve higher sum rate.
The problem is formulated as a Markov decision process, and they also propose a \gls{SAS} process for constraint violations.
An important justification for using \gls{DDPG} is that they consider a time-varying channel due to user mobility, instead of the common quasi-static channel model. 
The \gls{DDPG} is used to learn the \gls{CSI} in real time and maximize the sum rate in the long term.
Another justification is that, traditional optimization algorithms do not scale well as the number of \gls{RIS} elements increase.
It was observed that the sum rate increases linearly with $\log(N_R)$ using the proposed \gls{DDPG} algorithm.
They compare performance with a \acrfull{CE}-based scheme, a \gls{PPO}-based scheme, and a local search scheme over quantized \gls{RIS} phase shifts.

\subsection{Outage Analysis}
A cognitive-\gls{RSMA} system is explored in \cite{11LiuP2022,30You2023,47Liu2023cognitive} where an \gls{RIS} is used to facilitate the uplink transmission of the \gls{SU}.
An important objective is to maintain the \gls{QoS} of the \gls{PU}, thus, no interference should be introduced by the \gls{SU}.
For that, an interference threshold sent by the \gls{BS} is assumed, and three cases are detailed in \cite{11LiuP2022} based on that interference threshold.
An \gls{OP} expression is derived for the \gls{SU} for each of the three cases.
Through simulations, the authors show that this system has better outage performance compared to a \gls{NOMA}-based system, as well as a cognitive \gls{RSMA} system without \gls{RIS}.
A similar journal article also focuses on cognitive-\gls{RSMA} \cite{47Liu2023cognitive} where only one user (the \gls{SU}) does \gls{RS}, and a continuous range of \gls{RIS} phase shifts is considered. 
The authors derive the outage probability for the \gls{SU}. 
Furthermore, it is noted that considering more than two users as well as discrete \gls{RIS} phase shifts are possible future directions.
A quite similar article to the previous two (i.e. \cite{11LiuP2022,47Liu2023cognitive}) is \cite{30You2023}. 
Similar articles without an \gls{RIS} appeared earlier \cite{Liu2020,Liu2022a}.

\subsection{Lessons Learned}

The following summarizes the lessons learned for uplink RIS-aided RSMA systems.
\begin{itemize}
	\item The sum rates increase linearly with $\log(N_R)$ for an uplink system with the \gls{DDPG} algorithm proposed in \cite{60Hua2023}.
	\item For a 2 user uplink system where a single user does \gls{RS}, the \gls{RIS}-assisted \gls{NOMA} system is generally the second best after the \gls{RIS}-assisted \gls{RSMA} system, and the gap between them diminishes as the number of \gls{RIS} elements or the \gls{BS} antennas are increased.
	\item A cognitive \gls{RIS}-assisted \gls{RSMA} system can achieve better outage performance compared to a cognitive \gls{RIS}-assisted \gls{NOMA} system or a cognitive \gls{RSMA} system \cite{11LiuP2022,30You2023,47Liu2023cognitive}.
\end{itemize}
\section{Open Challenges and Research Directions} 
This section briefly discusses some directions for future research and highlights some challenges.
In particular, it discusses modelling the \gls{RIS}, the number of receive antennas, \acrfull{ISAC}, Holographic \gls{MIMO}, and \gls{SIC}-free decoding.

\subsection{RIS models}

Increasing the number of elements in the \gls{RIS} is expected to increase performance, e.g. the \gls{SNR}.
Nevertheless, larger surfaces require revised and detailed models to capture physical phenomena properly, which allows better optimization for the required objective.
Simple \gls{RIS} models may not be physically consistent.
In addition, they may overlook some characteristics that can be exploited by the multiple access technique.
More examination of \gls{2D} \glspl{RIS} \cite{Tsilipakos2020} models is expected, as they may provide a better chance of redirecting or refracting the incident beam into the desired direction.
The next paragraphs highlights mutual coupling, and nonlinearity of the \gls{RIS}.

Mutual Coupling is a natural phenomenon that can occur in mechanical or electrical systems such as antenna arrays.
When the elements of the \gls{RIS} are extremely close to each other on the sub-wavelength level, the response of an element cannot be unaffected by the response of its neighbours.

Frequency modulation through the \gls{RIS} surface remains an open direction, compared to the majority of studies that explore amplitude and phase modulation.
For instance, an incident signal at $\theta_1$ with a frequency $f_1$ can leave the surface at $\theta_2$ with a different frequency $f_2$ \cite{Tsilipakos2020}.

Channel estimation needs to be addressed in a way that does not require the use of two coherence blocks, as in \cite{09Weinberger2022csi} where the same common message is sent twice in two coherence blocks in the \gls{ORS} scheme.

Furthermore, traditionally, the \gls{RIS} is assumed to have a linear operation, whereas in practice it may not always behave linearly. 
Nonlinearity can be leveraged to modulate the incident signal before reflecting it \cite{Yuan2021}.
The effect of the \gls{RIS}'s nonlinearity on the performance of the communications system based on multiple access techniques like \gls{RSMA} remains to be explored.
This can be useful, for instance, in the uplink, where the \gls{BS} can distinguish the signal from the far-user because it arrives at two different frequencies due to the modulation incurred at the \gls{RIS} \cite{Yuan2021}. 
This implies bandwidth expansion which may introduce new challenges in multi-user interference management \cite{Yuan2021}.
A joint design scheme may be formulated to increase the \gls{SE} or \gls{EE}.
From here, the tunability of the interference-decoding level by the \gls{RS} can provide a synergistic gain in terms of \gls{SNR} at the receiver. 
In addition, it can be utilized to combat Doppler shifts resulting from user mobility \cite{Bjoernson2022} or the \gls{RIS}'s mobility if mounted on moving platform like a \gls{UAV}.

\subsection{Holographic Surfaces}
\gls{mMIMO} \cite{Marzetta2010}
and \gls{mmWave} \cite{Rappaport2013}
are envisioned to be fundamental in the next generation wireless systems, such as \gls{6G}. 
This may require densification of the cells \cite{Saad2020} 
which will introduce newer challenges such as managing inter-cell interference.
In addition, with large and/or dense apertures, it is essential to consider mutual-coupling between antenna elements in addition to near-field scattering.
Perhaps a more fundamental issue is that \gls{mMIMO} is costly and power-hungry due to the need of a large number of \gls{RF} chains.
A novel technology called \emph{Holography} may contribute to meeting the requirements of future communications systems \cite{Gong2023}.
Holography allows recording and reconstruction of \gls{EM} wavefronts \cite{Gong2023}.
Metamaterials \cite{Chen2016} 
can be a viable option to realize Holographic effects.
\gls{HMIMO} surfaces combine the capabilities of holography and \gls{mMIMO}, and is expected to be a more efficient implementation of large antenna systems \cite{Gong2023}.
\gls{HMIMO} can have elements with sub-wavelength spacing allowing pencil-like beams \cite{Gong2023}. 
Therefore, the surface provides a nearly-continuous aperture and can induce almost continuous phase changes to the \gls{EM} wave \cite{Gong2023}.
This analog-domain signal processing can be realized through \glspl{LWA} \cite{Araghi2021}
or \glspl{TCA} instead of the bulky and costly \gls{RF} devices \cite{Zhou2018}. 
Two works \cite{49Liu2023,61LiB2023} have examined a transmissive \gls{RIS} in an \gls{RSMA} system, but they do not examine the details of the interactions between the transmissive \gls{RIS} surface and the transmitter.
Further examination of Holographic surfaces whether at the transmitter, at the receiver, or in between is an open direction with its own challenges.

\subsection{RIS-Aided MIMO}
Most of the existing works on \gls{RIS}-aided \gls{RSMA} systems consider single-antenna users. 
We identified two papers that assumed multiple antennas at the users in the downlink \cite{29Soleymani2023workshop,58Ge2023}.
Further examination of the \gls{RIS}-assisted \gls{RSMA} multi-antenna users downlink systems, let alone uplink systems is still an open research direction.
Recent studies have examined \gls{MIMO}-\gls{BC} \gls{RSMA} systems \cite{Hao2017,GholamiDavoodi2020,Flores2020,Mishra2022}.
The \gls{DoF} are characterized in \cite{Hao2017, GholamiDavoodi2020} where the former establishes the achievability of the \gls{DoF} region while the latter proves the optimality.
Furthermore, Regularized Block Diagonalization (RBD) precoders are proposed in \cite{Flores2020}.
Moreover, \cite{Mishra2022} examines precoder design based on \gls{WMMSE} transformation, in addition to \gls{DoF} analysis and flexibility in the number of common streams that can be allocated.
Nevertheless, \cite{Mishra2022} assumes that all users have the same number of antennas.
These developments are yet to be studied in the \gls{RIS}-assisted \gls{RSMA} system.
In the \gls{RIS}-assisted \gls{RSMA} system, two papers \cite{29Soleymani2023workshop,58Ge2023} explored that direction where they assumed multi-antennas at the users.
Both utilize \gls{AO}. In addition, the former assumes that receivers suffer from \gls{IQI} and use \gls{IGS} to compensate for that;
While the latter \cite{58Ge2023} develops an \gls{MMSE} precoder and tests the performance in perfect and statistical \gls{CSI} scenarios.
However, the works do not support flexibility in the number of allocated common streams.

\subsection{Less Reliance on SIC}
This section might seem counter-intuitive, which it is.
The fact that \gls{RSMA} requires users to do \gls{SIC}, at least once, might lead to error propagation when the \gls{SIC} is imperfect.
If the \gls{SIC} step can be skipped in some scenarios or setups, then the expected loss in the gain of \gls{RSMA} due to imperfect \gls{SIC} can be lowered
One way to accomplish that is using a dual-polarized \gls{RSMA} and a dual-polarized \gls{RIS} \cite{38Sena2023}.
The reason is that orthogonal signals do not interfere unless depolarized by the channel or hardware errors.
The authors in \cite{38Sena2023} study this phenomena for the \gls{RIS}-aided \gls{RSMA} system and demonstrate its efficiency in mitigating the depolarization effect.
In particular, they utilize the  dual-polarized \gls{RIS} to nullify the depolarized streams. 
They assume that the \gls{RIS} is close enough to the users that no further depolarization is incurred in the channel from the \gls{RIS} to the users.
Applying this idea to \gls{STAR}-\gls{RIS} is still unexplored. In addition, other ideas to reduce the reliance on \gls{SIC} are still an open field.

\subsection{Cooperative Systems}
The cooperative \gls{RSMA} system was considered in \cite{Mao2020coopRS} and extended in \cite{48Khisa2023} to include the \gls{RIS} in the system.
However, as highlighted in \cite{48Khisa2023}, their investigation is limited to the two user scenario, and multi-user investigation is an open direction.
In particular, they proposed mapping users into pairs and assigning an \gls{RIS} to each pair.
At any rate, more possibilities of pairing or grouping exist and the use of the \gls{2L}-\gls{HRS} may also be investigated. 

\subsection{Integrated Sensing and Communications in an RIS-Aided RSMA System}
Sensing, for example by radar, has been in active development in the past, just like communications. 
Sensing and communications have been mostly developed independently despite sharing signal processing algorithms, and many equipment \cite{Zhang2022isac}.
For example, cognitive radio is a type of joint communications and sensing where secondary systems are installed with possibly complicated infrastructure to sense the spectrum and utilize it while the primary system is idle \cite{Zhang2022isac}, and despite the complications, the performance of secondary systems cannot be guaranteed.
cognitive radio systems have been explored in the \gls{RIS}-assisted \gls{RSMA} systems in \cite{11LiuP2022,30You2023,47Liu2023cognitive,49Liu2023} as mentioned previously.
Nevertheless, recent efforts aim to integrate communications and sensing while minimizing the interference among them.
Integration means using the same signal for both communications and sensing.
The latter approach of interference mitigation was found to be limited \cite{Zhang2022isac}. 
The \gls{RIS}-assisted \gls{RSMA} system could help in managing the interference in this approach.
In addition, it may also be useful in the integrated system. 
For instance, a simple scheme may involve a special design on the waveform of the common signal of the \gls{RS} for the purposes of sensing, while using that for communications at the same time.
The \gls{RIS} could play a role in distinguishing users or suppressing certain types or directions of interference.
This idea and similar ones are yet to be explored.
What is more, \gls{ISAC} can be further integrated with covert communications, where the direction of the Warden is sensed by the transmitter before transmitting to the legitimate user.
Then, the transmitter maximizes the interference at the warden \cite{Ma2023}.

\subsection{Large-Scale system Analysis for RIS-aided RSMA Systems}
Exploring the performance of the \gls{RIS}-assisted \gls{RSMA} system in large scale, perhaps using stochastic geometry \cite{Nguyen2007,Baccelli2010} might be informative.
Stochastic geometry provides an average view of the performance of the network from a macro level.
Although \gls{RSMA} is a good strategy for interference management in the cell, the interference of the common messages from neighbouring \glspl{BS} or from a large number of users in the same cell may need to be studied.
The interplay with multiple \glspl{RIS} is essential in providing a full understanding of the large-scale \gls{RIS}-assisted \gls{RSMA} system.
In \cite{56Mohamed2023}, the outage probability was found for a \gls{STAR}-\gls{RIS} system with help of stochastic geometry tools.
A comparison of the performance to \gls{CSMA} \cite{Alfano2014}, for example, or \gls{NOMA} \cite{Zhang2022stoc,Hou2019} may provide new insights.
In addition, there are other directions for realistic systems without stochastic geometry.
For instance, the current literature explored \glspl{UAV}, yet one direction is studying the movements of both users and the \glspl{UAV} simultaneously as well as dynamic positioning adjustments of the \glspl{UAV}.
Another direction would be to examine non-terrestrial systems or \gls{LEO} satellite systems.

	\section{Conclusion}
	The RIS-assisted RSMA system can improve the energy efficiency, spectral efficiency, and other utility functions while keeping user requirements in communications, power delivery, sensing, etc.,
	thanks to the robustness of RS against errors in SIC and channel estimation.
	However, the fact that SIC is required at the receivers, even in the simplest RSMA scheme (i.e. 1L-RS) might pose a challenge in its adoption in large-scale networks.
	This is because of the possibility of compounding and propagating errors, where the device will be in outage when it fails to decode the common rate.
	The current literature explores a wide range of possible system setups, mostly in the downlink; and span a range of resource allocation methods mostly built upon the alternating optimization framework, as well as fewer papers on machine learning and deep learning formulations.
	Comparisons to NOMA, SDMA, decode-and-forward relays, among others have been discussed.
	In addition, integration with paradigms such as SWIPT, ISAC look promising.
	Furthermore, newer directions start appearing like SIC-free RIS-assisted RSMA system.
	This paper concludes with many possible research directions that  would be useful in fully understanding the performance of RIS-assisted RSMA systems in real-world systems with more physically-consistent models.

	\IEEEpubidadjcol

\bibliographystyle{IEEEtran}

	\vfill
	
\end{document}